\documentclass{emsart}
\usepackage{makeidx}
\usepackage[dvips]{graphicx}

\usepackage{psfrag}

\makeindex

\usepackage{bm}

\contact[ettore.minguzzi@unifi.it]{E. Minguzzi, Dipartimento di
Matematica Applicata, Universit\`a di Firenze,  Via S. Marta 3,
I-50139 Firenze, Italy} \contact[sanchezm@ugr.es]{M. S\'anchez,
Departamento de Geometr\'{\i}a y Topolog\'{\i}a Facultad de
Ciencias, Avda. Fuentenueva s/n. E-18071 Granada, Spain}

\newtheorem{theorem}{Theorem}[section]

\newtheorem{lemma}[theorem]{Lemma}
\newtheorem{proposition}[theorem]{Proposition}

\theoremstyle{definition}
\newtheorem{definition}[theorem]{Definition}
\newtheorem{remark}[theorem]{Remark}

\newcommand{\R}{\mathbb{R}}

\newcommand{\lambdas}{s}

\newcommand{\p}{\partial}

\newcommand{\dd}{{\rm d}}

\newcommand{\be}{\begin{equation}}
\newcommand{\ee}{\end{equation}}
\newcommand{\noi}{\noindent}
\newcommand{\ben}{\begin{enumerate}}
\newcommand{\een}{\end{enumerate}}
\newcommand{\bit}{\begin{itemize}}
\newcommand{\eit}{\end{itemize}}
\newcommand{\edoc}{\end{document}}

\newcommand{\g}{\bm{g}}
\newcommand{\gam}{\bm{\gamma}}
\newcommand{\dl}{d}
\newcommand{\Tau}{{\cal T}}

\font\ddpp=msbm10  at 11 truept
\def\R{\hbox{\ddpp R}}

\def\L{\hbox{\ddpp L}}
\def\S{\hbox{\ddpp S}}
\def\Z{\hbox{\ddpp Z}}

\def\N{\hbox{\ddpp N}}

\title[The causal hierarchy of spacetimes]{The causal hierarchy of spacetimes\thanks{Contribution to the Proceedings of the ESI semester ``Geometry of
Pseudo-Riemannian Manifolds with Applications in Physics'' (Vienna,
2006) organized by D. Alekseevsky, H. Baum and J. Konderak. To
appear in the volume `Recent developments in pseudo-Riemannian
geometry' to be published as ESI Lect. Math. Phys., Eur. Math. Soc.
Publ. House, Z\"urich, 2008.}}

\author[E. Minguzzi and M. S{\'a}nchez]{E. Minguzzi and M. S\'anchez}

\begin{document}

\begin{abstract}
 The full causal ladder of spacetimes is constructed,
 and their updated main properties are developed.
Old concepts and alternative definitions of each level of the
ladder are revisited, with emphasis in minimum hypotheses. The
implications of the recently solved
``folk questions on smoothability'', and 
alternative proposals (as recent {\em isocausality}), are also
summarized.

\end{abstract}



\maketitle

{\footnotesize

\tableofcontents

 }

\section{Introduction}


Causality is an essential specific tool of Lorentzian Geometry,
which appears as a fruitful interplay between relativistic
motivations and geometric developments. Most of the goals of this
theory are comprised in the so-called {\em causal hierarchy of
spacetimes}: a ladder of spacetimes sharing increasingly better
causal properties, each level with some specific results. This
ladder and its main features were established at the end of the
70's, after the works of Carter, Geroch, Hawking, Kronheimer,
Penrose, Sachs, Seifert, Wu and others (essentially, the last
introduced level was in \cite{hawking74}) and were collected in the
first version of Beem-Ehrlich book (1981) ---later re-edited with
Easley, \cite{beem96}. Nevertheless, there are several reasons to
write this revision. A first one is that the ``folk questions on
smoothability'' of time functions and Cauchy hypersurfaces, which
were left open in that epoch, have been solved only recently
\cite{bernal03, bernal04, bernal06}. They affect to two levels of
the ladder in an essential way ---the equivalence between two
classical definitions of {\em stable causality} and the structure of
globally hyperbolic spacetimes. Even more, new results which fit
typically on some of the levels, as well as some new viewpoints on
the whole ladder,
have been developed in the last years. So, we think that the full
construction of the ladder from the lowest level to the highest
one, may clarify the levels, avoid redundant hypotheses and
simplify reasonings.

This paper is organized as follows. In Section \ref{s2} the
typical ingredients of Causality are introduced: time-orientation,
conformal properties, causal relations, maximizing properties of
causal geodesics... Most of this introductory material is
well-known and is collected in books such as \cite{beem96,
hawking73, oneill83, penrose72, wald84}. Nevertheless, some
aspects may be appreciated by specialists, as the introduction of
globally hyperbolic neighborhoods (Theorem \ref{fund}), the
viewpoint of causal relation $I^+, J^+, E^+$ in $M\times M$
(Subsect. \ref{subfurther}), or the conformal properties of
lightlike pregeodesics (Theorem \ref{tinvconf}). The conformal
invariance of some elements is stressed, even notationally (Remark
\ref{rconv}).

In Section \ref{s3} the causal ladder is constructed.  The nine
levels are  developed  in subsections, from the lowest
 (non-totally vicious) to the highest one (globally hyperbolic).
Essentially, our aims for each level are: (a) To give natural
alternative definitions of the level (see, for example,
Definitions \ref{ddist} or \ref{dccontinuous} and further
characterizations), with minimum hypotheses (see Definitions
\ref{dcsim} or \ref{dgh}, with Prop. \ref{s4ll}, Remark
\ref{rdiscgh-strongc}). (b) To check its strictly higher degree of
specialization, in a standard way. (c) To explain geometric
techniques or specific results of the level (for example, see
Theorems \ref{ttotvic}, \ref{tviciohiperbolico},
\ref{closedtimelikestatic} or Subsections \ref{sec-causcont},
\ref{secVolume}). In particular, we emphasize that only after the
solution of the folk questions on smoothability, the classical
characterization of causal stability in terms of the existence of
a time function can be regarded as truly equivalent to the natural
definition (see Theorem \ref{testcaus} and its proof). Even more,
we detail the consequences of these folk questions for the
structure of a globally hyperbolic spacetime (Theorem
\ref{tgerochsmooth}). Although the description of the smoothing
procedure lies out of the scope of the present review (see
\cite{sanchez05b}, in addition to the original articles
\cite{bernal03, bernal04, bernal06}), the main difficulties are
stressed, Remark \ref{r2cojones}.

Finally, in Section \ref{s4} we explain briefly the recent
proposal of {\em isocausality} by Garc\'{\i}a-Parrado and
Senovilla \cite{garciaparrado03}. This yields a partial ordering
of spacetimes which was expected to refine the total order
provided by the standard hierarchy. Even though, as proven later
in \cite{garciaparrado05b},  this ordering does not refine exactly
the standard one, this is an alternative viewpoint,  worth to be
born in mind.

\section{Elements of causality theory} \label{s2}
Basic references for this section are
\cite{beem96,hawking73,misner73,oneill83,penrose72,wald84}, other
useful references will be
\cite{candela06,carter71,garciaparrado05b,garciaparrado05,kronheimer67,leray52,sorkin96}.

\subsection{First definitions and conventions}
\begin{definition} A {\em Lorentzian manifold} \index{Lorentzian manifold} is a  smooth manifold $M$,
of  dimension $n_0\geq 2$, endowed with a non-degenerate
 metric $g:M \to T^{*}M\otimes T^{*}M$ of signature $(-,+,\dots,+)$.
\end{definition}
By {\em smooth}  $M$ we mean $C^{r_0}$, $r_0\in \{3, \dots,
\infty\}$. Except if otherwise explicitly said, the elements in $M$
will be also assumed {\em smooth}, i.e., as differentiable as
permitted by $M$ ($C^{r_0-1}$ in the case of $g$, and $C^{r_0-3}$
for curvature tensor $R$)\footnote{We will not care about problems
on differentiability (see the review \cite[Sect. 6.1]{senovilla97}).
But notice that, essentially, $r_0=2$ suffices throughout the paper
(the exponential map being only continuous), with the remarkable
exception of Subsect. \ref{sub-conj}. Moreover, taking into account
that {\em globally hyperbolic neighborhoods} make sense for $r_0=1$,
many elements are extendible to this case, see also
\cite{sorkin96}.}. Manifolds are assumed Hausdorff and paracompact,
even though the latter can be deduced from the existence of a
non-degenerate metric (recall that the bundle of orthonormal
references is always parallelizable; thus, it admits a -positive
definite- Riemannian metric, which implies paracompactness \cite[vol
II, Addendum 1]{spivak79}, \cite{marathe72}).

The following convention includes many of the ones in the
bibliography (the main discrepancies come from the causal character
of vector 0, which somewhere else is regarded as spacelike
\cite{oneill83}), and
can be extended for any 
indefinite scalar product:

\begin{definition} \label{type} \index{lightlike vector or curve} \index{timelike vector or curve}
\index{spacelike  vector or curve} \index{causal vector or curve}
\index{null vector} \index{nonspacelike vector}
 A tangent vector $v\in TM$ is classified as:
\begin{itemize}
\item {\em timelike}, if $g(v,v)<0$. \item {\em lightlike}, if
$g(v,v)=0$ and $v \ne 0$. \item {\em causal}, if either timelike or
lightlike, i.e., $g(v,v)\le 0$ and $v \ne 0$. \item {\em null}, if
$g(v,v)=0$. \item {\em spacelike}, if $g(v,v)>0$. \item {\em
nonspacelike},  if $g(v,v)\le 0$.
\end{itemize}
\end{definition}
At each tangent space $T_pM$, $g_p$ is  a (non-degenerate) {\em
scalar product} \index{scalar product}, which admits an
orthonormal basis \index{orthonormal basis} $B_p=(e_0,e_1,\dots ,
e_{n-1}), g_p(e_\mu,e_\nu)=\epsilon_\mu \delta_{\mu\nu}$, where
$\delta_{\mu\nu}$ is Kronecker's delta and $\epsilon_0=-1,
\epsilon_i=1$ (Greek indexes $\mu, \nu$  run in $0, 1, \dots ,
n-1$, while Latin indexes $i,j$ run in $1, \dots , n-1$). Each
 $(T_pM,g_p), p\in M$
contains two {\em causal} cones\index{causal cones}. Definition
\ref{type} is naturally extended to vector fields $X\in
\mathfrak{X}(M)$ and curves $\gamma: I \to M$ ($I\subset
\mathbb{R}$ interval of extremes,\footnote{ We use $\subset$ as a
reflexive relation as in \cite{penrose72}, that is, for any set
$A$, $A \subset A$.} $-\infty \leq a < b \leq \infty$).
Nevertheless, when $I=[a,b]$ we mean by {\em timelike}, {\em
lightlike} or {\em causal} curve any {\em piecewise smooth} curve
$\gamma: I \to M$, such that not only the tangent vectors are,
respectively, timelike, lightlike or causal, but also the two
lateral tangent vectors at each break lie in the same causal cone.
The notion of causal curve will be extended below non-trivially to
include  less smooth ones, see Definition \ref{continc}.

 A time-orientation at $p$ is a choice of one of the two causal
 cones at $T_pM$, which will be called {\em future} cone, in opposition of
the non-chosen one or {\em past} cone. In a similar way that for
usual   orientation in manifolds, a smooth choice of
time-orientations at each $p\in M$ (i.e., a choice which coincides
at some neighborhood $U_p$ with the causal cone selected by a
-smooth- causal vector field on $U_p$) is called a {\em
time-orientation}. The Lorentzian manifold is called {\em
time-orientable} \index{time-orientation} \index{time-orientable
manifold} when one such time-orientation exists; no more generality
is obtained either if smooth choices are weakened in $C^r$ ones,
$r\in \{0, \dots, r_0-1\}$, or if causal choices are strengthened in
timelike ones. As the causal cones are convex, a standard partition
of the unity argument yields easily:

\begin{proposition} \label{pot}
A Lorentzian manifold is time-orientable if and only if it admits a
globally defined timelike vector field $X$ (which can be chosen
complete\footnote{$X$ can be chosen complete because, given $X$ and
an auxiliary complete Riemannian metric (which exists due to a
theorem by Nomizu and Ozeki \cite{nomizu61}) it can be replaced by
the
 timelike vector field $X/|X|_R$, which is necessarily complete.}).
\end{proposition}
Recall that this vector field $X$ can be defined to be
future-directed at all the points and, then, any causal tangent
vector $v_p\in T_pM$ is future directed \index{future-directed
vector} \index{past-directed vector} if and only if $g(v_p,X_p)<0$.

\index{time-orientable double covering} Easily one has: (a) any
Lorentzian manifold admits a time-orientable double covering
\cite{penrose72}, \cite[Lemma 7.17]{oneill83}, and (b) let $g_R$ be
any Riemannian metric on $M$ and  $X\in \mathfrak{X}(M)$
non-vanishing, with $g_R$-associated 1-form $X^\flat$, then $$ g_L =
g_R - \frac{2}{g_R(X,X)} X^\flat \otimes X^\flat $$ is a
time-orientable Lorentzian metric.  As a consequence, the possible
existence of Lorentz metrics can be characterized
\cite[5.37]{oneill83}, \cite[Sect. 3.1]{beem96}, \cite[Sect.
1]{penrose72}:

\begin{theorem}\label{texislor}
For any connected smooth manifold, the following properties are
equivalent:

(1) $M$ admits a Lorentz metric.

(2) $M$ admits a time-orientable Lorentz metric.

(3) $M$ admits a non-vanishing vector field  $X$.

(4) Either $M$ is non-compact or its Euler characteristic
\index{Euler characteristic} is 0.
\end{theorem}

\begin{proof}  (3) $ \Leftrightarrow$
(4) Well-known result in algebraic topology.

(2) $\Leftrightarrow $ (3) To the right, Proposition \ref{pot}; to
the left, comment (b) above.

$(1) \Rightarrow (2)$ (The converse is trivial.) The time orientable
double covering $(\tilde{M},\tilde{g})$,  satisfies (3) and hence
(4). So, the latter is satisfied obviously by $M$.
\end{proof}

\begin{figure}[t]
\begin{center} \psfrag{H}{$x\!=\!0$}
\psfrag{J}{$x\!=\!1/2$}\psfrag{K}{$x\!=\!1$} \psfrag{F}{$x\!=\!3/2$}
\psfrag{G}{$X_1$} \psfrag{A}{$\!\!\! X_2$}
 \includegraphics[width=10cm]{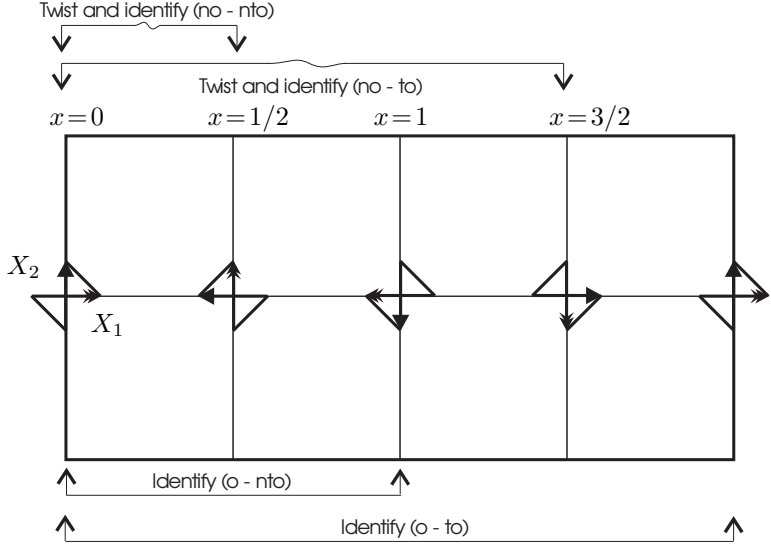}
\end{center}
\caption{Time-orientability and orientability are logically
independent. Here we use the short-hand notation: {\em
o}=orientable; {\em no}= non-orientable; {\em to}=
time-orientable; {\em nto}=non-time-orientable.
 Figures in which the causal cones are
explicitly displayed are standard in Causality Theory. In this
work, if the spacetime is time-oriented the past cones are
displayed in black, see figure \ref{ipair}.}\label{esempio}
\end{figure}

The relevant new ingredient of a spacetime is a time-orientation:
\begin{definition}
A  {\em spacetime} \index{spacetime} $(M,g)$  is a time-oriented
connected Lorentz manifold.
\end{definition}
The points of $M$ are also called {\em events}. \index{event}
Notice that the time-orientation is implicitly assumed in the
notation $(M,g)$ for a spacetime. In principle, $M$ is not assumed
to be an orientable manifold. Recall that orientability
\index{orientable manifold} and time-orientability
\index{time-orientable manifold} are logically independent. In
fact, one can construct easily time-orientable and
non-time-orientable Lorentz metrics on both a M\"obius strip (or
Klein bottle) and cylinder (or torus) by starting with the metric
$g$ on $\R^2$
$$
g(X_1,X_2) \equiv -1, \quad g(X_1,X_1) \equiv 0 \equiv g(X_2,X_2)
$$
$$ X_1= \cos \pi x \, \partial_x + {\rm sen} \pi x \partial_y ,
\quad  X_2= -{\rm sen} \pi x \, \partial_x + \cos \pi x
\partial_y, $$ by making natural quotients (see figure
\ref{esempio}).

\subsection{Conformal/classical causal  structure}
The following algebraic result (Dajczer et al. criterion)
has important consequences for the conformal structure  of
spacetimes, and has no analog in the positive definite case:

\begin{proposition}\label{cdn}  Let
$(V,g)$ be a real vector space with a non-degenerate indefinite
scalar product, and let $b$ be a bilinear symmetric form on $V$.
The following properties are equivalent:
\begin{enumerate}
\item $b=c\cdot g$ for some $c\in \R$, \item $b(v,v)=0$ if
$g(v,v)=0$,
\end{enumerate}
\end{proposition}
The proof can be seen also in \cite[Lemma 2.1]{beem96}, \cite[App.
D]{wald84}. Obviously, 1 $\Rightarrow 2$, and   the converse can be
proved in dimension 2 easily; for higher dimensions, the problem is
reduced to dimension 2, by grouping suitably the elements of a
$g$-orthonormal basis. By the way, it is also known that any of the
following conditions is equivalent to items 1, 2 (this yields bounds
on the possible curvatures): (a)
 $\exists a>0: \: b(v,v)/g(v,v) \leq a$ if $g(v,v)\neq 0$, (b)
$\exists a>0: \: b(v,v)/g(v,v) \geq a$ if $g(v,v)\neq 0$, (c)
$\exists a>0: \: |b(v,v)| \leq a |g(v,v)|$ if $g(v,v)<0$, (d)
$\exists a>0: \: |b(v,v)| \leq a |g(v,v)|$ if $g(v,v)>0$. In fact,
any of these items  implies item 2, by using that any lightlike
vector can be approximated by both, timelike and spacelike ones.
For some algebraic extensions to higher order tensors, see
\cite{bergqvist}.

Two Lorentzian {\em metrics} $g, g^*$  on the same manifold $M$ are
called {\em pointwise conformal} \index{pointwise conformal metrics}
if $g^*=e^{2u} g$ for some function $u:M \to \mathbb{R}$.
Proposition \ref{cdn} yields directly:
\begin{lemma} \label{confm}
Two Lorentzian metrics $g, g^*$ on a manifold $M$ of dimension
$n_0>2$ are pointwise conformal if and only if both have the same
lightlike vectors. \end{lemma} (The exceptional case $n_0=2$ appears
because a negative conformal factor keep   lightlike vectors
unchanged, while exchanges timelike and spacelike vectors.)

Two {\em spacetimes} on the same manifold $M$ are {\em pointwise
conformal} \index{pointwise conformal spacetimes} if both, their
metrics are pointwise conformal and their time-orientations agree
at each event. The spacetime $(M,g)$ is called {\em conformal}
\index{conformal spacetimes} to the spacetime $(M^*,g^*)$ if there
exists a diffeomorphism $\Phi: M\rightarrow M^*$ such that the
pull-back spacetime on $M$ obtained inducing the metric and the
time-orientation through $\Phi$ is pointwise conformal to $(M,g)$.
Two spacetimes which only differ in the time-orientation are by
definition {\em not} pointwise conformal and, moreover, they may
be also non-conformal (see, for example, Figure \ref{notimesim} at
the end). Clearly, the conformal relation is a relation of
equivalence in the class of all the spacetimes. The following
definition will be revisited in Section \ref{s4}, in order to
discuss what {\em causality} means.

\begin{definition} \index{causal structure}
The  {\em conformal} or {\em classical causal structure}  of the
spacetime $(M,g)$ is the equivalence class $[(M,g)]$ for the
conformal relation.
\end{definition}
Several concepts in Lorentzian geometry do depend on the full metric
structure of the  spacetime $(M,g)$. Examples are the length of a
curve, the time-separation between two events (see below), the
non-lightlike geodesics or the geodesic completeness of a spacetime.
Nevertheless, the conformal structure is  particularly
 rich by itself, and its interplay with  the metric becomes specially interesting.

\begin{remark} \label{rconv}
The elements which come only from the conformal structure will be
emphasized with the following conventions. \index{conventions} For
practical purposes, we will work with the {\em relation of
equivalence induced by the pointwise conformal relation} in the
spacetimes on the same $M$. For the spacetime $(M,g)$, its pointwise
conformal class \index{conformal class of metrics} will be denoted
as $(M, \bm{g})$ ($\bm{g}$ denotes the set of all pointwise
conformal metrics to $g$) where all the spacetimes in the class have
the same time-orientation. When we refer to a spacetime as $(M,
\bm{g})$, we emphasize that the considered properties hold for any
$g$ in the class and, thus, depend only on the conformal structure.
The boldface will be extended to equivalence classes of vectors and
curves. So, $\bm{v}$ denotes the equivalence class of vectors
$v'=\alpha v$, $\alpha>0$ and $\bm{g}(\bm{v},\bm{w})$ is just the
sign $(-1,0,+1)$ of the scalar product $g(v,w)$.

Analogously, if $\gamma: I\rightarrow M$ is a curve then $
\bm{\gamma}$ is the equivalence class  of curves coincident with
$\gamma$ up to a strictly increasing reparametrization. Note that if
$I$ is closed (or compact or open) for a representative of $
\bm{\gamma}$, the same holds for any representative $\gamma :
{I}\rightarrow M$. Analogously, we say that $\bm{\gamma}$ {\em
connects} $p$ with $q$ if for a representative, $I=[a,b]$,
$\gamma(a)=p$ and \index{connecting curve} $\gamma(b)=q$, or write
$p\in \bm{\gamma}$ if $p$ belongs to the image of $\gamma$.
If a future-directed  causal curve $\gamma$ satisfies
$\lim_{t\rightarrow b} \gamma(t) = q$ (resp. $\lim_{t\rightarrow a}
\gamma(t) = p$), where $a, b$ $(-\infty \leq a < b \leq \infty)$ are
the extremes of the interval $I$, the event $q$ (resp. $p$) is
called the future (resp. past) endpoint of $\gamma$ (and
\index{endpoint} the other way round if $\gamma$ is past-directed).
These concepts are obviously extended to $\bm{\gamma}$, so one can
assume that $I$ is bounded when dealing with the endpoints of
$\gamma$. A causal curve without future (resp. past) endpoint is
said future (resp. past) {\em inextendible}. \index{inextendible
curve}

\end{remark}

\subsection{Causal relations. Local properties} \index{causal relations}
Given a spacetime $(M,\bm{g})$ the event $p$ is {\em
chronologically} \index{chronologically related events}
\index{causally related events} \index{strictly causally related
events} \index{horismotically related events} (resp. {\em strictly
causally; causally; horismotically}) related to the event $q$,
denoted $p\ll q$ (resp. $p<q$; $p\leq q$; $p\rightarrow q$) if there
is a future-directed timelike (resp. causal; causal or constant;
causal or constant, but not timelike) curve connecting $p$ with $q$.
If $W \subset M$, given $p,q \in W$, the analogous relations for the
spacetime $(W,\g\vert_{W})$ will be denoted $p\ll_W q$, $p<_W q$,
$p\leq_W q$, $p\rightarrow_W q$.

From  the viewpoint of set theory, relations $\ll, \leq,
\rightarrow$ are written, regarded as subsets of $M \times M$, as:
$$
I^{+}=\{(p,q) : p\ll q  \} , \quad J^{+}=\{(p,q) : p\leq q  \},
\quad E^{+}=\{(p,q) : p\rightarrow q  \}.
$$
Clearly, $E^\pm={J^\pm}\backslash I^\pm$.

 {\em Note:} all the definitions and properties
extend naturally to the ``minus'' sign without further mention; for
example, the sets (and binary relations) $I^{-}$, $J^{-}$, and
$E^{-}$, are defined changing each $(p,q)$ above by $(q,p)$.

The chronological future \index{chronological future} of an event is
defined as:
$$
I^{+}(p)= \{q\in M: p\ll q\} = \pi_2(\pi^{-1}_1(p)\cap
I^{+})=\pi_1(\pi^{-1}_2(p)\cap I^{-})
$$
where $\pi_1$ and $\pi_2$ are the canonical projections to the
factors of $M\times M$. Analogous expressions hold for the causal
future $J^+(p)$ \index{causal future} and horismos
$E^+(p)$.\index{horismos} By using juxtapositions of curves, it is
obvious that the relations $\ll$ and $\leq$ are transitive, but
$\rightarrow$ is not (see Proposition \ref{nhu}).

Every point of a spacetime admits an {\em arbitrarily small} (i.e.
contained in any given neighborhood) \index{arbitrarily small} {\em
convex neighborhood} \index{convex neighborhood} $U$, that is, $U$
is a (starshaped) normal neighborhood \index{normal neighborhood} of
any of its points $p\in M$. This means that the the domain $\tilde U
\subset T_pM$ of the exponential map at $p$, is chosen starshaped
and yields a diffeomorphism onto $U$, $\exp_p: \tilde U \rightarrow
U$. Thus, for any $p,q \in U$, there exists a unique geodesic
$\gamma_{pq}:[0,1]\rightarrow U$ which connects $p$ with $q$. Notice
that one also has a diffeomorphism \cite[Lemma 5.9]{oneill83}
between $U\times U$ and and its image on $TU$, which sends $(p,q)
\to \exp_p^{-1}q = \overrightarrow{pq}\in TM$.
 Such a convex $U$ can be chosen {\em simple}, \index{simple neighborhood}
that is, with compact closure $\bar{U}$ included in another open
convex neighborhood \cite[p. 6]{penrose72}.

Given an open subset $U$ by $I^+(p,U)$, $J^+(p,U)$, $E^{+}(p,U)$,
will be denoted the corresponding future elements in $U$ regarded
as  spacetime. If $U$ is a convex neighborhood the causal
relations in $U$ are easily characterized \cite[Lemma
14.2]{oneill83}:

\begin{proposition}\label{plocal} Let $(M,g)$ be a spacetime, $\exp_p$ the exponential map at $p\in M$,
and $U$  a convex neighborhood. Regarding $U$ as a spacetime, given
$p\neq q$, $p,q \in U$:

\begin{enumerate}
\item $q\in I^{+}\left( p,U\right)$ (resp. $ q\in J^{+}\left(
p,U\right); q\in E^{+}\left( p,U\right)$) $ \Leftrightarrow
\overrightarrow{pq} = \exp_p^{-1}q $ is timelike (resp. causal;
lighlike) and future-pointing.

\item $I^{+}\left( p,U\right) $ is open in $U$ (and $M$).

\item   $J^{+}\left( p,U\right) $ is the closure in $ U$  of
$I^{+}\left( p,U\right) $.

\item  Causal relation $J^+$ is closed in  $U \times U$.

\item  Any  causal curve $\bm{\gamma}$  contained in a compact
subset of $U$ has two endpoints. 
\end{enumerate}
\end{proposition}
 Notice from the first item (which can be regarded
as a consequence of Theorem \ref{tmaximlocal} below) that the study
of the causal relations in $U$ is reduced to the study of the causal
character of tangent vectors type $\overrightarrow{pq}$.

Nevertheless, convex neighborhoods depend on the metric structure.
The following  concept is purely conformal.

\begin{definition} \label{dcausconvex}
Let $U, V$ be open subsets of a spacetime $(M,\g )$, with $V\subset
U$. $V$ is called {\em causally convex in $U$} \index{causally
convex neighborhood} if any causal curve contained in $U$ with
endpoints in $V$ is entirely contained in $V$.

In particular, when this holds for $U=M$, $V$ is called  {\em
causally convex}.
\end{definition}
\begin{remark} \label{rcausconvex}
Note that, in this case, if $\leq_U$, $\leq_V$ denote, resp., the
causal relations in $U, V$ regarded as  spacetimes, then the
restriction of $\leq_U$ to $V$ agrees with $\leq_V$ (this property
does {\em not} characterize causal convexity, as can be checked
from $U=\L^2, V=\{(t,x)\in \R^2: |t|,|x|<1\}$).
 There are spacetimes such that the only open subset $V\neq \emptyset$ of $U=M$ which is
causally convex  is $V=M$ (see the totally vicious ones in Sect.
\ref{subs-vicio}). Nevertheless, at least when $U$ is also convex,
 the existence of arbitrarily small such $V$ in $U$, even
with further properties, will be shown next.

Finally, note that if $V$ is causally convex in $U$ and $W$ is an
open set such that $V \subset W\subset U$, then $V$ is causally
convex in $W$.
\end{remark}
 It is  also possible to prove that any point of $(M,\g)$
admits a neighborhood with the best possible causal structure, i.e.,
which will belong to the top of the ladder, {\em global
hyperbolicity} (see Section \ref{globhyp}). Recall first the
following result (by $\bm{g} < \bm{g}'$ we mean that the causal cone
of $\bm{g}$ at each point $p$ is included in the timelike cone of
$\bm{g}'$ at $p$, see Section \ref{subs-stably}).

\begin{lemma} \label{gth} Let $(M,\g)$ be a spacetime. Given $p\in M$, and a neighborhood $U\ni p$,
there exists a neighborhood $V\ni p$, $V\subset U$ and two {\em
flat} metrics $g^-, g^+$ on $V$ such that $\g^- < \g < \g^+$.
\end{lemma}

\begin{proof}
Take a coordinate neighborhood $(V_\delta, (t=x^0,x^1, \dots ,
x^{n-1}))$, centered at $p$, $\sum_\mu (x^\mu)^2 <\delta^2$, such
that the tangent basis $B_p= (\partial_0,
\partial_1, \dots ,
\partial_{n-1})$ is orthonormal at $p$ according to a representative $g$ of $\bm{g}$.
Now, recall that the scalar product $g_p^-$ (resp. $g_p^+$) at
$T_pM$, such that $B^-_p= (2\partial_0, \partial_1, \dots ,
\partial_{n-1})$ (resp. $B^+_p= ((1/2)\partial_0, \partial_1, \dots ,
\partial_{n-1})$) is an orthonormal basis has the cones strictly
less (resp. more) open than the cones of $\g_p$. By continuity, this
property holds in a neighborhood $V_{\delta}$ of $p$, for $\delta$
sufficiently small, moreover given $U$, by taking $\delta$
sufficiently small we have $V_{\delta} \subset U$ and so the
required metrics are $g^-= -({1}/{4}) dt^2 + \sum_i (dx^i)^2$ and
 $g^+= -4 dt^2 + \sum_i (dx^i)^2$.
\end{proof}

Let $\L^n$ be Lorentz-Minkowski spacetime with natural coordinates
$(x^\mu)=(t=x^0,x^i)$, \index{Lorentz-Minkowski spacetime}
$p_{\epsilon},q_{\epsilon} \in \L^n$, $p_{\epsilon}=(-\epsilon, 0,
\dots , 0)$, $q_{\epsilon}=(\epsilon, 0, \dots , 0)$, $\epsilon>0 $.
The open neighborhood in $\L^n$, $V_\epsilon = I^+(p_\epsilon) \cap
I^-(q_\epsilon)$,  satisfies that $t\equiv 0$ is a spacelike {\em
Cauchy hypersurface} $S$ of $V_\epsilon$, that is, it is crossed
exactly once by any inextendible timelike curve contained in
$V_\epsilon$ (see Section \ref{globhyp}).  We will mean by a {\em
globally hyperbolic neighborhood} \index{globally hyperbolic
neighborhoods} of $p$ any coordinate neighborhood $(V, x^\mu)$ such
that $x^0\equiv 0$ is a Cauchy hypersurface of $V$.  \index{globally
hyperbolic neighborhood}
 The following result shows
that the local  structure of a spacetime fulfills all good
properties from the viewpoint of Causality (see also the study in
\cite[Sect. 2]{krasnikov02}).
\begin{theorem} \label{fund}
Let $(M,\g)$ be a spacetime. For any $p\in M$ and any neighborhood
$U\ni p$ there exists an open neighborhood $U'$, $p\in U'\subset
U$, and a sequence of nested globally hyperbolic neighborhoods
$(V_n, x^\mu)$, $V_{n+1}\subset V_n$,  $\{p\}=\cap_n V_n$, all
included in $U'$, such that each $V_n$ is causally convex in $U'$.
\end{theorem}
\begin{proof}
Consider the metric $g^+$ in Lemma \ref{gth} defined in some
neighborhood $U'\subset U$ of $p$. As it is flat, one can find the
required sequence of globally hyperbolic neighborhoods
$(V_n,x^\mu)$ for $\g^+$, $V_n \subset U'$, each one
$\g^+$-causally convex in $U'$. Nevertheless, any causal curve for
$\g$ will be timelike for $\g^+$ and, so, each $(V_n, x^\mu)$ will
be both, globally hyperbolic and causally convex for $\g$.
\end{proof}

\begin{remark}\label{rfund} 
(1) Of course, in Theorem \ref{fund} (which is formulated in a
conformally invariant way) we can assume $U'=V_1$. Nevertheless, it
is clear from the proof that, for any representative $g$ of the
conformal class, {\em $U'$ can be chosen simple}, which leads to the
strongest  local causal properties.

 (2) The sequence $\{V_n\}_n$ yields a
topological basis at $p$. Thus, an alternative formulation of
Theorem \ref{fund} would ensure the existence of a (simple)
$U'\subset U$ which admits arbitrarily small globally hyperbolic
neighborhoods of $p$, all of them causally convex at $U'$.

 (3) It also holds that each obtained neighborhood $V=V_n$ of the sequence
satisfies: each $q\in V$ admits an arbitrarily small neighborhood
which is causally convex in $V$. In fact, this property is one of
the alternative definitions of being {\em strongly causal}, see
Sect. \ref{strongsect}. Hence we have also proved that any spacetime
$(M,\g)$ is locally strongly causal, as any point $p$ admits an
arbitrarily small strongly causal neighborhood $V$. Also, due to the
last observation of Remark \ref{rcausconvex} any open set $W \subset
V$, $p \in W$, is a strongly causal neighborhood of $p$ as well. In
particular,{\em any spacetime $(M,g)$ admits arbitrarily small
simple strongly causal neighborhoods}.

\end{remark}


\subsection{Further properties of causal relations}\label{subfurther}
 None of the  properties in Proposition \ref{plocal}, but the second one, holds
globally. In fact, given a timelike curve $\bm{\gamma}$ connecting
the pair $(p,q)$ there are open neighborhoods $U\ni p$, $V \ni q$
such that if $\tilde p \in U$, $\tilde q \in V$, then there exists a
timelike curve $\tilde{\bm{\gamma}}$ connecting $\tilde p$ and
$\tilde q$ (say, $U, V$ can be chosen as $I^-(p_1)\cap U_p$,
$I^+(q_1)\cap U_q$, where $U_p, U_q$ are convex neighborhoods of $p,
q$ which contains $p_1, q_1$, resp., and these points are chosen
such that $ \bm{\gamma}$ runs consecutively $p, p_1, q_1, q$).
Summing up,
\begin{proposition} \label{pglobal}
The set $I^+$   is open in $M\times M$.
\end{proposition}
In what follows we claim that  $p \ll r$ and $r \le q$ (or the other
way round) implies  $p \ll q$ (see Proposition \ref{nhu} for a more
accurate result).
 In general, $J^+(p) \subset \bar
I^+(p)$ but the equality may not hold. Nevertheless, both closures
as well as both boundaries (denoted with a dot in what follows) and
interiors (denoted Int) coincide. Even more:

\begin{proposition} \label{padher}
It is $\bar{J}^+=\bar{I}^+$, $\textrm{Int}\, J^+=I^+$,
$\dot{J}^+=\dot{I}^+$.
\end{proposition}

\begin{proof}
Since ${I}^+\subset {J}^+$, it is $\bar{I}^+\subset \bar{J}^+$. Let
$(p,q) \in \bar{J}^{+}$ and let $U$ and $V$ be arbitrarily small
neighborhoods of respectively $p$ and $q$. There are events $p'\in
U$, $q' \in V$, such that $(p',q')\in J^{+}$. Take events $p''\in
U\cap I^{-}(p')$ and $q''\in V \cap I^{+}(q')$.
Then $p''$ can be connected to $q''$ with the composition of a
timelike, a causal, and finally a timelike curve, and, as claimed
above, 
it follows $p'' \ll q''$. Since $U$ and $V$ are arbitrary $(p,q)$ is
an accumulation point for points belonging to $I^{+}$. We conclude
that $\bar{I}^+= \bar{J}^+$.

Let us show that ${I}^+= \textrm{Int}{J}^+$ from which it follows
$\dot{J}^+=\dot{I}^+$. Since $I^{+}$ is open and included in $J^+$,
$I^{+} \subset \textrm{Int}{J}^+$. If $(p,q) \in \textrm{Int} J^+$,
then chosen normal convex neighborhoods $U \ni p$, $V \ni q$, such
that $U \times V$ is included in $\textrm{Int}{J}^+$, and taken
$p'\in U \cap I^{+}(p)$, $q' \in V \cap I^{-}(q)$, then $q' \in
J^{+}(p')$ and thus $q \in I^{+}(p)$, i.e., $(p,q) \in I^{+}$.
\end{proof}

\begin{definition}
An open subset $F$ (resp. $P$) is a future (resp. past) set if
$I^{+}(F)=F $ (resp. $I^{-}(P)=P $ ). \index{future set} \index{past
set}
\end{definition}

An example of future set is $I^{+}(p)$ for any $p\in M$. We have
the following 
characterization:

\begin{proposition}
If $F$ is a future set then $\bar{F}=\{p: I^{+}(p) \subset F\}$, and
analogously in the past case.
\end{proposition}

\begin{proof} ($\supset$). If $I^{+}(p) \subset F$ then $p \in \bar{I}^{+}(p)\subset
\bar{F}$.

($\subset$). Let $p \in \bar{F}$ and take any $q \in I^{+}(p)$. As
$I^{-}(q)\ni p$ is open, $I^{-}(q)\cap F \ne \emptyset$. Thus, $q
\in I^{+}(F)=F$, i.e. $I^{+}(p) \subset F$.
\end{proof}

\begin{remark} Even though the closure of $J^+$ in $M\times M$ induces a binary relation, this is not always transitive.
As closedness becomes relevant for different purposes (for example,
when one deals with limit of curves) Sorkin and Woolgar
\cite{sorkin96} defined the {\em $K$-relation} \index{K-relation} as
the smallest one which contains $\ll$ and is: (i) transitive, and
(ii) topologically closed. (That is, the corresponding set $K^+
\subset M\times M$ which defines the $K$-relation, is the
intersection of all the closed subsets $C$ which contain $I^+$
such that $(p,q), (q,r) \in C \Rightarrow (p,r)\in C$). Among the 
applications, some results on positive mass and globally hyperbolic
spacetimes with lower order of differentiability ($r_0=1$) have been
obtained.

Notice that, in particular, $\bar{I}^{+} \subset K^{+}$, but perhaps
there exists $(p,q) \in K^{+}\backslash \bar{J}^{+}$. In this case,
$q \notin \bar{I}^{+}(p)$ and hence there is a point $r \in
I^{+}(q)$ not contained in $I^{+}(p)$. As a consequence $(p,q) \in
K^{+}$ and $r \in I^{+}(q)$ do not imply $r \in I^{+}(p)$ (as
 happens when the causal relation $K^+$ is replaced with $J^{+}$,
see Prop. \ref{nhu}). In particular, the relation $K^+$ does not
define a causal space in the sense of Kronheimer and Penrose
\cite{kronheimer67}. Nevertheless, this cannot happen if
$(M,\bm{g})$ is causally simple, because then
$\bar{I}^{+}=\bar{J}^{+}=J^{+}=K^{+}$ (see Sect. \ref{simsec});
moreover, $(I^+,K^+)$ defines such a causal space if and only if
$(M,\bm{g})$ is causally continuous \cite{dowker00}.
\end{remark}

\begin{figure}[ht]
\begin{center} \psfrag{P}{$p$}
\psfrag{Q}{$q$}\psfrag{G}{$\gamma$}
 \includegraphics[width=8cm]{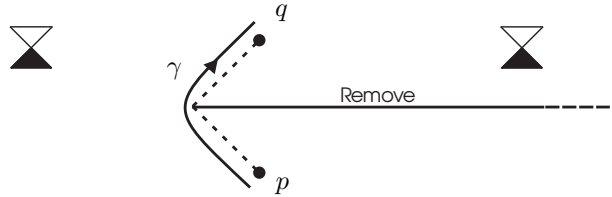}
\end{center}
\caption{Minkowski spacetime without a spacelike half-line is an
example of stably causal non-causally continuous spacetime (see
Sects. \ref{subs-stably}--\ref{ccontinuous}). Here $(p,q) \in
\bar{I}^{+}$, but neither $q\in \bar{I}^{+}(p)$, nor $p \in
\bar{I}^{-}(q)$.} \label{ipair}
\end{figure}

\begin{remark}
In general $(p,q) \in \bar{I}^{+}$ does not imply $q \in
\bar{I}^{+}(p)$ or $p \in \bar{I}^{-}(q)$ (see figure \ref{ipair}).
For this reason it may be more useful to regard the causal relations
as defined in  $M \times M$, although it is
customary to introduce them in $M$, 
that is, through $I^{\pm}(p)$, $J^{\pm}(p)$, $E^{\pm}(p)$.
\end{remark}

Recall that we have defined three binary relations $\ll, \leq,
\rightarrow$ (and trivially a fourth one $<$) on any spacetime and,
obviously, two of them determine the third. But starting with only
one of them, one can define naturally a second (and then, a third)
binary relation, which  will coincide with the other causal-type
relation in sufficiently
 well-behaved spacetimes:

\begin{definition} \label{binaryrelations}
Let $\ll, \leq, \rightarrow$ be the canonical binary relations of a
spacetime $(M, \bm{g})$. We define the associated relations
\begin{enumerate} \item starting at chronology $\ll$:
\begin{itemize}
\item[(a)]  $x \leq^{(\ll )}  y$ $\Leftrightarrow$ $
  I^+(y) \subset I^+(x)$ and $I^-(x)
 \subset I^-(y)$.
 \item[(b)] $x \rightarrow^{(\ll )}  y$ $\Leftrightarrow$ $x\leq^{(\ll )}
 y$ and not $ x \ll y$.
\end{itemize}
 \item starting at horismos $\rightarrow$:
\begin{itemize}
\item[(a)] $x \le^{(\rightarrow )} y$ $\Leftrightarrow$ $x=x_1 \rightarrow
x_2 \dots \rightarrow x_{n-1}\rightarrow x_n=y$ for some finite
 sequence $x_1, \dots  ,x_n \in M$.
\item[(b)] $x \ll ^{(\rightarrow )} y$ $\Leftrightarrow$ $x \le^{(\rightarrow )}
y$ and not $x \rightarrow y$.
\end{itemize}

\item starting at causality $\leq$ (and, thus, $<$):
\begin{itemize}
\item[(a)]
     $x\rightarrow^{(\le)} y$ $\Leftrightarrow$ $x\le y$ and $\leq$  is a
total linear order in $J^+(p)\cap J^-(q)$ for any $p, q$ such that
$x<p<q<y$ (i.e., the  topological space $J^+(p)\cap J^-(q)$, ordered
by $\leq$, is isomorphic to $[0,1]$ with its natural order; in
particular, each two distinct $p', q' \in J^+(p)\cap J^-(q)$ satisfy
either $p'\rightarrow q'$ or $q' \rightarrow p'$).
\item[(b)] $x\ll^{(\le)} y$ $\Leftrightarrow$ $x \le y$ and not
$x\rightarrow^{(\le)} y$.
\end{itemize}
\end{enumerate}
\end{definition}
As we will see, $\leq^{(\ll )} \; = \; \leq$ in causally simple
spacetimes (Theorem \ref{pk2}),
$\le^{(\rightarrow )} \; = \; \le$ in distinguishing spacetimes
(Theorem \ref{pk1}), and $\rightarrow^{(\le)} \; = \;\rightarrow$ in
causal spacetimes (Theorem \ref{pk3}). 

Some authors have studied the abstract properties of $\ll, \leq,
\rightarrow$ and defined spaces which generalize (well-behaved)
spacetimes with their canonical causal relations. Among them causal
spaces by Kronheimer and Penrose \cite{kronheimer67}, etiological
spaces by Carter \cite{carter71} and chronological spaces by Harris
\cite{harris98}. Among the applications to spacetimes, a better
insight on the meaning of {\em causal boundaries} \index{causal
boundary} (whose classical construction by Geroch, Kronheimer and
Penrose \cite{geroch72} relies on some types of future sets) is
obtained, see \cite{flores06b, garciaparrado05} and references
therein.

\subsection{Time-separation and maximizing geodesics}

Let  $\left( M,g\right) $ be a spacetime, fix $p,q\in M$ and let
$\hat C(p,q)$  be the set of the future-directed causal curves
which connect $p$ to $q$. The following concept is metric
(non-conformally invariant) as it depends on  the Lorentzian
length \index{lorentzian length} $L(\gamma)=\int_{t_p}^{t_q} \vert
\gamma'\vert \dd t$, $p=\gamma(t_p)$, $q=\gamma(t_q)$, $\gamma \in
\hat C(p,q)$. Nevertheless,  some of its properties will depend
only on the conformal structure.

\begin{definition} The {\em time-separation}  \index{time-separation} (or {\em Lorentzian
distance}) \index{Lorentzian distance} is the map $d :M\times
M\rightarrow \left[ 0,+\infty \right] $ defined as:
$$
d(p,q) = \left\{
\begin{array}{l}
0,\hbox{\ if }\hat C(p,q)=\emptyset  \\
\sup  \left\{ L\left( \alpha \right) ,\,\alpha \in \hat
C(p,q)\right\}  \hbox{, if }\hat C(p,q)\neq \emptyset
\end{array}
\right. $$
\end{definition}
Some simple properties are:
\begin{proposition}
  Let $p,q,r\in M$:

\begin{enumerate}
\item  $d(p,q) >0\Leftrightarrow p\in I^{-}\left(
q\right)$

\item  If there exists a closed timelike curve through $p$, $d(p,p) = +\infty $; otherwise: $d(p,p) = 0. $

\item  $0<d(p,q) <+\infty \Longrightarrow d(q,p) =0$ ($d $ is not symmetric)

\item   $p\leq q\leq r \Longrightarrow d(p,q) + d(q,r) \leq d(p,r) $.
\end{enumerate}
\end{proposition}
Most of the proof of this proposition is  straightforward; take into
account Th. \ref{t0} below.

\begin{figure}[h]
\begin{center} \psfrag{Q}{$q$}
\psfrag{P}{$p$}\psfrag{PN}{$p_n$}
\includegraphics[width=7cm]{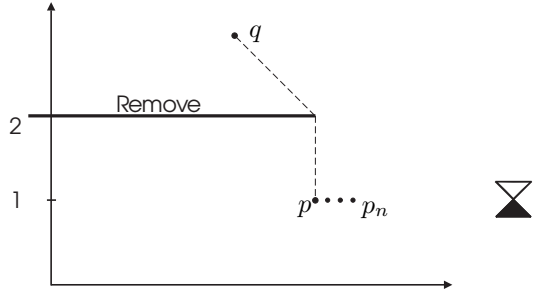}
\end{center}
\caption{A classical example of spacetime for which $d$ is not upper
semi-continuous. Here $\overline{\lim}_{n} \,
d(p_{n},q)=1>0=d(p,q)$.}\label{cont}
\end{figure}

Of course, $d$ is not a true distance, but the last property
suggests possible similitudes with the distance associated to a
Riemannian metric. A first one is:

\begin{proposition} In any spacetime, $d$ is \index{lower semi-continuous (Lor. distance)}
lower semi-continuous, that is, given $ p,q,p_{m},q_{m}\in M,
\left\{ p_{m}\right\}_m \rightarrow p$ , $\left\{ q_{m}\right\}_m
\rightarrow q$,  the lower limit satisfies:
$$\underline{\lim}_{m}d(p_{m},q_{m}) \geq d(p,q)
$$
\end{proposition}
Nevertheless, 
$d$ may be no upper semi-continuous (see Figure \ref{cont}).

The main Riemannian similarities come from the maximizing properties
of causal geodesics, which are consequences of an infinitesimal
application of reversed triangle inequality. \index{reversed
triangle inequality}
Concretely, the maximizing properties can be summarized in the
following two results (see, for ex., \cite[Lem. 5.34,
5.9]{oneill83}, 
or around \cite[Prop. 2.1]{senovilla97}), the first one local (see
also Proposition \ref{plocal}) and the second global:



\begin{theorem} \label{tmaximlocal} Let $U$ be a convex neighborhood of $(M,g)$, and $p, q\in U$.
Assume there exists a causal curve $\alpha :\left[ 0,b\right]
\rightarrow U$ from $p$ to $q$. Then, the radial segment
$\gamma_{pq} :\left[ 0,1\right] \rightarrow U$ from $p$ to $q$,
(which has initial velocity $\overrightarrow{pq}=\exp_p^{-1}(q)$ and
length $|\overrightarrow{pq}|=
\sqrt{|g(\overrightarrow{pq},\overrightarrow{pq})|}$), is causal and
(up to reparametrization) maximizes strictly the
length 
among all the causal curves in $U$ which connect $p$ to $q$.

In particular, if $\gamma_{pq}$ is lightlike then it is the unique
causal curve contained entirely in $U$ which connects $p$ to $q$.
\end{theorem}

\begin{theorem} \label{tmaximglobal}  Assume that there exists a causal curve $\alpha:\left[ 0,b\right] \rightarrow M$ which connects $p$ to $q$,
$p,q\in M$, with maximum length among all the causal curves which
connect $p$ to $q$ in the spacetime $(M,g)$. \index{conjugated
events} Then, $\alpha$ is, up to a reparametrization,  a causal
geodesic without conjugate points (Def. \ref{def-conjugate-point})
except, at most, the endpoints.

\end{theorem}
That is: (i) the length of a causal geodesic contained in a convex 
neighborhood is equal to the time-separation (computed in the
neighborhood as a spacetime), of its endpoints, and
 (ii) if a causal curve in the spacetime has a length equal to the
 time-separation of its endpoints, then it is, up to a
 parametrization, a causal geodesic without conjugate points,
 except at most its endpoints.

 Recall that if $p,q\in M$ satisfy
 $p < q$ and $d(p,q)=0$, then these two properties
  are conformally invariant. So, Theorem \ref{tmaximglobal} implies that any lightlike geodesic (and its first conjugate point) must be
  conformally
  invariant,  up to
 a reparametrization. Next, we will see that this can be
 made much more precise.

\subsection{Lightlike geodesics and conjugate events}
\label{sub-conj} \index{conjugated events} It is known (see in these
proceedings \cite[Sect. 2.3]{candela06}) that a curve $\gamma: I
\rightarrow M$ with non-vanishing speed $\gamma'$ is a {\em
pregeodesic} \index{pregeodesic} (i.e., it can be reparametrized as
a geodesic for the Levi-Civita connection $\nabla$ of the spacetime)
if and only if it satisfies
\begin{equation} \label{gcon}
\nabla_{\gamma'}\gamma'=f  \gamma'
\end{equation}
for some function $f: I \rightarrow \R$. Explicitly, the
reparametrization is  $\tilde{\gamma} (\tilde s) = \gamma (s(\tilde
s))$ where, for  constants $\tilde s_0\in \R,  s_0 \in I, \tilde
s'_0\neq 0$,
\begin{equation} \label{eparam}
 \tilde s(s)=\tilde s_0 + \tilde s'_0 \int^{s}_{
s_0}e^{\int_{ s_0}^{t} f(r)\dd r} \dd t.
\end{equation}
 If $\gamma$ is a lightlike geodesic for
$g$ then it satisfies (\ref{gcon}) for the Levi-Civita connection
$\nabla^*$ of any conformal metric $ g^* = e^{2u} g$, being $f= 2
\frac{d(u\circ \gamma)}{dt}$ (see \cite{candela06} or the proof of
Theorem \ref{tinvconf} below) and, thus, with the natural choice of
$\tilde s'_0$ in (\ref{eparam}):
\begin{equation}\label{epregeod}
\tilde{\gamma} (\tilde s) = \gamma (s(\tilde s)) \; \hbox{lightlike
geodesic with} \quad \tilde \gamma' = e^{-2u} \gamma'.
\end{equation}
 That is, lightlike pregeodesics are (pointwise) conformally
invariant and the following definition for the conformal class makes
sense.

\begin{definition}
Given $(M, \bm{g})$, a lightlike curve $\bm{\gamma}$ is a {\em
lightlike geodesic} \index{lightlike geodesic}
 if for a choice of  representatives (and hence for
any choice), $g$ and $\gamma$,  equation (\ref{gcon}) holds.
\end{definition}
 Note that although the concept of
lightlike geodesic makes sense given only the conformal structure,
the definitions of timelike and spacelike geodesics do not. The
fact that two events have a zero time-separation is also
conformally invariant and, thus, the following definition makes
also sense.

\begin{definition} \index{maximizing lightlike curve}
A lightlike curve $\bm{\gamma}$ connecting two  events $p$ and $q$
is  {\em maximizing} if there is no timelike curve connecting $p$
and $q$.
\end{definition}
Recall that this concept is a pure conformal one, but the notion of
{\em maximizing}  for timelike curves depends on the metric.

The following result is standard, and relies on the possibility to
deform any causal curve which is not a lightlike geodesic without
conjugate points in a timelike one (see, for example, \cite[Cor.
4.14]{beem96}, \cite[Prop. 4.5.10]{hawking73} or  \cite[Prop.
2.20]{penrose72}). 
As discussed below Theorem \ref{tmaximglobal}, all these elements
are conformally invariant, and the result is stated consequently.

\begin{theorem} \label{t0} Let $(M,\g)$ be a spacetime.
\begin{itemize}
\item[(i)]  Each two events $p,q \in M $ connected by a causal curve
$\bm{\gamma}$ which is not a maximizing lightlike curve are also
connected by a timelike curve.
\item[(ii)] Any maximizing lightlike curve
 is a  lightlike geodesic of $\g$ without conjugate points (i.e., when reparametrized as a lightlike geodesic for any $g\in \g$,
 does not have conjugate points) except
at most the endpoints.
\end{itemize}
\end{theorem}
In fact, the timelike curve in (i) can be chosen arbitrarily close
(in the $C^0$ topology) to $\bm{\gamma}$. As a straightforward
consequence, one has:

\begin{proposition} \label{nhu}
Two events $p, q$ are horismotically related \index{horismotically
related events} if and only if they can be joined by a maximizing
lightlike geodesic. Thus:
\begin{itemize}
\item[(i)] If $p \ll r$ and $r \le q$ then $p \ll q$ (analogously, if $p
\le r$ and $r \ll q$ then $p \ll q$).
\item[(ii)] If $r \in E^{+}(p)$ and $q \in E^{+}(r)$ then either $q \in
E^{+}(p)$ or $p \ll q$.
\end{itemize}
\end{proposition}

The conformal invariance of  conjugate points along lightlike
geodesics is not only a consequence of  maximizing properties,
(which would be applicable only in a restricted way, for example, it
would apply only for the first conjugate point) but a deeper one.
Next, our aim is to show that the definition of Jacobi field
\index{Jacobi field} in the lightlike case can be made independent
of the metric and indeed depends only on the conformal structure. As
a consequence the concept of conjugate point and its multiplicity,
depends also only on the conformal structure for lightlike
geodesics, while in the timelike case it requires the metric. We
begin with the metric-dependent definition of Jacobi field, and show
later that  it can be made independent of the conformal factor in
the lightlike geodesic case.

\begin{definition} \label{def-conjugate-point}
Let $\gamma:I \to M$ be a geodesic of a spacetime (or any
semi-Riemannian manifold), $(M,g)$. A vector field $J$ on $\gamma$
is a {\em Jacobi field} if it satisfies the {\em Jacobi
equation}\index{Jacobi equation}
\[
J''+R(J,\gamma')\gamma'=0
\]
where $R$  is the (Riemann) curvature tensor, \index{ curvature
tensor}  $R(X,Y)= [\nabla_X, \nabla_Y]- \nabla_{[X, Y]}$. The events
$p=\gamma(\lambdas_p)$ and $q=\gamma(\lambdas_q),
\lambdas_p<\lambdas_q$ are said to be {\em conjugate} (of
multiplicity $m$) if there exist $m>0$ independent Jacobi fields
such that $J(\lambdas_p)=0= J(\lambdas_q)$.
\end{definition}
As in the (positive-definite) Riemannian case, one has:
\begin{lemma} For any geodesic $\gamma: I \rightarrow M$ of $(M,g)$: 
\begin{itemize}
\item[(i)] The variation vector field $V$ of $\gamma$  by means of
a variation $(s,v)\rightarrow \gamma^{v}(s)$ with geodesic
longitudinal curves (at constant $v$), 
is a Jacobi field. \item[(ii)] If $J$ is a Jacobi field for $\gamma$
then $g(J,\gamma')(s)=a\lambdas +b$ for suitable constants $a$ and
$b$ and all $s\in I$. Thus:
\begin{itemize}
\item[(a)]  If $J$ vanishes at the endpoints, then
$g(J,\gamma')=0$. \item[(b)] The only Jacobi fields proportional to
$\gamma'$, $J(s)=f(\lambdas)\gamma'(s)$ satisfy $f=c\lambdas+d$ for
suitable constants $c$ and $d$, hence if they vanish at the
endpoints they vanish everywhere. \item[(c)] If $J_1$ and $J_2$ are
two Jacobi fields vanishing at the endpoints and $J_2=J_1+f \gamma'$
for some function $f$, then they coincide.
\end{itemize}
\end{itemize}
\end{lemma}
As two causal vectors cannot be orthogonal, a straightforward
consequence of the (a) part is:




\begin{proposition}
Let $\gamma$ be lightlike and let $J$ be a Jacobi field which
vanishes at the endpoints but not everywhere, then $J$ is spacelike
and orthogonal to $\gamma'$. In particular, no lightlike geodesic
$\gamma:I \to M$ in a 2-dimensional spacetime \index{2-dimensional
spacetime} admits a pair of conjugate events.
\end{proposition}
 Indeed, the last assertion follows because no spacelike vector
field $J$ exists which is orthogonal to $\gamma'$.

In what follows $\bm\gamma$ will  be always lightlike. We are
interested in the case of conjugate points. It is convenient to
introduce the space $N(\gamma')$ of vector fields over $\gamma$
orthogonal to $\gamma'$ and the quotient \index{quotient space of
vector fields $Q$} space $Q$ of vector fields of $N(\gamma')$
defined up to additive terms type $f \gamma'$. If $X \in N(\gamma')$
is a vector field orthogonal to $\gamma'$ then $[X] \in Q$ will
denote its  equivalence class. Let $\pi:N(\gamma') \to Q$,
$\pi(X)=[X]$ be the natural projetion. The covariant derivative,
also denoted $'=\nabla_{\gamma'}$, can be induced on $Q$ by making
it to conmute with $\pi$, i.e. $[X]'=[X']$. This definition is
independent of the representative because:
\begin{align*}
(i)& \qquad X' \in N(\gamma'), \textrm{ since }
g(X',\gamma')=g(X,\gamma')'=0 \\
(ii)&\qquad [X+f\gamma']'=[X'+(f\gamma')']=[X'+f'\gamma']=[X'],
\end{align*}
Even more, the curvature term in Jacobi equation can be projected to
the map $\mathcal{R}:Q\to Q$ defined as:
\begin{equation}\label{jq0} \mathcal{R}[X]=\pi
(R(X,\gamma')\gamma'),
\end{equation}
which, again, is independent of the chosen representative $X$.
\begin{lemma}
If $J\in N(\gamma')$ is a Jacobi field then $[J]\in Q$ is a Jacobi
class, \index{Jacobi class} that is, it solves the {\em quotient
Jacobi equation} \index{quotient Jacobi equation}
\begin{equation}
\label{jq}
 [J]''+\mathcal{R}[J]=0
\end{equation}
(where the zero must be understood  in  $Q$, that is, as the class
of any vector field pointwise proportional to $\gamma'$).

Conversely, if $[\bar J]\in Q$ is a Jacobi class in the sense of Eq.
(\ref{jq}) and $J_p, J_q\in TM$ are orthogonal to $\gamma$ at
$\gamma(s_p), \gamma(s_q), s_p<s_q,$ with $[\bar J]_p=[J_{p}]$,
$[\bar J]_q=[J_q]$,  then there exist a representative $J\in
N(\gamma'), [J]=[\bar J],$ which is a Jacobi field and fulfills the
boundary conditions $J(\lambdas_p)=J_p$, $J(\lambdas_q)=J_q$. In
particular if $[\bar J]$ vanishes at the endpoints then there exists
a representative $J$ which vanishes at the endpoints.
\end{lemma}

\begin{proof}
The first statement is obvious. For the converse,
$\tilde{J}''+R(\tilde{J},\gamma')\gamma'=h \gamma'$ for some
suitable function $h$. Let $J$ be another representative,
$\tilde{J}=J+f\gamma'$, with $f''=h$. Then $J$ is a Jacobi field,
and given the initial conditions, $f_p=f(\lambdas_p)$ and
$f_q=f(\lambdas_q)$, function
\[
f(\lambdas)=\int_{\lambdas_p}^{\lambdas}\left(\int_{\lambdas_p}^{\lambdas'}h
\dd \lambdas''\right)\dd \lambdas'
+f_p+\frac{\lambdas-\lambdas_p}{\lambdas_q-\lambdas_p}\left[
f_q-f_p-\int_{\lambdas_p}^{\lambdas_q}\left(\int_{\lambdas_p}^{\lambdas'}h
\dd \lambdas''\right)\dd \lambdas'\right],
\]
solves the problem.
\end{proof}
These lemmas imply that in order to establish whether two events $p$
and $q$ are conjugate along a lightlike geodesic (and its
multiplicity, \index{multiplicity of Jacobi fields} i.e., the
dimension of the space of Jacobi fields vanishing at the endpoints)
it is easier to look for Jacobi fields vanishing at the endpoints in
the quotient space $Q$, as the reduced equation (\ref{jq}) collects
the relevant information.

\begin{theorem}\label{tinvconf} The quotient Jacobi equation (\ref{jq}) is invariant under conformal
transformations, \index{invariant under conformal transformations}
that is: if $g^*=e^{2u}g$, the curve $\gamma$ is a lightlike
$g$-geodesic, $\tilde \gamma$ is its parametrization as a
$g^*$-geodesic (given by (\ref{epregeod})), $J\in N(\gamma')$, and
$\tilde J\in N(\tilde \gamma')$ is the corresponding
reparametrization of $J$ on $\tilde \gamma$, then $[J]$ satisfies
Eq. (\ref{jq}) on $\gamma$ (taking $\mathcal{R}[J]$ from
(\ref{jq0})) if and only if $[\tilde J]$ satisfies Eq. (\ref{jq}) on
$\tilde\gamma$ (where $\mathcal{R}^*[\tilde J]$ is defined as
$\mathcal{R}^*[\tilde X]=\pi (R^*(\tilde X,\tilde\gamma')
\tilde\gamma')$, and $R^{*}$ denotes the curvature tensor  of
$g^*$).
%

Thus, the concept of {\em conjugate events $p$ and $q$ along a
lightlike geodesic} $\bm{\gamma}$, and its multiplicity, is
well-defined for the conformal structure $(M,\bm{g})$.
\end{theorem}

\begin{proof} We will put
$[X]\,\tilde{}=[\nabla^*_{\tilde\gamma'}X]$, $X' = \nabla_{\gamma'}
X$, and  use index notation as  in \cite[App. D]{wald84},
$a,b,c,d=0,\ldots, n-1$, (see \cite{candela06} for more intrinsic
related computations). It is proved in that reference:
\[
\nabla^*_a X^c=\nabla_a X^c+C^c_{a b}X^b ,
\]
where $C^c_{a b}=2 \delta^{c}_{(a} \p_{b)} u-g_{a b} g^{c d} \p_d 
u$, which implies that if $X\in N(\gamma') (=N(\tilde \gamma')$ up
to reparametrizations),
\[[X]\,\tilde{}=e^{-2u}[X'+C^c_{a b}X^b(\gamma')^a]=e^{-2u}[X'+u'X ]=e^{-2u}([X]'+u'[X]) , \]
and in particular
\begin{equation} \label{fds}
[X]\,\tilde{}\,\,\,\tilde{}=e^{-4u}([X]''+u''[X]-(u')^2[X]) .
\end{equation}
We use the transformation of the Riemann tensor under conformal
transformations (see, for example, \cite{wald84})
\begin{align*}
(R^*)^{d}_{\ cab}=&\, {R}^{d}_{\ cab}-2 \delta^{d}_{[a}
\nabla_{b]}\p_c u+2g^{de}g_{c[a} \nabla_{b]}\p_e u-2(\p_{[a}u)
\delta^{d}_{b]} \p_c u\\
&+2(\p_{[a}u) g_{b]c}g^{df}\p_f u+2g_{c[ a} \delta^{d}_{b]}
g^{ef}(\p_e u)(\p_f u). \end{align*} Using $\gamma'_a J^a=\gamma'_a
\gamma'^a=0$,
\begin{align*}
(R^*)^{d}_{\ cab}(\gamma\,\tilde{}\,)^{c} J^{a}
(\gamma\,\tilde{}\,)^{b}=&\,e^{-4u} \{ {R}^{d}_{\ cab}-2
\delta^{d}_{[a} \nabla_{b]}\p_c u+2g^{de}g_{c[a} \nabla_{b]}\p_e
u-2(\p_{[a}u)
\delta^{d}_{b]} \p_c u\\
&+2(\p_{[a}u) g_{b]c}g^{df}\p_f u+2g_{c[ a} \delta^{d}_{b]}
g^{ef}(\p_e u)(\p_f u)\}(\gamma')^{c}
J^{a} (\gamma')^{b}\\
=&  \,e^{-4u} \{ {R}^{d}_{\ cab}- \delta^{d}_{a} \nabla_{b}\p_c
u+(\p_{b}u) \delta^{d}_{a} \p_c u\}(\gamma')^{c} J^{a}
(\gamma')^{b}+f (\gamma')^d,
\end{align*}
for a suitable function $f$. This equation reads (up to
reparametrizations)
\[
\mathcal{R}^*[J]=e^{-4u} ({\mathcal{R}}[J]-u''[J]+(u')^2[J]) ,
\]
which together with Eq. (\ref{fds}) for $X=J$, gives the thesis.
\end{proof}


%
%

\section{The causal hierarchy} \label{s3}
As explained in the Introduction, the aim of this section is to
construct the {\em causal ladder}, \index{causal ladder} a
hierarchy of spacetimes \index{hierarchy of spacetimes} according
to strictly increasing requirements on its conformal structure.
Essentially, some alternative characterizations of each level will
be studied,  as well as some of its main properties, checking also
that each level is strictly more restrictive than the previous
one. At the top of this ladder {\em globally hyperbolic}
spacetimes appear. Even though somewhat restrictive, this last
hypothesis is, in some senses, as natural as {\em completeness}
for Riemannian manifolds. Even more, according to the {\em Strong
Cosmic Censorship Hypothesis}, \index{Strong Cosmic Censorship
Hypothesis} the natural (generic) models for physically meaningful
spacetimes are globally hyperbolic ones. So, these spacetimes are
the main target of Causality Theory, \index{Causality Theory} and
it is important to know exactly the generality and role of their
hypotheses.


\begin{figure}[ht]
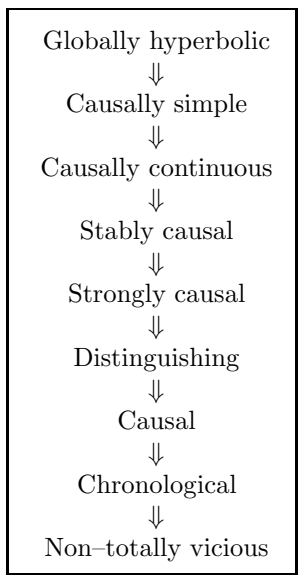

\begin{center} \fbox{
\begin{tabular}{c}
\vspace{-0.3cm}\\
{Globally hyperbolic} \\
$\Downarrow $ \\
{Causally simple} \\
$\Downarrow $ \\
{Causally  continuous} \\
$\Downarrow $ \\
{Stably causal} \\
$\Downarrow $ \\
{Strongly causal} \\
$\Downarrow $ \\
{Distinguishing} \\
$\Downarrow $ \\
{Causal} \\
$\Downarrow $ \\
{Chronological}\\
$\Downarrow $ \\
{Non--totally vicious}
\end{tabular}}
\end{center}
\caption{The causal ladder.} \index{causal ladder}
\end{figure}

\noindent Most of the levels are related to the non-existence of
travels to the past  either for observers travelling through
timelike curves (``grandfather's paradox''), \index{grandfather's
paradox} or for light beams, or for certain related curves. It is
convenient to distinguish between the following notions,
especially in the case of causal geodesics:

\begin{definition} Let $\gamma:[a,b]\rightarrow M$ be a piecewise-smooth curve with non-vanishing velocity at any point:

(a) $\gamma$ is a {\em loop (at $p$)} if $\gamma(a)=\gamma(b)=p$;
\index{loop}

(b) $\gamma$ is {\em closed} if 
it is smooth and $\bm{\gamma}'(a) = \bm{\gamma}'(b)$ (following our
convention in Remark \ref{rconv} for vectors). \index{closed curve}

(c) $\gamma$ is {\em periodic} if it is closed with
$\gamma'(a)=\gamma'(b)$ \index{periodic curve}
\end{definition}
Recall that if $\bm{\gamma}$ is a lightlike geodesic, the
properties of being closed or periodic are conformal invariant;
moreover, such a closed $\bm{\gamma}$ can be extended to a
complete geodesic if and only if it is periodic (see in these
proceedings \cite{candela06}). For non-lightlike geodesics, the
notions of closed and periodic become equivalent.

\subsection{Non-totally vicious spacetimes} \label{subs-vicio}
\index{non-totally vicious spacetimes}

Recall that if $p \ll p$ then there exist a timelike loop at $p$
and, giving more and more rounds to it, one finds $d(p,p)=\infty$.
Even more, if this property holds for all $p\in M$, then $I^+(p),
I^-(p)$ are both, open and closed. So, one can check easily the
following alternative definitions.

\begin{definition} \label{totvic} A spacetime $(M, \g)$ is called {\em totally vicious} if
it satisfies  one of the following equivalent properties:

\begin{itemize} \item[(i)] $\dl (p,q)=\infty$, $\forall p, q \in M$.
\item[(ii)] $I^+(p)=I^-(p)=M$, $\forall p \in M$. \item[(iii)]
Chronological relation is {\em reflexive}: $p\ll p$, $ \forall p
\in M$.
\end{itemize}
 Accordingly, a spacetime is {\em
non-totally vicious} if $p\not\!\ll p$ for some $p\in M$.
\end{definition}
Of course, it is easy to construct non-totally vicious spacetimes.
Nevertheless, totally vicious ones  are interesting at least from
the geometric viewpoint, and  sometimes even in physical
relativistic examples (G\"odel spacetime is the most classical
example). Let us consider an example. A spacetime $(M, g)$ is called
{\em stationary} if it admits a timelike Killing vector field
\index{Killing vector field} $K$; classical Schwarzschild, Reissner
Nordstr\"om or Kerr spacetimes (outside the event horizons) are
examples of stationary spacetimes. \index{stationary spacetime} This
definition depends on the metric $g$, but the fact that $K$ is {\em
conformal Killing} depends only on the conformal class $\g$.
Moreover, if $K$ is timelike and conformal Killing then it selects a
unique representative $g^K$ of \index{conformal Killing vector
field} $\g$ such that $g^K(K,K)\equiv -1$ (and then $K$  will be
Killing for $g^K$, as the conformal factor through the integral
lines of $K$ must be equal to 1; see also, for example, \cite[Lemma
2.1]{sanchez97}).

\begin{theorem} \label{ttotvic} \cite{sanchez06} Any  compact spacetime \index{compact spacetime} $(M,\g)$
which admits a timelike conformal Killing vector field $K$ is
totally vicious. \index{totally vicious}
\end{theorem}

\noi \begin{proof} (Sketch).  In order to prove that each $p\in M$
is crossed by a timelike loop, it is enough to prove that there
exists a timelike vector field $X$ with periodic integral curves.
Recall that $K$ is Killing not only for the selected metric $g^K=
-g/g(K,K)$ in the conformal class $\g$, but also for the associated
Riemannian metric $g_R$:
$$g_R(u,v) = g^K(u,v)+ 2 g^K(u,K)g^K(v,K), \forall u,v \in T_pM, p\in M. $$
Now, let $G$ be the  subgroup generated by $K$ of the isometry
group Iso$(M,g_R)$. Then its closure $\bar G$ satisfies:

\bit \item $\bar G$ is compact, because so is Iso$(M,g_R)$ (recall
that $g_R$ is Riemannian and $M$ is compact). \item $\bar G$ is
abelian, because so is $G$. \item As a consequence, $\bar G$ is a
$k-$torus, $k\geq 1$, and there exists a sequence of subgroups
$S_m$ diffeomorphic to $S^1$ which converges to $G$. \eit
 Finally, notice that the corresponding infinitesimal generator $K_m$ of
$S_m$ (which is Killing for $g_R$)  have periodic integral curves
and, for big $m$, are timelike for $g$. Thus, one can choose $X=K_m$
for large $m$.  \end{proof}

\begin{figure}[ht]
\begin{center} \psfrag{A}{$t=1/2$}
\psfrag{M}{$I^{+}(p)$}\psfrag{D}{$p$} \psfrag{G}{$t=-1/2$}
 \psfrag{J}{$x=1/3$} \psfrag{K}{$x=2/3$} \psfrag{F}{$x=1$}
\includegraphics[width=10cm]{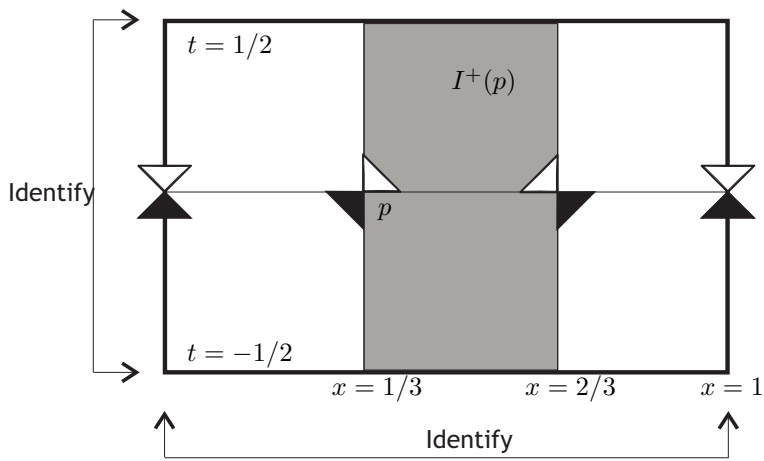}
\end{center}
\caption{Non-totally vicious and non-chronological torus with a
Killing vector field $K=\partial_t$. The vector field $K$ is
timelike everywhere except when $x=1/3, 2/3$, where it is
lightlike.}\label{f1}
\end{figure}

\begin{remark} The result is sharp: if  $K$ is allowed
to be lightlike in some points  there are counterexamples, see
Figure \ref{f1}.
\end{remark}


\subsection{Chronological spacetimes}
\index{chronological spacetimes}

\begin{definition} A spacetime $(M, \g)$ is called {\em chronological}  if
it satisfies  one of the following equivalent properties:
\begin{itemize}
 \item[(i)] No  timelike loop exists.
\item[(ii)] Chronological relation is {\em irreflexive}, i.e.,
$p\ll q \Rightarrow p\neq q$. \item[(iii)] $\dl (p,p)<\infty$ (and
then equal to 0) for all $p\in M$.
\end{itemize}
\end{definition}
A chronological spacetime is clearly non-totally vicious (see
Definition \ref{totvic}(iii)) but the converse does not hold, as
 Figure \ref{f1} shows. Notice that this example is compact and, in
fact, as a general fact:

\begin{theorem}  No  compact spacetime $(M,\g)$ \index{compact spacetime}
 is chronological.
\end{theorem}

\begin{proof} 
Recall the open covering of $M$: $\left\{ I^{+}\left( p\right)
,\,p\in M\right\}$. 
Take  a finite subrecovering $\left\{ I^{+}\left( p_{1}\right)
,I^{+}\left( p_{2}\right) ,\ldots ,I^{+}\left( p_{m}\right)
\right\}$ and, without loss of generality, assume that, if $i\neq
j$ then $ p_i\notin I^{+}\left( p_{j}\right) $ (otherwise,
$I^{+}\left( p_{i}\right) \subset I^{+}\left( p_{j}\right)$, and
 $I^{+}\left( p_{i}\right) $ can be removed).
Then, $ p_{1}\in I^{+}\left( p_{1}\right)$, as required.
\end{proof}

\subsection{Causal spacetimes}
\index{causal spacetimes}

\begin{figure}[h]
\begin{center} \psfrag{P}{$p$} \psfrag{A}{$C$}
\psfrag{B}{$\!\! C\backslash \{p\}$}
\includegraphics[width=10cm]{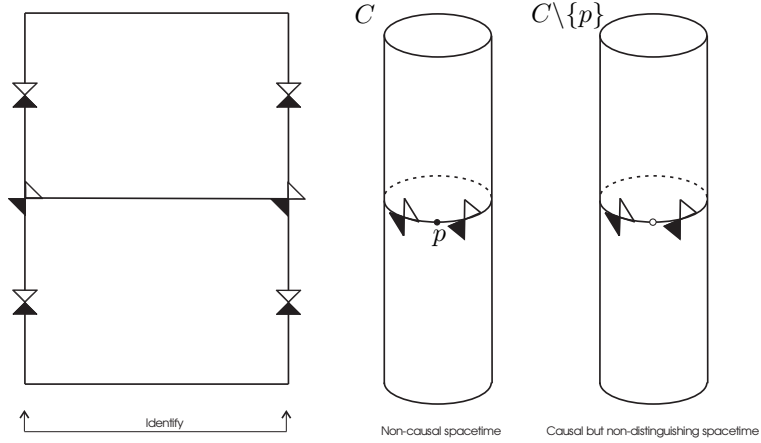}
\caption{Chronological non-causal cylinder, and causal but
non-distinguishing spacetime obtained by removing $\{p\}$. If one
also  removed   the vertical half line below $p$, a causal
past-distinguishing but non-future distinguishing spacetime would
be obtained.}\label{T3N8}
\end{center}
\end{figure}

\begin{definition} A spacetime $(M, \g)$ is called {\em causal} if
it satisfies  one of the following equivalent properties:

\begin{itemize}
\item[(i)] No  causal loop exists.
\item[(ii)] Strict causal relation is {\em irreflexive}, i.e.,  $p < q \Rightarrow p\neq q$.
\end{itemize}
\end{definition}
The following possibility is depicted in Figure \ref{T3N8}.
\begin{theorem} A chronological but non-causal spacetime $(M, \g)$
admits a closed lightlike geodesic.
\end{theorem}
 \begin{proof} Take a causal loop
 $\bm{\gamma}$
at some $p\in M$. If $\bm{\gamma}$ were not a lightlike geodesic
loop then  $p\ll p$ (Theorem \ref{t0}), in contradiction with
chronology condition. And if  $\bm{\gamma}$ were not closed, run
it twice to obtain the same contradiction. \end{proof}


%

Now, recall relation $\rightarrow^{(\le)}$ in Definition
\ref{binaryrelations}.

\begin{theorem} \label{pk3}
In any causal spacetime  $(M,\g)$: $x\rightarrow^{(\le)} y
\Leftrightarrow x \rightarrow y$.
\end{theorem}

\begin{proof} $(\Leftarrow)$.
If $x \rightarrow y$, $x\neq y$, then $x$ and $y$ are connected by a
(non-necessarily unique)  maximizing lightlike geodesic contained in
$J^{+}(x) \cap J^{-}(y)$. Taken $x<p<q<y$, the points $p$ and $q$
must lie on a unique maximizing lightlike geodesic $\bm{\gamma}$,
which will also cross $x$ and $y$ (otherwise, there would be a
broken causal curve joining $x$ with $y$, and hence $y \in
I^{+}(x)$). Thus, $J^+(p)\cap J^-(q)$ is nothing but the image of a
portion of $\bm{\gamma}$,  which can be either homeomorphic to a
segment joining $p$ to $q$, or to a circumference (the latter
excluded by the causality of $(M,\g )$).

$(\Rightarrow)$. If $x\rightarrow^{(\le)} y$, $x\neq y$,  there are
$(p_n, q_n)$, $x<p_n<q_n<y$, $p_n \to x$, $q_n \to y$, such that
$J^{+}(p_n) \cap J^{-}(q_n)$ is linearly ordered; in particular, $x
< y$. But clearly $x \not\!\ll y$ because, otherwise, as $I^{+}$ is
open,  $q_n \in I^{+}(p_n)$ for large $n$. That is,
 the open set $ I^{+}(p_n) \cap I^{-}(q_n)$ would be non empty,
which clearly makes $J^{+}(p_n) \cap J^{-}(q_n)$ non-isomorphic to
$[0,1]$.
\end{proof}

\subsection{Distinguishing spacetimes}

The {\em set of parts of} $M$, i.e.,  \index{set of parts ${\cal
P}(M)$} the set of all the subsets of $M$, will be denoted  ${\cal
P}(M)$. Here it is regarded as a point set, but it will be
topologized later (see Proposition \ref{linneroutercont}).

\index{distinguishing spacetime} The equivalence between some
alternative definitions of {\em distinguishing} is somewhat
subtler than in previous cases
 \cite{kronheimer67,senovilla97}. So, we need the following previous result, which is proved below.

 \begin{lemma} \label{ldist}
The following properties are  equivalent for $(M,\g)$:
\begin{itemize}
\item[(i)] $I^{+}(p)=I^{+}(q)$ (resp. $I^{-}(p)=I^{-}(q)$)
$\Rightarrow p=q$, \item[(ii)] The set-valued function $I^+$
(resp. $I^-$) $:M\rightarrow  {\cal P}(M)$, $p\rightarrow I^+(p)$
(resp. $p\rightarrow I^-(p)$), is one to one, \item[(iii)] Given
any $p\in M$ and any neighborhood $U\ni p$ there exists a
neighborhood $V\subset U$, $p\in V$, which {\em distinguishes $p$
in $U$ to the future} \index{distinguishes $p$ in $U$ (future,
past)} (resp. past) i.e. such that any future-directed (resp.
past-directed) causal curve ${\gamma}: I=[a,b]\rightarrow M$
starting at $p$ meets $V$ at a connected subset of $I$ (or,
equivalently, if $p={\gamma}(a)$ and ${\gamma}(b)\in V$ then
$\gamma$ is entirely contained in $V$).
\item[(iv)] Given any $p\in
M$ and any neighborhood $U\ni p$ there exists a neighborhood
$V\subset U$, $p\in V$, such that $J^{+}(p,V)=J^{+}(p)\cap V$ (resp.
$J^{-}(p,V)=J^{-}(p)\cap V$).
\end{itemize}

\end{lemma}

\begin{definition} \label{ddist}
A spacetime $(M, \g)$ is called {\em future (resp. past)
distinguishing} if \index{distinguishing spacetime} it satisfies
one of the equivalent properties in Lemma \ref{ldist}. A spacetime
is {\em distinguishing} if it is both, future and past
distinguishing.
\end{definition}

\begin{proof} (Lemma \ref{ldist} for the future case.) (i) $\Leftrightarrow$ (ii) and  (iii) $\Rightarrow$  (iv) Trivial.

No (i) $\Rightarrow$ no (iii).  Let $p\neq q$ but $I^+(p)=I^+(q)$,
take $U\ni p$ such that  $q\not\in \bar U$ and any $V\ni p$,  $V
\subset U$. Then, choose $p'\in V, p\ll_V p'$ and any $q'\not\in
U, q'\neq q,$ on a future-directed timelike curve $\bm{\gamma_1}$
which joins $q$ with $p'$. The required $\bm{\gamma}$ is obtained
by joining $p, q'$ with a future-directed timelike curve
$\bm{\gamma_0}$, and then $q'$ and $p'$ through $\bm{\gamma_1}$.

No (iii) $\Rightarrow$ no (i). Let $U\ni p$ be a neighborhood where
(iii) does not hold, that is,  every $V \subset U$ intersects a
suitable ($V$-dependent) future inextendible causal curve starting
at $p$ in a disconnected set of its domain $I$. Take the sequence
$\{V_n\}_n$ of nested globally hyperbolic neighborhoods in Theorem
\ref{fund}. They will be causally convex in some $U'\subset U$ and
we can assume $U=U'$ (if (iii) does not hold for the pair $(p,U)$
then it does not hold for the pair $(p,U')$, $U' \subset
U$), being $U$  also 
with closure contained in a simple neighborhood $W$. For each $V_n$,
the causal curve $\bm{\gamma}_n$ which escapes $V_n$ and then
returns $V_n$ also escapes $U$ (because of causal convexity) and
then returns to some point in the boundary $q_n \in \dot U$ which is
the last one outside $U$, and to another point $p_n\in V_n$. As $W$
was simple, $\{q_n\}\rightarrow q\in \dot U$, up to a subsequence.
Even more, $q\in J^-(p,W)$, because $q_n \in J^-(p_n,W)$, $(q_n,p_n)
\to (q,p)$, and $J^{-}$ is closed on any convex neighborhood (see
Prop.
\ref{plocal}).  
Thus, $I^+(p) \subset I^+(q)$. Moreover, let $q'\in I^+(q)$ then,
for large $n$, $(p \le) q_n \ll q'$, that is $q'\in I^+(p)$,
$I^{+}(q) \subset I^{+}(p)$.

No (iii) $\Rightarrow$  No (iv). Follow the reasoning in the last
implication, with the same assumptions on $U$, and assuming that
such a $V$ as in (iv) exists. Notice that connecting the obtained
$q\in J^-(p,W)$ (satisfying $I^+(p)=I^+(q)$) with $p$ by means of
the unique geodesic $\rho$ in $W$, one point $q_V\in (V \cap
\rho)\backslash \{p\}$ will also satisfy $I^+(p)=I^+(q_V)$. But,
as $U$ is convex, $J^+(p,U)\neq J^+(q_V,U)$ (use Prop.
\ref{plocal}) and, even more\footnote{These statements can be
strengthened, as any convex subset $U$ is, in fact, causally
simple and, thus, any open neighborhood of $U$ is not only
distinguishing, but stably (and strongly) causal.}, this  holds
arbitrarily close to $q_V$. Concretely, $(J^+(p)\cap V=) $
$J^+(p,V)\not\supset I^+(q_V,V)$ $ (\subset I^+(q_V)\cap V \subset
J^+(p)\cap V)$, a contradiction.
\end{proof}

\begin{remark}\label{rdist} (1) One can give easily another two alternative
characterizations of being distinguishing, say (iii'), (iv'), just
by replacing causal curves and futures in (iii), (iv) by timelike
curves and chronological futures.

(2) Notice that, Lemma \ref{ldist} also allows to define in a
natural way what means to be {\em distinguishing at $p$}. In this
case, for any neighborhood $U$ of $p$, a  neighborhood $V$ which
{\em distinguishes $p$ in $U$} \index{distinguishes $p$ in $U$
(future, past)} satisfies   (iii)  (and, thus, (iv)) for future
and past causal curves. Notice also that, given $U$, one can find
another neighborhood $U'$ and a sequence of nested neighborhoods
$V_n \subset U'$ such that $\cap_n V_n=\{p\}$ and each $V_n$ is
causally convex in $U'$ (see also Theorem \ref{fund}).

(3) Note that if $V$ future-distinguishes $p$ in $U$ then it also
future-distinguishes in $U$ any other point $q$ on a future directed
causal curve $\rho$ starting at $p$ contained in $V$.

(4) Obviously, any past or future distinguishing spacetime is
causal (if $p, q$ lie on the same closed causal curve then
$I^\pm(p)=I^\pm(q)$), but the converse does not hold (Fig.
\ref{T3N8}).
\end{remark}

A remarkable property of distinguishing spacetimes (complementary
to Prop. \ref{tdistinguishing} below) is the following
\cite{malament77b}.
\begin{proposition} \label{kjh}
Let $(M_1,\bm{g}_1)$, $(M_2,\bm{g}_2)$ be two spacetimes,
$(M_1,\bm{g}_1)$ distinguishing, and $f:M_1 \to M_2$ a
diffeomorphism which preserves $\leq$, that is, such that: $p \leq
q \Leftrightarrow f(p) \leq f(q)$. Then $(M_2,\bm{g}_2)$ is
distinguishing and $\g_1=f^{*}\g_2$.
\end{proposition}

\begin{proof}
Let us show that  $(M_2,\bm{g}_2)$ is distinguishing. First note
that since $f$ is bijective it preserves also $<$. Take $p_2\in
M_2$, $U_2 \ni p_2$ and let $p_1=f^{-1}(p_2)$, $U_1=f^{-1}(U_2)$.
Let $V_1 \subset U_1$ be a neighborhood which distinguishes $p_1$ in
$U_1$, and let us check that $V_2=f(V_1) \subset U_2$ distinguishes
$p_2$ in $U_2$. Otherwise, there would be a causal curve $\gamma_2$
intersecting $V_2$ in a disconnected set of its domain. In
particular one could choose points on the curve $p_2^1 < p_2^2 <
p_2^3$ such that $p_2^1, p_2^3 \in V_2$, $p_2^2 \notin V_2$, hence
$p_1^{i}=f^{-1}(p_2^i)$, $i=1,2,3$, would satisfy the same property
with respect to $V_1$ a contradiction.

Let $p\in M$, $g_1 \in \bm{g}_1$, $g_2 \in \g_2$, two metric
representatives, $U_1\ni p$, $U_2 \ni f(p)$ two simple
neighborhoods
 with respect to the metric structures $(M,g_1)$ and $(M,g_2)$, and
$V_1 \ni p$, $V_1 \subset U_1$ a neighborhood such that
$V_2=f(V_1) \subset U_2$ and both,  $V_1$ and $V_2$ distinguish
$p_1$, $p_2$ in $U_1$, $U_2$, respectively. Then:
\begin{eqnarray*}
f(J_1^{+}(p,V_1))&=&f(J_1^{+}(p)\cap V_1) =f(J_1^{+}(p))\cap f(V_1)\\
&=&J_2^{+}(f(p))\cap V_2=J_2^{+}(f(p),V_2)\subset U_2.
\end{eqnarray*} The causal cones on $T_pM_1$ for the conformal structure $\g_1$ are determined through the exponential
diffeomorphism from the knowledge of $J^{+}_{1}(p,V)$ and since it
coincides up to a pullback with $J^{+}_{2}(p,V)$ we conclude by
Lemma \ref{confm} that $\g_1=f^{*}\g_2$.
\end{proof}

\begin{remark} (1) Again, an obvious timelike version of this result holds.

(2) Particularly interesting is the case in which $f$ is the
identity map, as it states that,  {\em in a distinguishing
spacetime, $J^{+}$ (as well as $I^+$) determines the metric, up to
a conformal factor}.

\end{remark}

\subsection{Continuous causal curves} \label{sec-causcont}
When questions on convergences of curves are involved, the space
of piecewise smooth causal curves is not big enough. So, the
following extension of these curves (which becomes especially
interesting in strongly causal spacetimes) is used.

\begin{definition} \label{continc}
A continuous curve $\gamma:I \to M$ is  {\em future-directed
causal}  at $t_0 \in I$ if for any convex neighborhood $U \ni
\gamma(t_0)$ there exist an interval $G\subset I$,
$\gamma(G)\subset U$, such that $G$ is an open neighbourhood of
$t_0$ in $I$, and satisfies: if $t' \in G$ and $t'<t_0$ (resp.
$t_0<t'$) then $\gamma(t') <_U \gamma(t_0)$ ($\gamma(t_0) <_U
\gamma(t')$). The continuous curve $\gamma$ is said to be {\em
future-directed causal} if it is so at any $t \in I$.

The definition for past-directed is done dually.
\end{definition}
Recall that it is enough to check this definition for one convex
neighbourhood $U$ of $\gamma(t_0)$ and, if $t_0$ is not an extreme
of $I$, then $G$ is just some  open neighbourhood of $t_0$
included $I$. In fact:

\begin{proposition}
A continuous curve $\gamma:I \to M$ is a  future-directed causal
if and only if for each convex neighbourhood $U$, given $t,t' \in
I, t<t'$ with ${ \gamma}([t,t']) \subset U$, it is ${ \gamma}(t)
<_U { \gamma} (t')$.
\end{proposition}
\begin{proof} For the left is trivial. For the converse,  let
$\gamma([t_0,t_1]) \in U$, and assume by contradiction that ${
\gamma}(t_0) \not <_U { \gamma} (t_1)$. Then there is a maximal
$\bar{t} \in (t_0,t_1)$, such that $\gamma(t_0) <_U \gamma(t)$ for
any $t \in (t_0,\bar{t}]$. But  since $\gamma$ is also
future-directed causal at $\bar t$ (and the causal relation is
transitive) a contradiction with maximality is obtained.
\end{proof}
 Note that several causal properties of piecewise
smooth curves hold naturally in the continuous case. For instance,
if $\gamma$ is a continuous causal curve which connects $p$ to $q
\in E^{+}(p)$ then $\gamma$ is a maximal lightlike pregeodesic.

\begin{remark}
This definition could be extended naturally to continuous timelike
curves. Nevertheless, recall that a possibility of confusion appears
here. The curve in $\L^2$, $\gamma(t)=(\tan t, t)$, regarded as a
continuous curve, would be  future-directed timelike. Nevertheless,
regarded as a (piecewise) smooth curve, it is not timelike at $t=0$,
as $\gamma'$ is lightlike there.
\end{remark}

\begin{remark} \label{lipschitz}  A (future-directed, continuous) causal curve
${ \gamma}$ must be locally Lipschitzian \index{Lipschitzianity of
causal curves} (when suitably reparametrized, for the distance
associated to any auxiliary Riemannian metric)  and, thus, almost
everywhere differentiable \cite[2.26]{penrose72}. If the interval
$I$ is compact, then ${\gamma}$ becomes Lipschitzian with finite
integral of its length. Therefore, it is absolutely continuous, and
contained in a Sobolev space $H^1$, see \cite[Appendix]{candela06}
for a detailed study. \index{Sobolev space of causal curves}

Notice that a continuous curve ${ \gamma}$, a.e. differentiable,
with timelike gradient (in the same time-orientation at each
differentiable point) and finite integral of its length, is not
necessarily a continuous causal curve. A counterexample in
Lorentz-Minkowski spacetime $L^2$ can be constructed as follows.
Consider a Cantor curve $t\rightarrow x(t), t\in[0,1]$ which is
continuous, with 0 derivative a.e., and connects $x(0)=0, x(1)=2$.
Now, the curve in natural coordinates of $L^2$, $\gamma(t)=
(x(t),t)$ satisfies all the required properties, but connects the
non-causally related points $(0,0), (2,1)$. In order to be a
causal curve, ${ \gamma}$ must be additionally (locally)
Lipschitzian. In fact, it is possible to prove \cite{candela06}:
{\em let $(M, g)$ be a spacetime, and $\gamma: [a,b] \rightarrow
M$ a continuous curve. Then $\gamma$ is
 future-directed causal if and only if $\gamma$ is $H^1$ (up to a reparametrization)  and $
\gamma'(s)$ is a future-directed causal vector for $s\in I$  a.e. }
\end{remark}

Continuous causal (and timelike) curves can be characterized in
distinguishing spacetimes as follows:

\begin{proposition} \label{tdistinguishing}
Let $(M,\g)$ be a  distinguishing spacetime. A continuous curve
$\gamma: I\rightarrow M$ is causal (either past or future
directed) if and only if it is
 totally ordered by $<$ (i.e.  for
any pair $t_1,t_2 \in I$, $t_1 < t_2 $,  $p= \gamma(t_1)$,
$q=\gamma(t_2)$, either $p < q$ or $q < p$).
\end{proposition}
A timelike version of the previous theorem (in fact easier to
prove), where $<$ and {\em causal} are replaced with $\ll$ and {\em
timelike}, also holds (for smooth versions see
\cite{garciaparrado03}).

\begin{proof} To the right is trivial. For the converse, we claim first that the causal relation
$<$ is consistent along $I$, that is, either $t<t' \Rightarrow
\gamma(t) < \gamma(t')$, or $t<t' \Rightarrow \gamma(t') <
\gamma(t)$ on all $I$ (the first possibility will be assumed
below). In fact, let us check that, in the case $t_1<t_2$ and
$\gamma(t_1) < \gamma(t_2)$, then $t_1<t_3 \Rightarrow \gamma(t_1)
< \gamma(t_3)$. Otherwise, since the spacetime is distinguishing,
defined $L^{\pm}(p)=J^{\pm}(p)\backslash\{p\}$, it is, $
\bar{L}^{+}(p)\cap {L}^{-}(p)={L}^{+}(p)\cap
\bar{L}^{-}(p)=\emptyset$. Thus, putting $p=\gamma(t_1)$, there is
$\bar{t}$ included in either $[t_2,t_3]$ or $[t_3,t_2]$ such that
$r=\gamma(\bar{t}) \in \bar{L}^{+}(p)\cap \bar{L}^{-}(p)$, which
is impossible because either $r \in L^{+}(p)$ or $r \in L^{-}(p)$.
Using a similar reasoning, $t_3 < t_1$ implies $\gamma(t_3) <
\gamma(t_1)$, and the claim follows.

Now, let $t_0 \in I$, $p_0=\gamma(t_0)$, $U$ a convex neighborhood
of $p_0$ and $V\ni p_0$, $V\subset U$ a neighborhood which
distinguishes $p_0$ in $U$. If $t_0<t$ we have $p_0<\gamma(t)$
and, thus, $p_0<_V\gamma(t)$, $p_0<_U\gamma(t)$, and the result
follows.
\end{proof}

\begin{remark}
This property does not characterize exactly distinguishing
spacetimes. Indeed, the reader may convince him/herself that in
the causal past-distinguishing but non-future distinguishing
spacetime obtained from Figure \ref{T3N8} by removing a vertical
half-line, any continuous curve $\gamma$ totally ordered by $<$ is
either a future or a past directed causal continuous curve.
\end{remark}


\subsection{Strongly causal spacetimes} \label{strongsect}
\index{strongly causal spacetimes} The following equivalence is straightforward by using Theorem \ref{fund}: 
\begin{lemma} \label{lstr}
For any event $p$ of a spacetime $(M,\g)$, the following sentences
are equivalent:

\begin{itemize}
\item[(i)]  Given  any neighborhood $U$ of $p$ there exists a
neighborhood $V\subset U$, $p\in V$ (which can be chosen globally
hyperbolic), such that $V$ is causally
convex in $M$ -and thus in $U$. 
\item[(ii)] Given any neighborhood $U$ of $p$ there exists a
neighborhood $V\subset U$, $p \in V$, such that any
future-directed (and hence also any past-directed) causal curve
$\bm{\gamma}: I\rightarrow M$ with endpoints at $V$ is entirely
contained in $U$.
\end{itemize}
\end{lemma}

\begin{definition} A spacetime $(M, \g)$ is called {\em strongly causal at $p$}  if
it satisfies one of the  equivalent properties in Lemma
\ref{lstr}. A spacetime is {\em strongly causal} if it is strongly
causal at $p$, for any $p\in M$.
\end{definition}

\begin{figure}[h]
\begin{center} \psfrag{V}{$V$}\psfrag{G}{$\bm{\gamma}$}
\psfrag{P}{$\, p$}
\includegraphics[width=12cm]{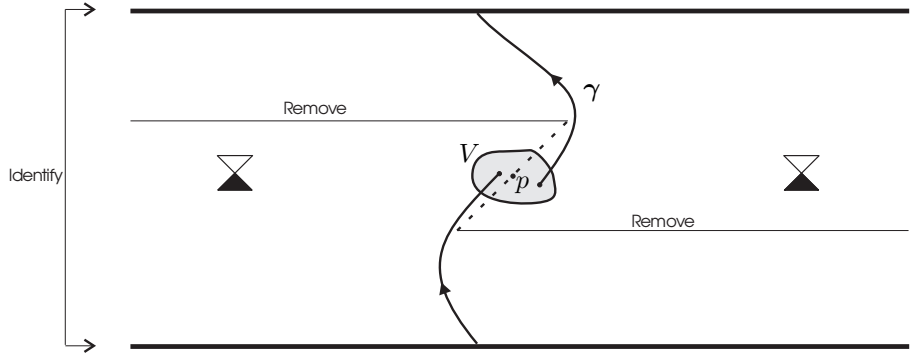}
\caption{Distinguishing non-strongly causal spacetime.}\label{T3N10}
\end{center}
\end{figure}

\begin{remark} \label{rstcaus}
(1)  Item (i) in Lemma \ref{lstr} collects the intuitive idea
that, in a strongly causal spacetime, no causal ``almost closed
curve or loop'' exist. Moreover, it also shows that strongly
causal spacetimes are distinguishing (but the converse does not
hold, see Fig. \ref{T3N10}). Item (ii) is more frequently used as
definition.

(2)  Notice that $V$ in item (i) can be assumed included in a
normal neighbourhood, and a nested sequence $\{V_n\}_n$ of
globally hyperbolic neighborhoods as in Theorem \ref{fund} can be
taken. Then, chosen any representative $g \in \g$, the causal
relations and time-separation function on each $V_n$ regarded as a
spacetime, agrees with the time-separation on $M$ restricted to
$V_n$.
\end{remark}

Finally, recalling the binary relations in Definition
\ref{binaryrelations}:
\begin{theorem} \label{pk1}
In a strongly causal spacetime, $x\le ^{(\rightarrow )}y
\Leftrightarrow x \le y$.
\end{theorem}

\begin{proof} ($\Rightarrow$). It holds trivially in any
spacetime.

($\Leftarrow$) Recall first the following claim: {\em in any
spacetime, if $p < q$ then $p$, $q$ can be connected by means of a
piecewise smooth future-directed lightlike curve $\gam$, such that
each unbroken piece is a geodesic without conjugate points.} In
particular, the required implication holds trivially in any convex
neighborhood $U$ regarded as a spacetime.

Thus, let $\gamma: [0,1]\rightarrow M$ be one such one such unbroken
future-directed geodesic piece of a  curve connecting $x$ and $y$.
Choose for each $p\in \gam$ a convex neighborhood $U$ and a causally
convex neighborhood $V\subset
U$. 
Taking  $\epsilon=1/m$ for some large integer $m$ (a  Lebesgue
number of the covering), each consecutive $\gamma(k\epsilon),
\gamma((k+1)\epsilon)$, $(k=0, \dots, m-1)$ lie in one such $V$ and,
thus, satisfy $J^{+}(q,V) =J^{+}(q)\cap V$ for any $q \in
\gamma([k\epsilon, (k+1)\epsilon])$ (see Remark \ref{rdist}(3)).
Therefore, $\gamma(k\epsilon) \le ^{(\rightarrow )}
\gamma((k+1)\epsilon)$, as required.

Finally, in order to prove the claim is sufficient to check that,
given a timelike curve $\rho$, for any point $p=\rho(t_0)$ and a
sufficiently small $\delta >0$ the events $p=\rho(t_0)$ and
$p_\delta =\rho(t_0+\delta)$ can be connected by means of one such
$\gam$ with one break. This can be checked by taking a convex
neighborhood $W$ of $p$ and noticing that, for small $\delta$, any
past directed lightlike geodesic starting at $p_\delta$ will cross
$E^+(p,W)$.

\end{proof}

\noi The properties below justify that strong causality is one of
the most important assumptions on causality.

\subsubsection{Characterization with Alexandrov's topology}
\index{Alexandrov's topology} The following topology can be
defined in any set with a binary relation type $\ll$.

\begin{definition} Let $(M, \g )$ be a spacetime.
Alexandrov's topology ${\cal A}$  on $M$ is the one which admits
as a base
 the subsets:

$$  B_{\cal A} = \left\{ I^{+}\left( p\right) \cap
I^{-}\left( q\right) : \,p,q\in M\right\}
$$
\end{definition}

\begin{remark} \label{ralex} Its easy to check that $B_{\cal A}$ is always a base for some topology. Notice also that, for any $p,q \in M$, $I^{+}(p)\cap I^{-} (q)$ is
open, thus the manifold topology is finer than  Alexandrov's.
\end{remark}

\begin{theorem} For a spacetime $(M,\g )$, the following properties are equivalent:
\ben \item[(i)] $(M,\g )$ is strongly causal. \item[(ii)]
Alexandrov's topology ${\cal A}$ is equal to the original topology
on $M$. \item[(iii)] Alexandrov's topology is Hausdorff. \een
\index{Hausdorff (topology)}
\end{theorem}

\begin{proof}
$(i) \Rightarrow (ii)$. From Remark \ref{ralex}, we have just to
show that for any open set $U$ and $x\in U$, there are $p,q \in
M$, such that $x \in I^{+}(p)\cap I^{-}(q) \subset U$. To this end
let $V\subset U$, $x\in V$ such that $\ll_V$ agrees $\ll$ on $V$
(Remark \ref{rstcaus}(2)), and any pair $p\ll_V x$, $ q \gg_V x$
suffices.


$(ii) \Rightarrow (iii)$. Trivial.

No $(i) \Rightarrow$ No $(iii)$. Assume that strong causality
fails at $p\in M$. Reasoning as  Lemma \ref{ldist} (implication No
$(iii) \Rightarrow$ No $(i)$), take a simple neighborhood $U\ni
p$, $W\supset U$ convex, a sequence of nested globally hyperbolic
neighborhoods $\{V_n\}_n$ causally convex in $U$, and a sequence
of future-directed causal curves $\{\bm{\gamma_n}\}_n$, each one
with endpoints $p_n, p'_n\in V_n$, and such that $\bm{\gamma_n}$
escapes $W$ and comes back at some last  point $q_n\in \dot U$,
$\{q_n\}\rightarrow q \in \dot U$ up to a subsequence. So,
$q_n\leq_W p'_n$, $q \leq_W p$ (since $J^{+}$ is closed in $W$),
and hence $q \le p$.

Now, recall that, if $q_1 \ll q \ll q_2$ then $q_1 \ll p$ and $p_n
\ll q_2$ for large $n$. Thus, $p$ is an accumulation point of the
Alexandrov open set $I^+(q_1)\cap I^-(q_2)$ and, so, this open set
is intersected by any (Alexandrov) open set which contains $p$, as
required.

\end{proof}

 \subsubsection{Non-imprisoning spacetime} \index{non-imprisoning spacetime} Strongly causal
 spacetimes will be non-imprisoning, that is, they will not
 contain any type of (partially) imprisoned causal curves,
 according to the following definitions.

\begin{definition} Let $(M,\g)$ be a spacetime, and let $\gamma: {[a,b)} \rightarrow M$, be a causal
curve with no endpoint at $b$. Then:
\begin{itemize}
\item[(i)]
 $\gamma$ is imprisoned (towards $b$) if, for some $\delta (\in
(0,b-a))$, then $\gamma \left( \left[ b-\delta ,b\right) \right)
\subset K$, for some compact subset $K$. \index{imprisoned curve} 
\item[(ii)] $\gamma$ is partially imprisoned (towards $b$) if, for
some sequence $\left\{ t_{m}\right\} \nearrow b$, then $\gamma
\left( t_{m}\right) \in K,\,\forall m\in \mathbb{N}$, for some
compact subset $K$. \index{partially imprisoned curve}
\end{itemize}
\end{definition}
(Analogous definitions holds for $\gamma$ when defined on $(a,b]$.)
The following result is easy to prove.

\begin{proposition} \label{pimpr} In a strongly causal spacetime, any causal curve $\gamma: {[a,b)} \rightarrow M$, with no endpoint at $b$
is a proper
function (i.e., if $K\subset M$ is compact then $\gamma^{-1}(K)$
is compact).
\end{proposition}

\begin{remark} Classical alternative statements of Proposition
\ref{pimpr} are:

(a) If $\gam $ crosses the compact subset $K$, it leaves $K$ at
some point and never returns.


(b) Curve $\gam$ 
is not partially
imprisoned (nor imprisoned) in any compact $K$. 

\end{remark}

\subsubsection{Limits of causal curves}\label{subsubseclimit} In the remaining of this subsection we
will consider continuous causal curves\footnote{An alternative
approach to the present study of limits of curves is developed by
O'Neill \cite{oneill83} by using the notion of  {\em
quasi-limit}\index{quasi-limit} Here, we follow essentially
\cite{beem96} and \cite{penrose72}, where we refer for detailed
proofs.}, according to Definition \ref{continc}.

\begin{definition} \label{dlimitscurves}
Let $\{\gam_k\}_k$ be a sequence of causal curves
in a spacetime $(M,\g)$. \bit \item
 A curve $\gam$ is a limit curve
 of
 $\{ \gam _k\}_k$ if there exists a subsequence
 $\{\gam_{k_m}\}_m$ which {\em distinguishes $\gamma$}, i.e., such that:

for all $p\in \gam$,  any neighborhood of $p$ intersects all
$\{\gam_{k_m}\}_m$ but a finite number of indexes.
\index{distinguishing subsequence (limit of curves)}

 \item Assume that all the $\gam _k$'s can be reparametrized in a compact interval $I=[a,b]$.
 A curve $\gamma:I \rightarrow M$ is a limit in the $C^0$ topology of
 $\{\gamma_k\}_k$ if:

(i)$\{\gamma_k(a)\}\rightarrow \gamma(a)$,
$\{\gamma_k(b)\}\rightarrow \gamma(b)$

(ii)  any neighborhood $U$ of $\gamma$ contains all
$\gamma_{k}$'s,  but  a finite number of $\gamma_k$. \eit

\end{definition}

\begin{remark} In general, these limits may be very bad behaved. For
example, consider the quotient torus $T^2=\L^2/\Z^2$ and the
projection $\gamma$ of the timelike curve $t\rightarrow (t,rt) \in
\L^2$, where $r$ is an irrational number, $|r|<1$. Then, any other
curve $\rho$ in $T^2$ is a limit curve of the sequence
$\{\gamma_n\}_n$ constantly equal to $\gamma$.
\end{remark}
The properties of these limits are well-known (see for example
\cite[Ch. 3]{beem96}, \cite[Ch. 6,7]{penrose72}), and remarkable
ones appear in the strongly causal case. Summing up:
 \ben \item
Any sequence of causal curves $\{\gamma_k\}_k$ without endpoints
which admits a point of accumulation $p$, admits an inextendible
causal limit curve $\gamma$ which crosses $p$. This result can be
obtained by applying  Arzela's theorem \cite[p. 76]{beem96}. \item
In strongly causal spacetimes: \bit \item All limit curves are
causal \cite[Lemma 2.39]{beem96} (and no inextendible limit curve
$\gamma$ can be contained in a compact subset). \item Given a
sequence $\{\gamma_k: I \rightarrow M\}, I=[a,b]$ which satisfies
$\{\gamma_k(a)\}\rightarrow \gamma(a)$,
$\{\gamma_k(b)\}\rightarrow \gamma(b)$, one has: $\gamma$ is a
limit curve of $\{\gamma_k\} \Leftrightarrow $ $\gamma$ is the
limit of a subsequence $\{\gamma_{k_m}\}_{m}$ in the {$C^0$
topology \cite[Prop. 3.34]{beem96}.

In this case the length $L$ for any metric $g$ in $\g$ satisfies:
$L(\gamma)\geq $ $\overline{\lim}_m$ $L(\gamma_{k_m})$}
\cite[Remark 3.35]{beem96} \cite[p. 54]{penrose72}. \eit \een

These properties have many applications for the geometry of the
spacetime, for example \cite[Th. 8.10]{beem96}:

\begin{proposition} If $(M,g)$ is a strongly causal spacetime then, for any $p\in M$,
 a {\em future-directed  geodesic ray} \index{geodesic
ray} $\gamma:[0,b)\rightarrow M$ starts at $p$ (that is, $\gamma$
is a f.-d. maximazing causal geodesic, $L( \gamma|_{[0,t]}) =
d(p,\gamma(t))$, for all $t \in [0,b)$, with $p=\gamma(0)$ and no
future endpoint).
\end{proposition}

\subsubsection{Isometries} Finally, it is worth pointing out that,
in the class of strongly causal spacetimes, the time-separation
$d$ determines the metric (as the distance function of a
Riemannian manifold determines the metric). Concretely (see
\cite[Th. 4.17]{beem96}, which is extended to characterize
homothetic maps and totally geodesic submanifolds):

\begin{theorem} Let $(M,g), (M',g')$ be two spacetimes with the same dimension, and let
$(M,g)$ be  strongly causal. If  $f: M\rightarrow M'$ is an onto
map (non-necessarily continuous) and  $f$ preserves de
time-separations $d, d'$, i.e.,
$$
d(p,q)=d'(f(p),f(q)), \forall p,q\in M $$ then $f$ is a
diffeomorphism and a metric isometry.

In particular, when $M=M'$ ($f=$ identity), it holds: the
time-separations of $g$ and $g'$ coincide if and only if $g=g'$.
\end{theorem}

\subsection{A break:  volume functions, continuous $I^\pm$, reflectivity} \label{secVolume}
\index{volume functions} \index{admissible measures}

\subsubsection{Admissible measures} A pair of functions constructed from the volumes of
$I^\pm(p), p\in M$ becomes very useful to study Causality.
Nevertheless, for such a purpose the volumes must be finite and,
thus, the natural measure of a spacetime associated to the metric
may not be useful (even more,  the representative $g$ in the
conformal class $\g$ of the spacetime must be irrelevant for the
definition of the functions). An appropriate choice of the {\em
Borel} measure (i.e. a measure on the $\sigma-$algebra generated
by the open subsets of $M$) is:
\begin{quote}
The measure $m$ associated to any auxiliary (semi-)Riemannian
metric $g_R$ with finite total volume $m(M)$.
\end{quote}
Without loss of generality, it can be completed in the standard
way, by adding to the Borel sigma algebra all the subsets of any
subset of measure 0 (which are regarded as new subsets of measure
0); by Sard's theorem, the subsets of measure 0 are  intrinsic to
the differentiable structure of $M$. It is worth pointing out:

\bit \item {\em Construction of  $m$.} Without loss of generality,
we can assume that $M$ is orientable (otherwise,  reason with the
orientable Lorentzian double-covering $\Pi: \tilde M \rightarrow
M$, and define the measure of any Borelian $A\subset M$ as --one
half of-- the measure of $\Pi^{-1}(A)$). Choose an orientation,
and let $\omega_g$ be the oriented volume element associated to
 the metric of the
spacetime $g$ (or any other semi-Riemannian metric). Fix any
covering of $M$ by open subsets with $\omega_g$-measure smaller
than 1, and take a partition of the unity $\{\rho_n\}_{n\in \N}$
subordinated to the covering. Define the measure $m$ as the one
associated to the volume element \be \label{eomegas} \omega_R=
\sum_{n=1}^\infty 2^{-n} \rho_n \omega . \ee It is easy to check
that $\omega_R$ is the measure associated to some pointwise
conformal metric for any auxiliary (semi-)Riemannian metric (see
\cite{sanchez05b} for more details).

\item {\em Relevant properties of the measures}. The so-defined
measure $m$ satisfies:

 \ben
\item  Finiteness: $m(M)<\infty$.

This is straightforward from (\ref{eomegas}) and one can
 normalize $m(M)=1$.
\item For any non-empty open subset $U$, $m(U)>0$.

\item The boundaries $\dot{I}^+(p), \dot{I}^-(p)$ have
measure 0, for any $p\in M$.

This holds for $m$ because $\dot{I}^+ (p)$, $\dot{I}^- (p)$ are
closed, embedded, achronal topological hypersurfaces
\cite[Proposition 6.3.1]{hawking73}; thus, for any (differentiable)
chart, they can be written as Lipschizian graphs, which have  0
measure.

\een Abstract measures satisfying these three  properties were
called {\em admissible} by Dieckmann \cite{dieckmann88},
\cite[Definition 3.19]{beem96}; these properties are the only
relevant ones for the applications below. Measure $m$ constructed
in (\ref{eomegas}) satisfies other interesting properties, as {\em
regularity} (see  \cite{sanchez05b} for a technical discussion). \index{regular measures} 

Obviously, the third property cannot be deduced from the first and
the second ones (choose a point $q\in M$, and construct a new
measure $m'$ regarding $q$ as an {\em atom}, say: $m'(A)=m(A)+1$
if $q\in A$, $m'(A)=m(A)$ if $q\not\in A$, for all measurable
subset $A$).  Note that this third property implies
$m(I^+(p))=m(\bar{I}^+(p))= m(J^+(p))$ for all $p$, and
analogously for $I^-$.  \eit

\subsubsection{Volume functions and their continuity}
In what follows, an admissible measure $m$ on $M$ is fixed. We
have already regarded $I^+$ (and analogously $I^-$) as a
set-valued map in the set of parts of $M$, $I^+: M \rightarrow
{\cal P}(M)$. As each $I^+(p)$ is an open set, it lies in the
$\sigma$-algebra of  $m$. So, one essentially takes the
composition of $m$ and $I^+$ in the following definition.
\begin{definition} Let $(M,\g)$ be a spacetime with an admissible
measure $m$. The   {\em future $t^-$ and past $t^+$ volume
functions} associated to $m$ are defined as:
$$t^-(p)= m(I^-(p)), \quad  \quad t^+(p)= -m(I^+(p)), \quad  \quad \forall p \in M.$$
\end{definition}

\begin{remark} \label{rvol} Clearly, $t^\pm$ satisfy:
\begin{enumerate}
\item  They  are both non-decreasing on any future-directed causal
curve (in fact, the sign -  is introduced for $t^+$ because of this
reason). But perhaps they are not strictly increasing; in fact, they
are constant on any causal loop.

\item They are not necessarily continuous (even though they are
semi-continuous, see below). Fig. \ref{fig1} shows a distinguishing
counterexample.
\end{enumerate}
\end{remark} The continuity of $t^\pm$ is closely
related to the continuity of $I^\pm$. Nevertheless, we have to
give an appropriate notion of what this means for a set-valued
function. (In what follows, when there is no possibility of
confusion we will make definitions and proofs for $I^-, t^-$, and
the reasonings for $I^+, t^+$ will be analogous.)

\begin{definition} \label{dinncont}
Function $I^-$, is {\em inner (resp. outer) continuous} at some
\index{inner continuity} \index{outer continuity} $p\in M$ if, for
any compact subset $K\subset I^-(p)$ (resp. $K\subset M\backslash
\bar{I}^-(p) $), there exists an open neighborhood $U \ni p$ such
that $K \subset I^-(q)$ (resp. $K\subset M\backslash
\bar{I}^-(q)$) for all $q \in U$.
\end{definition}
As usual, $I^\pm$ is (inner, outer) continuous when so is at each
event $p\in M$, and the spacetime is accordingly (future, past)
inner or outer continuous. In order to understand better Definition
\ref{dinncont}, consider the following topology in ${\cal P}(M)$.
For any compact $K\subset M$, the subsets of $M$ not intersecting
$K$ form a subset of ${\cal P}(M)$ which we define as {\em open}.
These open sets are a base for the topology on ${\cal P}(M)$
considered in what follows.
\begin{proposition} \label{linneroutercont}
The set valued maps $I^\pm: M \rightarrow {\cal P}(M)$ satisfy:

(i) $I^\pm$ are always inner continuous

(ii) $I^\pm$ are outer continuous if and only if they are
continuous as maps between topological spaces.
\end{proposition}

\begin{proof}
(i) Let $K\subset I^{-}(q)$ be any compact subset. It is covered
by the open sets $\{I^{-}(p): p \in I^{-}(q) \}$, and admits a
finite subcovering $\{I^{-}(p_1),\ldots,I^{-}(p_n)\}$. So, the
neighborhood of $q$, $U=\bigcap_{i=1}^{n} I^{+}(p_i)$, has the
required property.

(ii) Just check the definitions.
\end{proof}

Nevertheless,  it is easy to construct  non-outer continuous
examples
 (Fig. \ref{fig1}). And, in fact, this is related to the
continuity of $t^\pm$.

\begin{figure}[ht]
\centering \psfrag{U}{$u$} \psfrag{V}{$\!\!\! v$}
\psfrag{P}{$\!\!\! p$} \psfrag{PN}{$p_n$} \psfrag{Q}{$q$}
\psfrag{J}{$J^{+}(q)$}
\includegraphics[width=5cm]{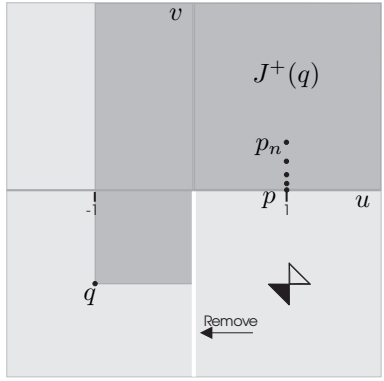}
\caption{$M \subset \L^2$, (in coordinates $u,v$, $g=-2dudv$),
$M=\{(u,v)\in \L^2: |u|, |v| <2\}\backslash \{(u,v)\in \L^2: u=0, -2
< v \le 0\}$ is stably causal, $I^+$ is outer continuous but $I^-$
is not. Correspondingly (for the canonical measure $m$ of $g$),
$t^+$ is continuous but $t^-$ is not as the sequence $p_n=(1,1/n)$
shows.  As a consequence, $M$ is non-causally simple (see Sect.
\ref{simsec}), indeed, for instance, $J^+(q)$ is not closed for
$q=(-1,-1)$} \label{fig1}.
\end{figure}


\begin{lemma}
The inner continuity of $I^-$ (resp. $I^+$) is equivalent to the
lower (resp. upper) semi-continuity of $t^-$ (resp. $t^+$). Thus,
it holds always.
\end{lemma}

\begin{proof}
As $I^-$ is always inner continuous, only the
implication to the right must be proved. Thus, let
$\{p_n\}\rightarrow p$, fix $\epsilon>0$ and let us prove $t^-(p_n)
>t(p)-\epsilon$ for large $n$.   There exists a compact subset $K \subset I^-(p)$
such that\footnote{One can check this for any admissible measure
(and it is obvious for any regular measure, as the explicitly
constructed $m$), see \cite[Lemma 3.7]{sanchez05b} for details.}
$m(K)>m(I^-(p))-\epsilon =t^-(p)-\epsilon$ and, by inner
continuity, $K \subset I^-(p_n)$ for large $n$. Thus, $t^-(p_n)
\geq m(K) >t^-(p)-\epsilon$, as required.
\end{proof}

\begin{lemma} \label{l33a} The following properties are equivalent:
(i) $I^-$ (resp. $I^+$) is outer continuous  at $p$,
and (ii) volume function $t^-$ (resp. $t^+$) is  upper (resp.
lower) semi-continuous at $p$.
\end{lemma}
\begin{proof}
(i) $\Rightarrow$ (ii) Completely analogous to the previous case,
taking now $K$ as a compact subset of $M\backslash \bar{I}^-(p)$
with $m(K) > m(M\backslash \bar{I}^-(p))-\epsilon$ and, then, for
large $n$: $t^-(p_n) \leq m(M)-m(K) <t^-(p)+\epsilon$.

(ii) $\Leftarrow $ (i) If $I^-$ is not outer continuous, there
exists a compact $K\subset M\backslash \bar{I}^-(p)$ and a sequence
$\{p_n\}\rightarrow p$ such that each $\bar{I}^-(p_n) \cap K$
contains at least one point $r_n$. Thus, $r_n\rightarrow r \in K$,
up to a subsequence, and choose $s\ll r$ in $M\backslash
\bar{I}^-(p)$ ($s$ exists otherwise $I^{-}(r) \subset I^{-}(p)$ thus
$r \in \bar{I}^{-}(p)$ a contradiction). As the chronological
relation is open, there exist neighborhoods $U,V \subset M\backslash
\bar{I}^-(p)$ of $s, r$, resp., such that $U \subset \cap_{r'\in V}
I^-(r')$, and, thus, $U \subset I^-(p_n)$ for large $n$. Now, choose
a sequence $\{q_j\}\rightarrow p$ satisfying
$$ p\ll q_j \ll q_{j-1} , \quad \hbox{for all}\; j .$$
Then, $U \subset I^-(q_j)$ for all $j$ and, putting $\epsilon =
m(U)>0$:
$$ t^-(q_j) = m (I^-(q_j)) \geq m(I^-(p)) + m(U) = t^-(p) + \epsilon .$$
\end{proof}

Thus, the previous two lemmas yields directly:
\begin{proposition} \label{pcaracterizI+1}
The following  properties are equivalent for a spacetime:
\begin{itemize}
\item[(i)] The set valued map $I^-$ (resp. $I^+$) is (outer)
continuous.

\item[(ii)] Volume function $t^-$ (resp. $ t^+$) is continuous.

\end{itemize}
\end{proposition}

\subsubsection{Reflectivity} Continuity of $I^\pm$ (and, thus, $t^\pm$) can be also
characterized in terms of reflectivity.

\begin{lemma} \label{lreflect} Given any pair of events
$(p,q) \in M \times M$ the following logical statements are
equivalent:
\begin{itemize}
\item[(i)] $ I^+(p) \supset I^+(q) \Rightarrow I^-(p) \subset
I^-(q)$,  $\ $  (resp.   $I^-(p) \supset I^-(q) \Rightarrow
 I^+(p) \subset I^+(q)\,$),
\item[(ii)] $q \in \bar{I}^{+}(p) \Rightarrow p \in
\bar{I}^{-}(q)$, $\ $ (resp.   $q \in \bar{I}^{-}(p) \Rightarrow p
\in \bar{I}^{+}(q)\,$) \item[(iii)] $q \in \dot{I}^{+}(p)
\Rightarrow p \in \dot{I}^{-}(q)$, $\ $ (resp.   $q \in
\dot{I}^{-}(p) \Rightarrow p \in \dot{I}^{+}(q)\,$).
\end{itemize}
\end{lemma}
\begin{proof} (Equivalence in the past case). (i) $\Leftrightarrow$ (ii). Trivial
from the equivalences: (a) $I^{+}(q) \subset I^{+}(p)
\Leftrightarrow q \in \bar{I}^{+}(p)$, and (b) $I^{-}(p) \subset
I^{-}(q) \Leftrightarrow p \in \bar{I}^{-}(q)$. \\
(ii) $\Leftrightarrow$ (iii). To the right, recall: $q \in
\dot{I}^{+}(p) \Rightarrow q \in \bar{I}^{+}(p)$ but $(p,q) \notin
I^{+}$ $\Rightarrow$ $p \in \bar{I}^{-}(q)$ but $(p,q) \notin
I^{+}$ $\Rightarrow$ $p \in \dot{I}^{-}(q)$. For the converse:
$q \in \bar{I}^{+}(p) \Rightarrow q \in
\dot{I}^{+}(p)$ or $(p,q) \in I^{+}$ $\Rightarrow$ $p \in
\dot{I}^{-}(q)$ or $(p,q) \in I^{+}$ $\Rightarrow$ $p \in
\bar{I}^{-}(q)$.
\end{proof}

\begin{definition} \label{dreflect} \index{future reflecting spacetimes} \index{past reflecting
spacetimes} \index{reflecting spacetimes} A spacetime $(M,
\bm{g})$ is {\em
 past (resp. future)  reflecting at} $q \in M$ if any of the
corresponding equivalent items (i), (ii), (iii) in Lemma
\ref{lreflect} holds for the pair $(p,q)$ for every $p\in M$. A
spacetime is {\em past (resp. future) reflecting} if it is so at any
$q\in M$, and {\em reflecting} if it is both, future and past
reflecting.
\end{definition}

\begin{remark} Notice that if the items of Lemma \ref{lreflect} are required for
 $(p,q)$, for every $q\in M$, a  different property, say {\em (past)
pseudo-reflectivity at} $p$, would be obtained. Even though
pseudo-reflectivity and reflectivity would be equivalent as
spacetime properties (i.e. with no reference to a single point),
they are different as  properties for a single event (as can be
checked in Figure \ref{fig1}), the former not to be considered in
what follows.
\end{remark}

Another characterization of reflectivity is the following.

\begin{proposition}\label{preflect}
A spacetime $(M, \bm{g})$  is past reflecting at $q$ (resp. future
reflecting at $p$) if and only if  \[  (p',q) \in \bar{I}^{+}
\Rightarrow p' \in \bar{I}^{-}(q), \quad (\textrm{resp. } (p,q')
\in \bar{I}^{+}   \Rightarrow q' \in \bar{I}^{+}(p)) .
\]
An analogous  result holds with $\bar{I}$ replaced with $\dot{I}$.
\end{proposition}

\begin{proof} (Past case). Assume the spacetime is past
reflecting at $q$ and let $(p,q) \in \bar{I}^{+}$, then there are
sequences $p_n \to p$, $q_n \to q$, $q_n \in I^{+}(p_n)$. Take any
$s \in I^{-}(p)$, so that $p \in I^{+}(s)$ and for large $n$, $q_n
\in I^{+}(s)$ which implies $q \in \bar{I}^{+}(s)$. By using past
reflectivity at $q$, $s \in \bar{I}^{-}(q)$ and taking the limit $s
\to p$, $p \in \bar{I}^{-}(q)$.

Conversely, assume that  $(p',q) \in \bar{I}^{+} \Rightarrow p'
\in \bar{I}^{-}(q)$ and consider any $p$ such that $q \in
\bar{I}^{+}(p)$. Then, $(p,q) \in \bar{I}^{+}$ which implies $p
\in \bar{I}^{-}(q)$, that is, the spacetime is past reflecting at
$q$.
\end{proof}

\begin{lemma} \label{l33abis} $\empty$
The following  properties are equivalent:
(i) $I^-$ (resp. $I^+$) is outer continuous  at $p$, and
(ii) the spacetime is past (resp. future) reflecting at $p$.
\end{lemma}

\begin{proof}
(i) $\Rightarrow$ (ii).  Let $I^{-}$ be outer continuous at $q$,
and  assume there is a $p$ such that $q \in \bar{I}^{+}(p)$ but $p
\notin \bar{I}^{-}(q)$. By outer continuity there is a
neighborhood $V \ni q$ such that for every $q' \in V$, $p \notin
\bar{I}^{-}(q')$, but since $q \in \bar{I}^{+}(p)$ there is $q'
\in V$ such that $(p,q') \in I^{+}$, a contradiction. \\
(ii) $\Rightarrow$ (i). Let the spacetime be past reflecting at $p$
and assume by contradiction that $I^{-}$ is not outer continuous.
Then there is a compact $K$, $K \cap \bar{I}^{-}(p)=\emptyset$, and
a sequence $p_n \to p$ such that $ K \cap \bar{I}^{-}(p_n)\ne
\emptyset$. Taken $r_n \in K \cap \bar{I}^{-}(p_n)$, up to a
subsequence $r_n \to r \in K$, and for any $s \in I^{-}(r)$ we have
for large $n$, $p_n \in I^{+}(s)$, which implies $p \in
\bar{I}^{+}(s)$. By using reflexivity at $p$, $s \in
\bar{I}^{-}(p)$, and making $s\rightarrow r$,
 $r \in \bar{I}^{-}(p)$, a contradiction because $r \in K$.
\end{proof}

The set $R$ of points that do not comply 
these conditions has been studied in detail. The set $R$ is a
suitable union of null geodesics without past or future endpoint
\cite[Prop. 1.7]{dieckmann88}. Moreover, no point of $R$ is isolated
\cite{vaya86}, and optimal bounds for its dimension are known
\cite{hawking74,clarke88}. From Lemma  \ref{l33abis}, obviously:

\begin{proposition} \label{p33a}
The following  properties are equivalent for $(M, \bm{g})$:
\begin{itemize}
\item[(i)] The set valued map $I^-$ (resp. $I^+$) is (outer)
continuous.


\item[(ii)] The spacetime is past (resp. future)  reflecting.

\end{itemize}
\end{proposition}

\subsection{Stably causal spacetimes}\label{subs-stably}

Volume and time functions are essential in this and following
levels. We start  discussing their relations with previous ones.

\subsubsection{Time-type functions and characterization of some levels} 

\begin{definition} \label{dtime}
Let $(M,\g ) $ be a spacetime. A (non-necessarily continuous) function $t: M\rightarrow \R$ is:
\bit
 \item A {\em generalized time function} if  $t$ is strictly
increasing on any future-directed causal curve $\gamma$.
\index{generalized time function}

\item A  {\em  time function} if  $t$ is a continuous generalized
time function. \index{time function}

\item A  {\em  temporal function} if  $t$ is a smooth function
with past-directed timelike gradient $\nabla t$. \index{temporal
function} \eit

\end{definition}
\noindent Notice that a temporal function is always a time
function ($d(t\circ \gamma(s)/ds)$ $ =$ $ g(\dot \gamma(s), \nabla
t)$ $
>$ $0$), but even a smooth time function may be non-temporal. From
Remark \ref{rvol}, volume functions are not far from being
generalized time ones. In fact, the next two theorems
characterize this property.

\begin{theorem}\label{voldis} A spacetime $(M,\g)$ is
chronological if and only if $t^-$ (resp. if and only if $t^+$) is
strictly increasing on any future-directed timelike curve.
\end{theorem}

\begin{proof} ($\Leftarrow$). Obvious.
($\Rightarrow$). If $p \ll q$ but $t^-(p) = t^-(q)$, necessarily
almost all the points 
in the
open subset $I^+(p)\cap I^-(q)$ lie in $I^-(p)$. Thus, any point
$r$ in $I^+(p)\cap I^-(q) \cap I^-(p)$ satisfies $p \ll r \ll p$.
\end{proof}

\begin{remark}
Notice that, as  $t^-$ is also constant on any  causal loop,
causal spacetimes cannot be  characterized in this way. Figure
\ref{T3N8} gives an example of causal non-distinguishing spacetime
for which $t^{-}$ is constant along a causal curve (the central
almost closed circle).
\end{remark}

\begin{theorem}\label{tdist-timef}
A spacetime $(M,\g)$ is past (resp. future) distinguishing if and
only if $t^-$ (resp. $t^+$) is a generalized time function.
\end{theorem}
\begin{proof}
($\Rightarrow$). To prove that $t^-$ is strictly increasing on any
future-directed causal curve, assume that $p< q$, $p\neq q$, but
$t^-(p)=t^-(q)$. Then, almost all the points of $I^-(q)$ are
included in $I^-(p)$. Choose a sequence $\{q_n\}_n \subset
I^-(p)\cap I^-(q)$ converging to $q$.
Recall that, necessarily then $I^-(q_n) \subset I^-(p)$ for all
$n$, and $I^-(q) = \cup_n I^-(q_n)$. But this implies $I^-(q)
\subset I^-(p)$ and, as the reversed inclusion is obvious, the
spacetime is non-past distinguishing.

($\Leftarrow$). If $I^-(p) = I^-(q)$ with $p\neq q$, choose a
sequence $\{p_n\} \subset I^-(p)$ which converges to $p$, and a
sequence of timelike curves $\gamma_n$ from $q$ to $p_n$. By
construction, the limit curve $\gamma$ of the
sequence 
starting at $q$ is a (non-constant) causal curve and $I^-(p)\subset
I^-(\gamma(t)) \subset I^-(q)$ for all $t$. Thus, the equalities in
the inclusions hold, and $t^-$ is constant on $\gamma$.
\end{proof}

\subsubsection{Stability of causality and chronology}\label{subsubs-stably} \index{stably
causal spacetimes} Stable causality is related with the simple
intuitive ideas that the spacetime must remain causal after
opening slightly its lightcones, or equivalently,  under small
($C^0$ fine) perturbations of the metric. Surprisingly, this is
equivalent to the existence of time and temporal functions.

 More precisely, let Lor$(M)$ be the set of all the  Lorentzian metrics on $M$
(which will be assumed time-orientable  in what follows, without
loss of generality). A partial (strict) ordering $<$ is defined in
Lor$(M)$:
\begin{quote}
$ g<g' $ if and only if all the causal vectors for $g$ are
timelike for $g'$. \end{quote} Notice that this ordering is
naturally induced in the set Con$(M)$ of all the classes of
pointwise conformal metrics on $M$. Even more, it induces
naturally a topology in Con$(M)$, the {\em interval} topology,
which admits as a subbasis the subsets type \index{interval
topology on the metric space}
$$
\bm{U}_{\bm{g_1}, \bm{g_2}} = \{\bm{g}: \bm{g_1} < \bm{g} <
\bm{g_2}\}
$$
where $\bm{g_1}, \bm{g_2} \in$ Con$(M)$, $\bm{g_1} < \bm{g_2}$.

Remarkably, the interval topology coincides with the topology
induced in Con$(M)$ from the $C^0$ fine topology on Lor$(M)$.
\index{fine topology on the metric space} Roughly, the $C^0$
topology on Lor$(M)$ can be described by fixing a locally finite
covering of $M$ by open subsets of coordinate charts with closures
also included in the chart.  Now, for any positive continuous
function $\delta: M\rightarrow \R$ and $g\in $Lor$(M)$ one defines
$U_\delta(g) \subset \hbox{Lor}(M)$ as the set containing metrics
$\tilde g$ such that, in the fixed coordinates at each $p$,
$|g_{ij}(p)-g_{ij}(p)| <\delta(p)$ (in order to define
 the $C^r$ topology on Lor$(M)$, this inequality is also required for the partial derivatives of
$g_{ij}$ up to order $r$). A basis for the $C^0$-fine topology is
defined as the set of all such $U_\delta(g)$ constructed for any
$\delta$ and $g$ (see \cite{sanchez05b,beem96,rendall88} for more
detailed descriptions of this topology). Then, the (quotient)
$C^0$-topology in Con$(M)$ is defined as the finer one such that
the natural projection Lor$(M)\rightarrow $Con$(M), g\rightarrow
\g$ is continuous.

A way to define directly the $C^0$-topology on Con$(M)$ which shows
the relation with the interval one is as follows
\cite{beem96,lerner73}. Fix an auxiliary Riemannian metric $g_R$,
and, for each $\g\in$ Con$(M)$, define the $g_R$-unit lightcone at
$p\in M$ as:
$$
C^{(R)}_p=\{v\in T_pM: g(v,v)=0, g_R(v,v)=1\}.
$$
Now, if $|\cdot |_R$ is the natural $g_R$-norm, one  define
naturally the distance of any vector $w\in T_pM$ to $C^{(R)}_p$ as
usual:
$$
d_R(w,C^{(R)}_p)= \hbox{Min}\{|w-v|_R: v\in C^{(R)}_p\}.
$$
Given a second $\tilde \g\in$Con$(M)$ with associated $g_R$-unit
lightcone $\tilde C^{(R)}_p$, the maximum and minimum distances
between the lightcones are, respectively:
$$
|\g-\tilde \g|^M_R(p)= \hbox{Max}\{d_R(v,\tilde C^{(R)}_p): v\in
C^{(R)}_p\},\quad  |\g-\tilde \g|^m_R(p)= \hbox{Min}\{d_R(w,\tilde
C^{(R)}_p): w\in C^{(R)}_p\}.
$$
Notice that
$$ 0< |\g-\tilde \g|^m_R \quad \Leftrightarrow \hbox{either}\; \g <
\tilde \g \; \hbox{or} \;  \tilde \g < \g .
$$
Now, for any positive continuous function $\delta: M\rightarrow
\R$, let $\bm{U}_\delta(\g)=\{\tilde \g\in \hbox{Con}(M): |\g-
\tilde \g|^M_R<\delta\}$. The sets $U_\delta(\g)$ yields a basis
for the $C^0$ topology.
%

\begin{definition} \label{dsc}
A spacetime $(M,\g)$ is stably causal if it satisfies,
equivalently:

\bit \item[(i)] There exists $\tilde \g \in $Con$(M)$ such that
$\g< \tilde \g$ and $\tilde \g$ is causal. \item[(ii)] There
exists a neighborhood $\bm{U}$ of $\g$ in the quotient $C^0$
topology such that all the metrics in $\bm{U}$ are causal. \eit
\end{definition}
\begin{remark}\label{rsc}
(1) The equivalence of both definitions is clear because, if
$\tilde \g$ is causal, then so are all the spacetimes with smaller
lightcones, and these spacetimes constitute a $C^0$ neighbourhood.

(2) A property of a metric $g$ is called {\em $C^r$ stable}
\index{stable spacetime property} ($r=0,1, \dots, \infty$) if it
holds for a $C^r$ neighborhood of $\g$. As the $C^r$ topologies for
$r>0$ are finer than the $C^0$ one,  {\em stable causality} means
that the metric of the spacetime is not only causal, but also that
this property is stable in all the $C^r$ topologies.
\end{remark}

\begin{proposition}
($C^0$) stable chronology and stable causality   are equivalent
properties for any spacetime $(M,\g )$. \index{stably
chronological spacetimes}
\end{proposition}

\begin{proof}
Obviously, the latter implies the former. Let us show than
non-stably causal implies non-stably chronological. Indeed, if the
spacetime is non-stably causal,  any $\bm{g}_1>\bm{g}$ admits a
closed causal curve $\bm{\gamma}_1$. But since this  is also true
for any $\bm{g}_2$ such that $\bm{g}<\bm{g}_2<\bm{g}_1$, then the
corresponding $\bm{\gamma}_2$ is a closed timelike curve with
respect to $\bm{g}_1$. Thus, any $\bm{g}_1>\bm{g}$ admits a closed
timelike curve.
\end{proof}

 A nice property of bidimensional spacetimes is the following.

\begin{theorem} Any simply connected 2-dimensional
spacetime $(M,\g)$ is stably causal. \index{2-dimensional
spacetime}
\end{theorem}

\begin{proof} As $M$ has 0 Euler characteristic \index{Euler characteristic} (Th. \ref{texislor}),  necessarily $M$  must be homeomorphic to
$\R^2$. Obviously, it is enough to prove that any  spacetime
constructed on $\R^2$ is causal. Otherwise, by closing if necessary
the lightlike cones in a tubular neighborhood of $\gamma$, we can
assume that there exists a lightlike closed curve $\gamma$, which
(regarded as Jordan's curve) bounds a domain $D$. Thus, taking any
timelike vector field $X$, we have $g(X,\gamma')$ never vanishes,
i.e., $X$ must point out either outwards or inwards $\gamma \equiv
\dot{D}$. Thus, a standard topological argument says that $X$ must
vanish on some point of $D$, a contradiction.
\end{proof}

\subsubsection{Time and temporal functions} The following
characterization of stable causality in terms of time-type
functions (see Definition \ref{dtime})
 becomes specially useful. Nevertheless, it has been
 proved with rigor only recently  \cite{bernal04, sanchez05b}.

\begin{theorem} \label{testcaus} For a spacetime $(M,\g)$ the following properties are equivalent:

\bit \item[(i)] To be stably causal. \item[(ii)] To admit a time
function $t$
 \item[(iii)] To admit a temporal
function ${\cal T}$ \eit
\end{theorem}

\begin{proof} (Sketch with comments; see \cite[Sect. 4]{sanchez05b} for  detailed proofs and discussions).
(iii) $\Rightarrow$ (i) As causality is a conformally invariant,
choose the representative $g$ of $\g$ with $g(\nabla \Tau, \nabla
\Tau)=-1$.  Now, the metric can be written as
$$ g= -d\Tau^2 + h$$
where $h$ is the restriction of $g$ to the bundle orthogonal to
$\nabla \Tau$ (up to natural identifications). Then, consider the
one parameter family of metrics
$$ g_\lambda = -\lambda d\Tau^2 + h ,  \quad \quad \lambda >0 .$$
Clearly, $\Tau$ is still a temporal function for each $g_\lambda$.
Thus, $g_\lambda$ is always causal,  and  $g=g_1 < g_2$, as
required.

 (i) $\Rightarrow$ (ii) (Hawking \cite{hawking69}, see also \cite[Prop.
6.4.9]{hawking73} or \cite[Theorem 4.13]{sanchez05b}).  The
fundamental idea is that, even though the past volume function
$t^-$ may be non-continuous (it is only a generalized
time-function), an ``average'' of such functions for a 1-parameter
family of metrics $g_\lambda$ will work if $g_\lambda$ satisfies:
(i) $g_0=g$, (ii) $g_\lambda$ is causal, for all $\lambda \in
[0,2]$, and (iii) $\lambda < \lambda' \Rightarrow g_\lambda <
g_{\lambda'}$. Concretely, one checks that the following function
is a time function:
$$t(p)= \int_0^1 t_\lambda^-(p)
d\lambda ,$$ where $t_\lambda^-$ is the  past volume function for,
say, $g_\lambda = g+(\lambda/2) (\tilde g-g)$, $\lambda \in [0,2]$
($\tilde g$ is chosen causal with $g<\tilde g$).

(ii) $\Rightarrow$ (iii) This has been one of the ``folk
questions'' on smoothability of the theory of Causality until its
recent solution \cite{bernal04}. It becomes crucial because,
otherwise, the implication (ii)$\Rightarrow$ (i) was also open. We
refer to the detailed exposition in \cite[Sect. 4.6]{sanchez05b}
(see also the comments on smoothability for globally hyperbolic
spacetimes below, especially Remark \ref{r2cojones}).
\end{proof}

\begin{proposition}
Stable causality implies strong causality.
\end{proposition}

\begin{proof}
Let $t$ be a time function and let us see that condition (ii) in
Lemma \ref{lstr} holds at any $p\in M$. Let $U\ni p$ a
neighborhood and assume, without loss of generality,  that $U$ is
simple, its closure is included in another simple neighborhood
$\tilde U$, and $t(p)=0$. For any $q \in U$ put $\epsilon_q^+=$
Min$\{ t(r): r\in J^+(q,\tilde U)\cap \dot{U}\}, \epsilon_q^-=$
Min$\{ -t(r): r\in J^-(q,\tilde U)\cap \dot{U}\}$; the variation
of $\epsilon_q^\pm$ with $q$ is continuous because $\tilde U$ is
convex. As $\dot{U}$ is compact, $-\epsilon_q^- < t(q) <
\epsilon_q^+$, in particular, $\epsilon_p^- , \epsilon_p^+
>0$. Thus, for a small neighborhood $W\ni p, \bar W \subset U$,
one has $\epsilon_q^- , \epsilon_q^+ >  0$ for all $q\in \bar W$.
From the compactness of $\bar W$, necessarily $\epsilon_W :=$
Min$\{\epsilon_q^\pm: q\in \bar W \}>0$. The required neighborhood
is $V=W\cap t^{-1}(-\epsilon_W/2, \epsilon_W/2)$. In fact, if a
future-directed timelike curve starts at some $q\in V$ and leaves
$U$ at some point $q_U$, then $t(q_U) \geq  \epsilon_W$; thus,
$\gamma$ cannot
return to $V$. 
\end{proof}

\begin{figure}[ht]
\centering
\includegraphics[width=9cm]{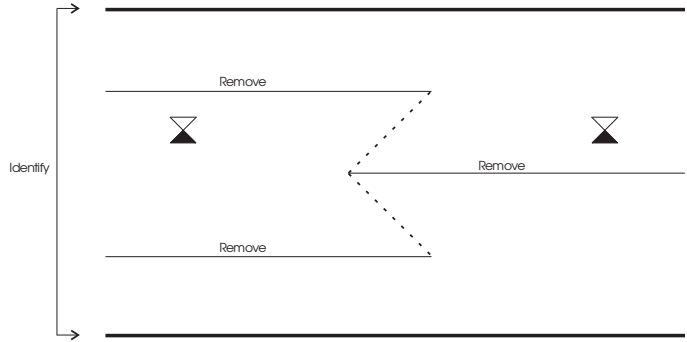}
\caption{An example of strongly causal non-stably causal
spacetime. By opening slightly the causal cones there appear
closed causal curves.} \label{stronglycausal} \label{strong}
\end{figure}

\begin{remark} (1) Stable causality implies strong causality but the converse
does not hold (see figure \ref{strong}).

(2) Between strong and stable causality, an infinite set of levels
can be defined by using Carter's ``virtuosity'' \cite{carter71}.
\index{Carter's ``virtuosity''}

\end{remark}

\subsection{Causally continuous spacetimes} \label{ccontinuous}
\index{causally continuous spacetimes} Taking into account the
characterizations of the continuity of $I^\pm$ (Prop.
\ref{pcaracterizI+1}, \ref{p33a}) as well as the behavior of
$t^\pm$ in distinguishing spacetimes (Th. \ref{tdist-timef}), the
following definitions of causal continuity (which can be also
combined with the characterizations of reflectivity, Lemma
\ref{lreflect}, Prop. \ref{preflect}) hold.

\begin{definition} \label{dccontinuous} A spacetime $(M,g)$ is
causally continuous if (equivalently, and for any admissible
measure):

\ben \item[(i)] Maps $I^\pm: M\rightarrow {\cal P}$
are: 
(a) one to one, and (b) continuous (i.e., $(M,\g )$ is reflecting,
Lemma  \ref{l33abis}).

\item[(ii)] $(M,\g)$ is: (a) distinguishing, and (b) with
continuous volume functions $t^\pm$.
 \item[(iii)] The volume
functions $t^\pm$ are time functions \een
\end{definition}

\begin{remark} \label{rconsistencia-causcont} Trivially, totally vicious spacetimes have continuous
 $I^\pm$. Even more, they are also continuous in the causal non-distinguishing spacetime
of Fig. \ref{T3N8} (notice that the removed point in the circle does
not affect to function $t^\pm$).
 Thus, the injectivity of these maps (i.e., the hypotheses
 ``distinguishing'')
 is truly necessary for this level of the ladder.
\end{remark}
 Recall that a causally continuous spacetime not only admits
a time function, but also the past and future volume functions are
time functions. In particular:

\begin{proposition}
Any causally continuous spacetime is stably causal.
\end{proposition}

 \begin{remark} (1) The converse does not hold, as the example in Fig. \ref{fig1} shows.

(2)  Until stable causality, all the levels in the hierarchy of
causality, except non--totally vicious, were inherited by open
subsets\footnote{A counterexample for total-viciousness can be
obtained from Figure \ref{f1}, taking the open region determined
by $1/3 < x <2/3$.}. This is not the case neither for causal
continuity (as the counterexample in  figures \ref{fig1}  shows,
being obtained from an  open subset of $\R^2$) nor for the
remaining levels of the ladder.
\end{remark}

\subsection{Causally simple spacetimes} \label{simsec}
\index{causally simple spacetimes} There are different
characterizations of causal simplicity (Prop. \ref{pcaraccauscont}),
we will start by the simplest one.

\begin{definition} \label{dcsim} A spacetime $(M,g)$ is {\em causally simple} if
it is: \bit \item[(a)] causal,  and
 \item[(b)]  $J^+(p), J^-(p)$ are closed for every  $p\in M$.
\eit
\end{definition}
Tipically, the condition of being distinguishing is imposed
directly in the definition of causal simplicity instead of
causality, but the former can be deduced from the latter
\cite[Sect. 2]{bernal07}, as proven next. Nevertheless,
``causality'' cannot be weakened in ``chronology'', see Remark
\ref{rdiscgh-strongc}(1).

\begin{proposition} \label{s4ll}
Conditions (a) and (b) in Definition \ref{dcsim} imply that the
spacetime is distinguishing.
\end{proposition}

\begin{proof}
Otherwise, if $p\neq q$ and, say $I^+(p)=I^+(q)$, any sequence
$\{q_n\}\rightarrow q$, with $q\ll q_n$ shows $q\in \bar{I}^{+}(q)
= \bar{I}^+(p) = \bar{J}^+(p) = J^+(p)$. Thus, $p<q$ and,
analogously, $q<p$, i.e., the spacetime is not causal.
\end{proof}
Condition (b) has also the following  consequence.
\begin{proposition} \label{s4l}
If a spacetime satifies that $J^+(p)$ (resp. $J^-(p)$) is closed
for every $p$, then $I^-$ (resp. $I^+$) is outer continuous.
 Thus, condition (b) in Definition \ref{dcsim} implies
the reflectivity of $(M,\g )$.
\end{proposition}

\begin{proof}
Recall first the equivalence between outer continuity and
reflectivity (Prop \ref{p33a}), and let us prove the
characterization of Lemma \ref{lreflect}(ii) (for the future
case). As now $\bar I^\pm (p)= J^\pm(p)$, we have: $q\in \bar
I^+(p) \Rightarrow p\in J^-(q) = \bar I^- (p)$.
\end{proof}

\begin{figure}[ht]
\centering \psfrag{P}{$p$} \psfrag{Q}{$q$} \psfrag{T}{$t$}
\psfrag{X}{$x$} \psfrag{A}{$\alpha > \pi/4$} \psfrag{B}{$r$}
\psfrag{S}{$s$} \psfrag{G}{$\gamma$}
\includegraphics[width=6cm]{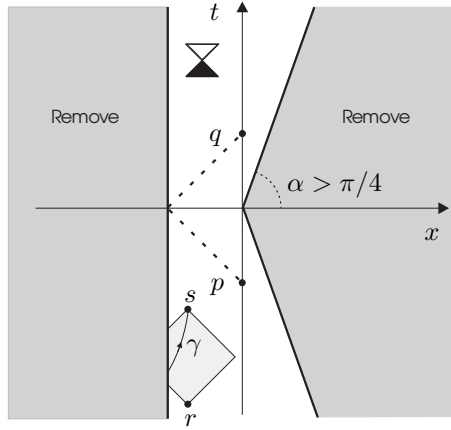}
\caption{An example of causally simple non-globally hyperbolic
spacetime, with a general metric conformal to the usual one,
$g=\Omega^2(t,x)(-\dd t^2+\dd x^2)$, $p=(0,-1)$, $q=(0,1)$,
$\Omega>0$. If $\Omega=1$ then $d(p,q)=2$ but no geodesic connects
them (Remark \ref{poi}, case 2b) while if $\Omega^2=1/(t^2+x^2)$,
$d(p,q)=+\infty$ (case 2a). If $\Omega^2=1/(x+1)^2$ then $d$ is
discontinuous (case 2c) as $d(p,q)<+\infty$ but $d(p,q')=+\infty$
for $q' \gg q$ (because the connecting causal curves can approach
a finite segment on the left-hand side border). The causal diamond
$J^{+}(r)\cap J^{-}(s)$ is not compact and there are inextendible
causal curves which, being ``created by the naked singularity'',
pass through $s$.}\label{example1}
\end{figure}

\begin{remark} $\empty$ \label{poi}
(1) By Propositions \ref{s4ll} and \ref{s4l}, causally simple
implies causally continuous, but the converse does not hold. A
counterexample can be obtained just by removing a point to $\L^2$.
On the other hand, a spacetime may have closed $J^-(p)$ for every
$p$ but non-closed $J^+(q)$ for some $q$ (Fig. \ref{fig1}).

(2) Even though these spacetimes are almost at the top
of the causal hierarchy, a metric in the pointwise conformal class
of a causally simple spacetime may have a time-separation $d$ with
undesirable properties (see Fig. \ref{example1}). For example:
\begin{itemize}
\item[(a)] For some $p,q$, perhaps $d(p,q) = \infty$. \item[(b)]
Even if $0<d(p,q)<\infty$, perhaps no causal geodesic connects $p$
and $q$. \item[(c)] $d$ may be discontinuous.

\end{itemize}
\noindent This will be remedied in the last step of the hierarchy.

 \end{remark}

Property (b) of Definition \ref{dcsim} can be also characterized in
different ways.
\begin{lemma}
 Let  $J^{+}(p)$ and $J^{-}(p)$ be closed for every $p \in M$,
 then:

(1) $J^{+}(K)$ and $J^{-}(K)$ are closed for every compact
$K\subset M$.

(2) $J^+$ (and hence $J^{-}$), regarded as a subset of $M\times
M$, is closed.
\end{lemma}

\begin{proof}
(1) Otherwise if, say, $q\in \bar J^+(K)\backslash J^+(K)$ there
exists sequences $q_n \rightarrow q$, $p_n \ll q_n$, $p_n\in K$,
where, up to a subsequence, $p_n \rightarrow p\in K$. Thus,
$(p,q)\in \bar{I}^+$ and, by using Proposition \ref{preflect}
(recall Prop. \ref{s4l}), $q\in \bar I^+(p)= J^+(p) \subset J(K)$.

(2) Obviously, $J^+ \subset \bar I^+$ and, for the converse, use
again $(p,q)\in \bar{I}^+ \Rightarrow q\in \bar I^+(p)= J^+(p)$.
\end{proof}

Thus, on the basis of these results, we have the following
characterization.

\begin{proposition} \label{pcaraccauscont} A spacetime $(M, \g)$ is {\em causally simple} if
it is causal and satisfies  one of the following equivalent
properties:
\begin{itemize}
\item[(i)] $J^{+}(p)$ and $J^{-}(p)$ are closed for every $p \in M$.
\item[(ii)] $J^{+}(K)$ and $J^{-}(K)$ are closed for every compact set $K$.
\item[(iii)] $J^{+}$ is a closed subset of $M\times M$.
\end{itemize}
\end{proposition}

Finally, notice that causal relations can be obtained now
``starting at chronology'' (Def. \ref{binaryrelations})
\begin{theorem} \label{pk2}
In a causally simple spacetime\footnote{Notice that, for this
result, one can define $x \leq^{(\ll )}  y$ $\Leftrightarrow$ {\em
either } $ I^+(y) \subset I^+(x)$ {\em or} $I^-(x)
 \subset I^-(y)$.}, $x\le^{(\ll)} y \Leftrightarrow x \le
y$.
\end{theorem}

\begin{proof} To the left, it is trivial in any spacetime. So, let $x\le^{(\ll)} y$. Since $I^{+}(y) \subset I^{+}(x)$, $y
\in \bar{I}^{+}(x)=\bar{J}^{+}(x)=J^{+}(x)$, where $J^+$ is the
usual causal relation.
\end{proof}

\subsection{Globally hyperbolic spacetimes} \label{globhyp}
\index{globally hyperbolic spacetimes} There are at least four ways
to consider global hyperbolicity: (1) by strengthening the notion of
causal simplicity, (2) by using Cauchy hypersurfaces, (3) by
splitting orthogonally the spacetime  and (4) by using the space of
causal curves connecting each two points. We will regard (1) as the
basic definition and will study subsequently the other approaches,
as well as some natural results under them.

\subsubsection{Strengthening causal simplicity}

\begin{definition} \label{dgh}
A spacetime $(M,\g )$ is {\em globally hyperbolic} if: \bit
\item[(a)] it is causal, and  \item[(b)] the intersections
$J^+(p)\cap J^-(q)$ are compact for all $p,q \in M$.\eit
\end{definition}
Following \cite[Sect. 3]{bernal07}, the next result yields
directly that a globally hyperbolic spacetime (according to our
Definition \ref{dgh}) is causally simple.

\begin{proposition} \label{pghcs}
Condition (b) implies both, $J^+(p)$ and $J^-(p)$ are closed for
all $p\in M$.
\end{proposition}
\begin{proof}
Assume that $J^+(p)$ is not closed and choose $r\in
\bar{J}^+(p)\backslash J^+(p)$ and $q\in I^+(r)$. Take a sequence
$\{r_n\}\rightarrow r$ with $r_n\in I^+(p)$ for all $n$ (Prop.
\ref{padher}), and notice that $r_n \ll q$ up to a finite number
of $n$ (Prop. \ref{pglobal}). Thus, $\{r_n\}_n \subset J^+(p)\cap
J^-(q)$, but  converges to a point out of this compact subset, a
contradiction.
\end{proof}

\begin{remark} \label{rdiscgh-strongc} 
(1) As stressed in \cite{bernal07}, the full consistency of the
causal ladder yields that any globally hyperbolic spacetime is not
only causally simple but also strongly causal. This last
hypothesis is usually imposed in the definition of global
hyperbolicity, instead of  causality, 
but becomes  somewhat redundant. Notice that causality does not
follow from property (b) and cannot be weakened. Indeed, there are
chronological non-causal spacetimes which satisfy it (see Fig.
\ref{T3N8}). 

(2) The open subset $M=\{(t,x)\in \L^2: 0<x \}$ shows that a
causally simple spacetime may be non-globally hyperbolic.
\end{remark}

Notice that the two conditions in Definition \ref{dgh} are natural
from the physical (even philosophical) viewpoint:

\ben \item 
Causality avoids paradoxes derived from trips to
the past (grandfather's paradox). \index{grandfather's paradox}
For example, one cannot ``send a laser beam which describes a
causal loop in the spacetime and kills him/herself''.

\item 
The compactness of the diamonds \index{causal
diamonds} $J^+(p)\cap J^-(q)$ can be interpreted as ``there are no
losses of information/energy in the spacetime''. In fact,
otherwise one can find a sequence $\{r_n\}_n \subset J^+(p)\cap
J^-(q)$ with no converging subsequence. Taking a sequence of
causal curves $\{\bm{\gamma}_n\}_n$, each one joining $p, r_n, q$,
the limit curve
$\bm{\gamma}_p$ starting at $p$ cannot reach $q$.
  This can be interpreted as something which is suddenly lost or created in
the boundary of the spacetime (see Fig. \ref{example1}). That is,
a {\em singularity} (this sudden loss/creation) is visible from
$q$ --there are ``naked singularities''. \index{naked
singularities}\een

\subsubsection{Cauchy hypersurfaces and Geroch's theorem}
\index{Cauchy hypersurface} \index{Cauchy temporal function}


Recall that a subset $A\subset M$ is called {\em achronal}
\index{achronal subset} (resp. acausal) \index{acausal subset} if it
is not crossed twice by any timelike (resp. causal) curve. The
following notions are useful in relation to Cauchy hypersurfaces.

\begin{definition}  Let  $A$ be an achronal
subset  of a spacetime $(M,\g)$. \bit \item The {\em domain of
dependence of $A$} \index{domain of dependence} is defined as
$D(A)=D^+(A)\cup D^-(A)$, where $D^+(A)$ (resp. $D^-(A)$) is
defined as the set of points $p\in M $ such that every past (resp.
future) inextendible causal curve through $p$ intersects $A$.

\item {\em The Cauchy horizon \index{Cauchy horizon} of $A$} is
defined as $H(A)=H^+(A)\cup H^-(A)$, where $H^+(A)= \bar
D^+(A)\backslash I^-(D^+(A)) =\{ p\in \bar D^+(A): I^+(p)$ does
not meet $D^+(A)\}$, and $H^-(A)$ is defined dually. \eit
\end{definition}
One can check that, if $A$ is a closed subset, then $\dot{D}^+(A)=
A\cup H^+(A)$. Recall that $D(A)$ can be interpreted as the part of
the spacetime predictable from $A$. A Cauchy hypersurface is defined
as an achronal subset from where the full spacetime is predictable:

\begin{definition}
 A {\em Cauchy hypersurface} of a spacetime $(M,\g)$ is, alternatively:
 \bit \item[(i)] A subset $S\subset M$ which is
intersected exactly once by any inextendible timelike curve.
\item[(ii)] An achronal subset $S$, with $D(S)=M$. \item[(iii)] An
achronal subset $S$, with $H(S)=\emptyset$. \eit
\end{definition}
Some properties of any such Cauchy hypersurface $S$ are the
following: \ben \item  Necessarily, $S$ is a closed subset and an
embedded topological hypersurface. \item The spacetime $M$ is the
disjoint union $M= I^-(S) \cup S \cup I^+(S)$.  \item Any
inextendible causal curve $\gamma$ crosses $S$ and, if $S$ is
spacelike (at least $C^1$) then $\gamma$ crosses $S$ exactly once
(in general $S$ may be non-acausal because $\gamma$ may intersect
$S$  in a segment, i.e., in the image of an interval $[c,d],
c<d$). \item If $K$ is compact then $J^\pm(K)\cap S$ is
compact.\een In what follows, a function $t:M\rightarrow \R$ (in
particular, a time or temporal one, according to Definition
\ref{dtime}) will be called {\em Cauchy} \index{Cauchy function}
if its levels $S_c= t^{-1}(c)$ are Cauchy hypersurfaces; without
loss of generality, we can assume that Cauchy functions are onto.
Notice that the levels of a Cauchy time function \index{Cauchy
time function} are necessarily {\em acausal} Cauchy hypersurfaces.

The characterization of global hyperbolicity in terms of Cauchy
 hypersurfaces comes from the following celebrated Geroch's theorem
 \cite{geroch70}.

\begin{theorem} \label{tgeroch}
$(M,\g $) is globally hyperbolic if and only if it admits a Cauchy
hypersurface $S$.

Even more, in this case: (i) the spacetime admits a Cauchy time
function.
 (ii)  all Cauchy hypersurfaces are homeomorphic to $S$,
and $M$ is homeomorphic to $\R \times S$.
\end{theorem}
The implication to the left is a (non-trivial) standard
computation written in many references (for example,
\cite{oneill83, wald84}).
 For the
implication to the right and the last assertion, recall first the
following result:

\begin{lemma}\label{l516}
In a globally hyperbolic spacetime, the continuous function \be
\label{ft} t(p)=\log \left(-\frac{t^-(p)}{t^+(p)}\right) =\log
\left(\frac{m(I^-(p))}{m(I^+(p))}\right)\ee satisfies: \be
\label{elt} \lim_{s\rightarrow a}t(\gamma(s))=-\infty, \quad \quad
\lim_{s\rightarrow b}t(\gamma(s))=\infty \ee for any inextendible
future-directed causal curve $\gamma: (a,b)\rightarrow M$.
\end{lemma}
\begin{proof} It is sufficient to check:
$$ \lim_{s\rightarrow a}t^-(\gamma(s))= 0, \quad  \lim_{s\rightarrow b}t^+(\gamma(s))=0.$$
Reasoning for the former,  it is enough to show that, fixed any
compact subset $K$,  then $K\cap I^-(\gamma(s_0)) =\emptyset $ for
some $s_0\in (a,b)$ (and, thus, for any $s<s_0$), see
\cite{sanchez05b} for details.  Choose any point on the curve, $q=
\gamma(c)$ for some $ c\in (a,b)$, and assume by contradiction the
existence of a sequence $p_j=\gamma(s_j), s_j\rightarrow a, s_j\in
(a,c)$, with an associate sequence $r_j \in K \cap I^-(p_j)$. Up to
a subsequence, $\{r_j\}\rightarrow r$, and choosing $p\ll r$, one
has $p\ll p_j\le q$, and $\gamma|_{(a,c]}$ lies in the compact
subset $J^+(p)\cap J^-(q)$. That is, $\gamma$ is totally imprisoned
to the past, in contradiction with strong causality, (see
Proposition \ref{pimpr}).
\end{proof}


\begin{proof} (Only Th. \ref{tgeroch} $\Rightarrow$.) As $t$ in Lemma \ref{l516} is a time function, each level  $S_c$ is an acausal
hypersurface. In order to check that any inextendible timelike
curve $\gamma$ crosses $S_c$ (thus proving (i)), recall that
$\gamma$ can be reparametrized on all $\R$ with $t$, and
(\ref{elt}) will also hold under any increasing continuous
reparametrization of $\gamma$. Thus, assuming that this
reparametrization has been carried out, $\gamma(c) \in S_c$.

For  assertion (ii),  it is enough to choose a complete timelike
vector field $X$, (Prop. \ref{pot}) and project the full spacetime
onto $S$ by using its flow.
\end{proof}

\subsubsection{The folk questions on smoothability and the global orthogonal splitting}
The statements of the results in Geroch's theorem and its proof,
suggest obvious problems on the  smoothability of $S$ and $t$.
\index{smoothability problems} In fact, these questions were
regarded as ``folk problems'' because, on one hand, some proofs were
announced and rapidly cited (see \cite[Section 2]{bernal04b} for a
brief account) and, on the other, smoothability results yield useful
simplifications and applications commonly employed. Nevertheless,
they have remained fully open until very recently.

\begin{remark} \label{r2cojones}
The solution to the problems on smoothability in \cite{bernal03,
bernal04, bernal06} involves  technical procedures very different
to the expected approaches in previous attempts. These approaches
can be summarized as:

(a) To smooth the Cauchy hypersurface $S$ or the (Cauchy) time
function $ t$ by using covolution \cite{seifert77}. The difficulty
comes from the fact that, even when $S, t$ are smooth, the tangent
to $S$ or the gradient of $t$ may be degenerate, that is, close
hypersurfaces or functions to $S, t$ may be non-Cauchy or non-time
functions. Therefore, $S, t$ must be smoothed by taking into account
that a $C^\infty$ approximation may be insufficient.

(b) To choose an admissible measure $m$ such that the volume
functions $t^+, t^-$ are directly not only continuous but also
smooth \cite{dieckmann88b}. Nevertheless, notice that {\em those
stably causal spacetimes which are not causally continuous, cannot
admit continuous $t^+, t^-$}, but they do admit temporal time
functions (Th. \ref{testcaus}).
\end{remark}
As a summary on these questions, assume that $(M,\g)$ be globally
hyperbolic:

\smallskip

\noindent {\em 1.- Must a (smooth) spacelike Cauchy hypersurface
${\cal S}$ exist?} This is the simplest smoothability question,
posed explicitly by Sachs and Wu in their review \cite[p.
1155]{sachs77}. One difficulty of this problem (which makes useless
naive approaches based on covolution) is the following. Even if a
Cauchy hypersurface $S$ is smooth at some point $p$, the tangent
space $T_pS$ may be degenerate; so, the smoothing procedure of $S$
must ``push'' $T_pS$ in the right spacelike direction.

The existence of one such ${\cal S}$ implies that the spacetime is
not only homeomorphic but also diffeomorphic to $\R \times {\cal
S}$. Physical applications appear because spacelike Cauchy
hypersurfaces are essential for almost any global problem in
General Relativity (initial value problem for Einstein equation,
singularity theorems, mass...), see \cite{sanchez06b}. For
example, from the foundational viewpoint, they are necessary for
the well-posedness of the initial value problem, as there is no a
general reasonable way to pose well these conditions if the Cauchy
hypersurface is not spacelike (or, at least, smooth).

\begin{figure}[ht]
\centering \psfrag{A}{$t$} \psfrag{D}{$S$} \psfrag{F}{$S_1$}
\psfrag{B}{$S_2$} \psfrag{Y}{$\!\!\!$(A)}
\psfrag{W}{$\!\!\!\!\!$(B)} \psfrag{C}{$N$} \psfrag{E}{$N$}
\psfrag{L}{$\mathbb{L}^2$} \psfrag{H}{$\!\!\!\!\!\!D(N)$}
\includegraphics[width=10cm]{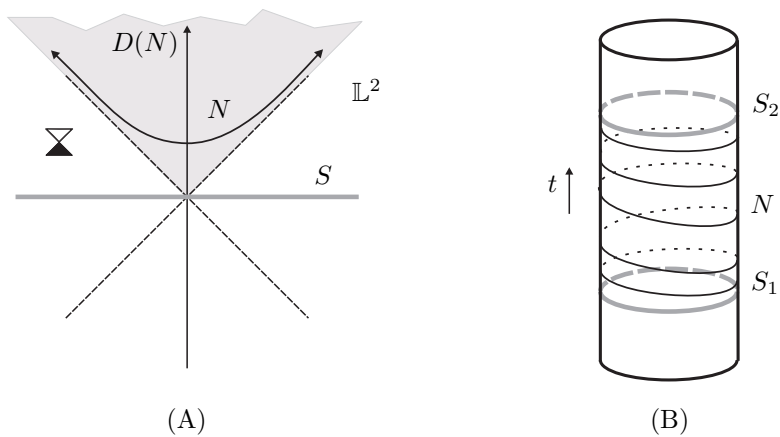}
\caption{(A). The embedded spacelike hypersurface $N$ is achronal.
because it lies in $I^+(S)$. But it is not (extendible to) a
spacelike Cauchy hypersurface.  (B). Now, as $N\subset M$ lies
between two disjoint Cauchy hypersurfaces $S_1, S_2$, it would be
a Cauchy hypersurface if it were closed.
 } \label{viena6}
\end{figure}

This smoothability problem was solved in \cite{bernal03}. The idea
starts recalling  the following result, interesting in its own
right (see also \cite{galloway85}):
\begin{quote}
Let $S$ be a Cauchy hypersurface. If a closed subset $N \subset M$
is a embedded spacelike (at least $C^1$) hypersurface which lies
either in $I^+(S)$ or in $I^-(S)$ then it is achronal. If $N$ lies
between two disjoint Cauchy hypersurfaces $S_1, S_2$ ($N\subset
I^+(S_1) \cap I^-(S_2)$) then it is a Cauchy hypersurface (see
Fig. \ref{viena6}).
\end{quote}
Thus, as Geroch's theorem ensures the existence of such $S_1, S_2$,
the crux is to find a smooth function $t$ with a regular value $c$
such that $S_c=t^{-1}(c)$ lies between $S_1$ and $S_2$, and $\nabla
t$  is timelike on $S_c$.

\smallskip

\noindent {\em 2.- Must a  Cauchy temporal function ${\cal T}$
exist?} This question is  relevant not only as a natural extension
of Geroch's, but in much more depth, because in the affirmative
case the smooth splitting $\R \times {\cal S}$ of the spacetime
can be strengthened in such a way that the metric has no cross
terms between $\R$ and ${\cal S}$  (see (\ref{ghsplit}) below for
the explicit expression). This splitting is useful from practical
purposes and also to introduce different techniques (Morse theory
\cite{uhlenbeck75}, variational methods \cite{masiello94},
quantization...)

\begin{figure}[ht]
\centering \psfrag{S}{$S$} \psfrag{V}{$\!\!v$} \psfrag{U}{$\!\! u$}
\psfrag{P}{$\! p$} \psfrag{PE}{$\! p_{\epsilon}$}
\psfrag{J}{$\!\!\!\!\!\! J^{+}(p)$}
\includegraphics[width=6cm]{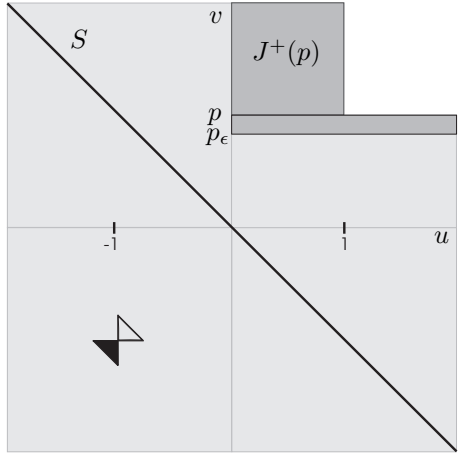}
\caption{$M \subset \L^2$, (coord. $u,v$). 
$M=\{(u,v)\in \L^2: |u|, |v| <2\}\backslash \{(u,v)\in \L^2: u, v
\geq 1\}$; $p=(0,1), p_\epsilon=(0,1-\epsilon)$. Diagonal $S$ is a
Cauchy hypersurface. For the natural $g$-measure, $t^+(p_\epsilon
) = 2 \epsilon + t^+(p)$ when $\epsilon>0$, and $t^+$  is not
smooth. } \label{fig3}
\end{figure}

Notice that the constructive proof of Geroch's Cauchy time
function \index{Geroch's Cauchy time function} may yield a
non-smooth one (Fig. \ref{fig3}). The freedom to choose an
admissible measure $m$ may suggest that, perhaps, a wise choice of
$m$ will yield directly a smooth Geroch's function. Nevertheless,
the related problem of smoothability in stably causal spacetimes
(Th. \ref{testcaus}) suggest that this cannot be the right
approach (in this case, even $t^\pm$ may be non-continuous). The
problem was solved affirmatively in \cite{bernal04} by different
means, based on the construction of ``time step functions''. We
also refer to \cite{sanchez05b} for a sketch of these ideas.

\smallskip

\noindent {\em 3.- If a spacelike Cauchy hypersurface  ${\cal S}$
is prescribed, does a Cauchy temporal function ${\cal T}$ exist
such that one of its levels is ${\cal S}$?} This question has
natural implications in classical General Relativity (even though
was proposed explicitly by B\"ar, Ginoux and Pfaffle in the
framework of quantization). For example, for the initial value
problem, one poses initial data on a prescribed hypersurface which
will be, a posteriori, a Cauchy hypersurface ${\cal S}$ of the
solution spacetime. Now, in order to solve Einstein equation, one
may assume that the spacetime will admit an orthogonal splitting
as (\ref{ghsplit}) below, with ${\cal S}$ one of the slices and
being $\beta$, and the evolved metric $g_\Tau$, the unknowns.

This problem was solved affirmatively in \cite{bernal06}. Notice
that even a non-smooth Cauchy (resp. acausal Cauchy) hypersurface
$S$ can be regarded as a level of a time (resp. Cauchy time)
function $t$ as follows. $I^+(S)$ and $I^-(S)$, regarded as
spacetimes, are globally hyperbolic and, thus, we can take  Cauchy
temporal functions ${\cal T}_{S^\pm}$ on $I^\pm(S)$. Now, the
required function is:

$$
t(p)= \left\{
\begin{array}{ll}
\exp({\cal T}_{S^+}(p)), & \forall p\in I^+(S) \\
0, & \forall p\in S \\
-\exp(-{\cal T}_{S^-}(p)), & \forall p\in I^-(S).
\end{array}
\right.
$$
Function $t$ is also  smooth (and a a Cauchy temporal function)
everywhere except at most  in $S= t^{-1}(0)$. Nevertheless
(replacing, if necessary, $t$ by a function obtained technically by
modifying $t$ around $S$), one can assume that $t$ is smooth even if
$S$ is not. Nevertheless, in this case the gradient of $t$ on $S$
will be 0 and, thus, $t$ will not be a true Cauchy temporal
function. Now, the crux is to show that, if $S$ is spacelike, then
it is possible to modify $t$ in a neighborhood of $S$, making its
gradient everywhere timelike, and maintaining its other properties.

\smallskip

\begin{figure}[ht]
\centering \psfrag{A}{$A$} \psfrag{P}{$p$} \psfrag{Q}{$q$}
\psfrag{P}{$p$}
\includegraphics[width=2.8cm]{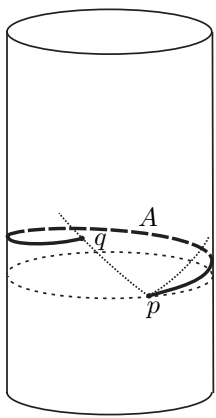}
\caption{The canonical Lorentzian cylinder ($\mathbb{R} \times
S^1$, $g = -\dd t^2 + \dd \theta^2$) with the spacelike
hypersurface $A = \{(\theta/2, \theta): \theta \in [0, 4\pi/3]\}$.
The spacelike achronal (but non-acausal) hypersurface $A$ can not
be extended to a smooth Cauchy hypersurface, although by adding
the null geodesic segment between $p$ and $q$ one obtains a
continuous Cauchy hypersurface $S_A$.} \label{cilindro}
\end{figure}

\noindent {\em 4.- Under which circumstances a spacelike
submanifold $A$ (with boundary) can be extended to a spacelike
(or, at least, smooth) Cauchy hypersurface?} As the previous
question, this one is solved in \cite{bernal06} and has a natural
classical meaning (but it was posed by Brunetti and Ruzzi
motivated by quantization).  Notice that an obvious requirement
for $A$ is achronality; moreover, compactness becomes also natural
(the hyperbola $t= \sqrt{x^2+1}$ would yield a counterexample, see
Fig. \ref{viena6}). Even more: {\em  any compact achronal  $K
\subset M$,  can be extended to a Cauchy hypersurface}. In fact,
$M'= M\backslash (I^{+}(K)\cup I^{-}(K))$ would be a (possibly
non-connected) globally hyperbolic spacetime and, then, would
admit a Cauchy hypersurface $S'$; the required Cauchy hypersurface
of $M$ would be $S_K=S' \cup K$. Nevertheless, the corresponding
Cauchy hypersurface $S_A$ for the (smooth, compact, achronal)
submanifold $A$, may be non-smooth and even non-smoothable, see
Figure \ref{cilindro}. But it is possible to prove that, {\em if
$A$ is not only achronal but also acausal, then $S_A$ can be
modified in a neighborhood of $\dot{A}$ to make it not only smooth
but also spacelike.}

\noindent As a summary of all these problems, it is possible to
prove:

\begin{theorem} \label{tgerochsmooth}
A spacetime $(M,\g)$ is  globally hyperbolic   if and only if it
admits a (smooth) spacelike Cauchy hypersurface ${\cal S}$.

In this case it  admits a Cauchy temporal function ${\cal T}$ and,
thus, it is isometric to the smooth product manifold
\begin{equation} \label{ghsplit} \R \times {\cal S}, \quad \langle
\cdot , \cdot \rangle = - \beta\,d\Tau^2 + g_\Tau
\end{equation}
where $\beta:\R \times {\cal S} \rightarrow \R$ is a positive smooth
function, $\Tau: \R\times {\cal S} \rightarrow \R$ the natural
projection, each   level  at constant $\Tau$, ${\cal S}_\Tau$, is  a
spacelike Cauchy hypersurface, and $ g_\Tau$ is a Riemannian metric
on each $S_\Tau$, which varies smoothly with $\Tau$.

Even more, if $S$ a prescribed (topological) Cauchy hypersurface
then there exists a {\em smooth} Cauchy function $\tau:
M\rightarrow \R$ such that $S$ is one of its levels ($S=S_0$). If,
additionally: \bit \item $S$ is also acausal  then function $\tau$
becomes a smooth Cauchy time function. \item  If $S$ is spacelike
(and thus smooth and acausal), then $\tau$ can be modified to
obtain a Cauchy temporal function ${\cal T}: M\rightarrow \R$ such
that $S={\cal T}^{-1}(0)$. \eit

Finally, if $A\subset M$ is a compact achronal subset then it can
be extended to a Cauchy hypersurface. If, additionally, $A$ is
acausal and a smooth spacelike submanifold with boundary, then it
can be extended to a spacelike Cauchy hypersurface ${\cal S}
\supset A$.
\end{theorem}

\subsubsection{The space of causal curves} \index{space of causal curves}
The first definition of global hyperbolicity was given by Leray
\cite{leray52}, and involves the compactness of the space of
causal curves which connects any two points. More precisely,
consider two events $p, q$ of the spacetime $(M,g)$, and let
$C(p,q)$ be
the set of all the continuous curves which are future-directed and
causal (according to Definition \ref{continc}) and  connect $p$
with $q$, under the convention in Remark \ref{rconv} i.e., two
such curves are regarded as equal if they differ in a strictly
monotonic reparametrization. For simplicity, $(M,g)$ will be
assumed to be causal, and we will consider the $C^0$
topology\footnote{This sense of $C^0$ topology agrees with the
$C^0$-limit of curves, described in Definition
\ref{dlimitscurves}. Even though this notion of limit had
specially good properties for strongly causal spacetimes,  we will
not need a priori this hypothesis but only causality (recall also
that the two extremes of the curves are fixed). Nevertheless,  a
posteriori, we will work with globally hyperbolic spacetimes,
where strong causality  holds.} on $C(p,q)$, that is, a basis of
open neighborhood of $\bm{ \gamma} \in C(p,q)$ is constructed by
taking all the curves in $C(p,q)$ contained in an open
neighborhood $U$ of the image of $\gamma$.

\begin{theorem}
A spacetime $(M,\g)$ is globally hyperbolic if and only if:

(i) it is causal,
 and

(ii) $C(p,q)$ is compact for all $p,q \in M$.
\end{theorem}
\begin{proof}
$(\Leftarrow )$ Let $\{r_n\}_n$ be a sequence in $J^+(p) \cap
J^-(q)$ and ${\bm \gamma}_n$ be a causal curve from $p$ to $q$
trough $r_n$ for each $n$. Up to a subsequence $\{{\bm
\gamma}_n\}_n$ converges to a curve ${\bm \gamma} \in C(p,q)$. So,
chosen any neighborhood $U\subset M$ of ${\bm\gamma}$ with compact
closure $\bar U$, all ${\bm \gamma}_n (\ni r_n)$ lie in $U$ for
large $n$ and, up to a subsequence, $\{r_n\}\rightarrow r \in \bar
U$. But necessarily $r\in {\bm \gamma}
(\subset J^+(p)\cap J^-(q))$, 
as required.

$(\Rightarrow )$ See for example \cite[p. 208-9]{hawking73}.
\end{proof}

\begin{remark} 
In fact, hypothesis (i) is somewhat redundant, because it is
possible to define a natural topology on $C(p,q)$ even if the
spacetime is not causal. But in this case, if there were a closed
causal curve ${\bm \gamma}$, parametrizing ${\bm \gamma}$ by
giving more and more rounds, a sequence of (non-equivalent) causal
curves would be obtained, and the compactness assumption of
$C(p,q)$ would be violated for this natural topology.

\end{remark}

With this notion of global hyperbolicity at hand, it is not
difficult to prove the main properties of the time-separation $d$ of
a globally hyperbolic spacetime. Recall that $d$ is not conformally
invariant, but the properties below will be so.

\begin{lemma} \label{ldgh} Let $(M,g)$ be
globally hyperbolic and $p < q$. Consider sequences:
$$\{p_k\} \rightarrow p , \quad \{q_k\}  \rightarrow q , \quad p_k \leq
q_k$$ Then, for any sequence ${\bm \gamma_k}$ of causal curves,
each one from $p_k$ to $q_k$, there exists a limit  in the $C^0$
topology $\gamma$ which joins $p$ to $q$. \end{lemma}
\begin{proof}
Choose $p_1\ll p,$ and $q \ll q_1$ and, for large $n$, construct a
sequence  of causal curves $\{\rho_n\}_n$  starting at $p_1$,
going to $p_n$, running $q_n$ and arriving at $q_1$. Then, use the
compactness of $C(p_1,q_1)$.
\end{proof}

\begin{remark} From the properties in subsection
\ref{subsubseclimit}, $\gamma$ is also a {\em limit curve} of the
sequence, and $L(\gamma)\geq \overline{\lim}_m L(\gamma_k)$.
\end{remark}

\begin{theorem} In any globally hyperbolic spacetime $(M,g)$:
\bit \item[(1)]  $d$ is finite-valued \item[(2)] (Avez-Seifert
\cite{avez63, seifert67}).\index{Avez-Seifert result} Each two
causally related points can be joined by a causal geodesic which
maximizes time-separation.
 \item[(3)] $d$ is continuous.
  \eit

\end{theorem}

\begin{proof} (1) Cover $J^+(p)\cap J^-(q)$ with a
finite number $m$ of convex neighbourhoods $U_j$ such that each
causal curve which leaves $U_j$ satisfies: (i) it never returns to
$U_j$, (ii) its lenght is $\leq 1$. Then $d(p,q)\leq m$.

(2)  Take a sequence of causal curves $\gamma_k$ with lengths
converging to $d(p,q)$ and use Lemma \ref{ldgh} (this also yields
an alternative proof of (1)).

(3)  Otherwise (taking into account that $d$ is always lower
semi-continuous) there are sequences $\{p_k\} \rightarrow p ,
\{q_k\} \rightarrow q , p_k \leq q_k$ with $$d(p_k,q_k) \geq
d(p,q) + 2\delta$$  for some $\delta>0$. Choose causal curves
$\gamma_k$ from $p_k$ to $q_k$ satisfying $$L(\gamma_k)\geq
d(p_k,q_k)-\delta .$$ Then the limit $\gamma$ yields the
contradiction:
$$L(\gamma)\geq \hbox{lim sup} L(\gamma_k) \geq d(p,q) +\delta >
 d(p,q). $$
 \end{proof}

\begin{remark}
(1) The finiteness of $d$ holds for all the time-separations of
metrics in $\g$. In fact, the following characterization is classical: 
{\em a strongly causal spacetime $(M,g)$ is globally hyperbolic if
and only if the time-separation $d^*$ of any metric $g^*$
conformal to $g$ is finite. }
To check it, notice that
when $(M,g)$ is not globally hyperbolic, there is a sequence
$\{\gamma_k\}_k \subset C(p,q)$ which has a limit curve $\gamma$
starting at $p$ with no final endpoint. The conformal factor must
be taken diverging  fast along a neighborhood of $\gamma$ (see
\cite[Th. 4.30]{beem96} for details).

(2) The existence of connecting causal geodesics in Avez-Seifert
\index{Avez-Seifert result} result can be made more precise: there
exists a $d-$maximizing geodesic in  each causal homotopy class and,
if $p\ll q$, there is also a maximizing timelike geodesic in  all
the timelike homotopy classes included in each causal homotopy
class, see the detailed study in \cite[Sect. 2]{minguzzi06}.
\end{remark}

\subsubsection{An application to closed geodesics and static spacetimes}

Next, we will see some simple applications of the properties of
globally hyperbolic spacetimes for the geodesics of some
spacetimes. We refer to \cite{sanchez05} for more results and
extended proofs, especially regarding  static spacetimes.

\begin{proposition}\label{ppp} If the universal covering $(\tilde M, \tilde g)$ of a totally vicious spacetime $(M,g)$ is globally
hyperbolic, then $(M,g)$ is geodesically connected through
timelike geodesics (i.e., each  $p,q \in M$ can be connected
through a timelike geodesic).
\end{proposition}
\begin{proof}
By lifting to $\tilde M$ any timelike curve $\rho$ which connects
$p, q $, one obtains two chronologically related points $\tilde p,
\tilde q \in \tilde M$. So, they are connectable by means of a
(maximizing) timelike geodesic $\tilde \gamma$, which projects in
the required one.
\end{proof}
Now, recall that a {\em static} spacetime is a stationary one such
that the orthogonal distribution to its timelike Killing vector
field $K$ is integrable. Locally, any static spacetime
\index{static spacetime} looks like  a {\em standard static}
spacetime i.e., the product $\R \times S$ endowed with the warped
metric $g=-\beta dt^2 + g_S$, where $g_S$ is a Riemannian metric
on $S$ and $\beta$ is a function which depends only on $S$. If $K$
is complete, any simply connected static spacetime is standard
static, in particular:

\begin{lemma} The universal covering $(\tilde M, \tilde \g )$ of a compact static spacetime is
standard static. \end{lemma}

These spacetimes have a good causal behaviour:

\begin{proposition} \label{t3a}
Any standard static spacetime $(M,g)$ is causally continuous, and
the following properties are equivalent:
\begin{itemize}
\item[(i)] $(M,g)$ is globally hyperbolic. \item[(ii)] The
conformal metric $g_S^* = g_S/\beta$ is complete. \item[(iii)]
Each slice $t=$constant is a Cauchy hypersurface.
\end{itemize}
In particular, the universal covering of a compact static
spacetime is globally hyperbolic.
\end{proposition}
\begin{proof} For the first assertion,  it is enough to prove past (and analogously
future) reflectivity $I^+(q) \subset I^+(p)  \Rightarrow I^-(p)
\subset I^-(q)$. Put $p=(t_p,x_p), q=(t_q,x_q)$. Assuming the
first inclusion, it is enough to prove $p_{-\epsilon} =
(t_p-\epsilon,x_p) \in I^-(q)$, for all $\epsilon >0$. As
$q_\epsilon := (t_q + \epsilon , x_q) \in I^+(p)$, there exists a
future-directed timelike curve $\gamma(s)=(s,x(s)), s\in [t_p,t_q
+ \epsilon]$ joining $p$ and $q_\epsilon$. Then, the
future-directed timelike curve
$\gamma_{-\epsilon}(s)=(s-\epsilon,x(s))$ connects $p_{-\epsilon}$
and $q$, as required.

The equivalences (i)---(iii) follows from standard computations
valid for warped product spacetimes \cite[Theorems 3.67,
3.69]{beem96}.

In particular,   a standard static spacetime will be globally
hyperbolic if $g_S$ is complete and $\beta$ is bounded (or at most
quadratic). These conditions hold in the universal covering of a
compact static spacetime, proving the last sentence.\end{proof}

 Thus, Proposition \ref{ppp}, \ref{t3a}, and Theorem
\ref{ttotvic} yields \cite{sanchez06}:

\begin{theorem}
Any compact static spacetime is geodesically connected through
timelike geodesics.
\end{theorem}

For closed geodesics, let us start with the following well-known
result by Tipler \cite{tipler79} (in Beem's formulation
\cite{beem96}), later extended by Galloway \cite{galloway84}.

\begin{theorem} \label{tviciohiperbolico} Any compact spacetime \index{compact spacetime}
$(M,g)$, regularly covered by a spacetime $(\tilde M, \tilde g)$
which admits a  compact Cauchy hypersurface $S$, contains a
periodic timelike geodesic. \index{periodic timelike geodesic}
\end{theorem}

\begin{proof}
 Take a  timelike loop $\gamma$
in $M$ and a lift $\tilde \gamma:[0,1]\rightarrow \tilde M$. Let
$\psi:\tilde M \rightarrow \tilde M$ be a deck transformation
which maps $\tilde\gamma(0)$ in $\tilde \gamma(1)$. The function
$f: S\rightarrow \R$ $p\rightarrow d(p,\psi(p))$ admits a maximum
$p_0$ (necessarily, $f(p_0)>0$). The  maximizing timelike geodesic
from $p_0$ to $\psi(p_0)$ projects not only onto a geodesic loop,
but also to a closed one (otherwise, a closed curve with bigger
length could be obtained by means of a small deformation).
\end{proof}

\begin{remark} The compactness of  $S$ cannot be removed (Guediri's counterexample, see \cite{guediri03} and references therein). Nevertheless, it can be replaced by the existence of a class
of conjugacy ${\cal C}$ of the fundamental group which contains a
timelike curve and satisfies one of the following two conditions
(see \cite{sanchez06}):

(a) ${\cal C}$ is finite, or

(b) The deck transformations satisfy a technical property of
compatibility with an orthogonal globally hyperbolic splitting
(roughly, $\phi(t,x)= (t+T_\phi , \phi^S(x))$ for some $T_\phi \in
\R$ and some automorphism $\phi^S$ of $S$), which is always
satisfied in the case of compact static spacetimes.
\end{remark}
Thus, this possibility (b) yields \cite{sanchez06}:

\begin{theorem} \label{closedtimelikestatic}
Any compact static spacetime admits a closed timelike geodesic.
\index{closed timelike geodesic}
\end{theorem}

\section{The ``isocausal'' ladder} \label{s4}
\index{isocausal ladder}

\subsection{Overview}
Up to now, the {\em causal structure} of a spacetime is related to
two notions: (a) its conformal structure, and (b) its position in
the
 causal hierarchy. Nevertheless, in order to understand ``when two
spacetimes share the same causal structure'' one can argue that
the first one is too restrictive, and the latter too weak. For
example: (a) most  modifications of a Lorentzian metric around a
point (say, any non-conformally flat perturbation of Minkowski
spacetime
 in a small neighbourhood) imply a different conformal structure; but,
one may have a very similar structure of future and past sets for
all points, and (b) all globally hyperbolic spacetimes belong to the
same level of the hierarchy, but clearly the causality of, say,
Lorentz-Minkowski and Kruskal spacetimes  behave in a very different
way. It is not easy to find an intermediate notion, because ``same
causal structure'' suggests ``same causal relations $\ll, <$'' and,
in any distinguishing spacetime, the  conformal structure is
determined by these relations (Prop. \ref{kjh}, Th. \ref{pk3}).

A fresh viewpoint was introduced by Garc\'{\i}a-Parrado and
Senovilla \cite{garciaparrado03, garciaparrado05} by taking into
account the following two ideas: (i) the definition  of most of
the levels of the standard causal hierarchy prevents a bad
behavior of some types of causal curves; thus, if the timecones of
a metric $g$ on $M$ are included in the timecones of another one
$g'$ ($g\prec g'$), then the causality of $g$ will be at least as
good as the causality of $g'$, and (ii) perhaps for some
diffeomorphisms $\Phi, \Psi$ of $M$ the pull-back metrics satisfy
$\Psi^*g \prec g' \prec \Phi^*g$; in this case (as the causality
of $g, \Psi^*g$, $\Phi^*g$ must be regarded equivalent), one says
that $g$ and $g'$ are ``isocausal''.

In this way, one introduces a partial (pre)order  in the set of
all the spacetimes, which was expected to refine the standard
causal ladder. Nevertheless, this new order was carefully studied
by Garc\'{\i}a-Parrado and S\'anchez \cite{garciaparrado05b}, who
observed that two of the levels of the standard ladder (causal
continuity and causal simplicity) were not preserved by it. Thus,
one obtains an alternative hierarchy of spacetimes, with common
elements but also with relevant differences and complementary
viewpoints. Next, we sketch this approach.

\subsection{The ladder of isocausality}

\begin{definition}  Let $V_i= (M_i,\g_i), i=1,2$, be two spacetimes. A diffeomorphism $\Phi: M_1\rightarrow M_2$  is a  \index{causal mapping} {\em causal mapping} if the
timecones of the pull-back metric $\Phi^*\g_2$ include the cones
of $\g_1$, and the time-orientations are preserved by $\Phi$. In
this case, we write $V_1\prec_\Phi V_2$, and $V_1\prec V_2$ will
mean that $V_1\prec_\Phi V_2$ for some $\Phi$.

The two spacetimes are {\em isocausal}, \index{isocausal
spacetimes} denoted $V_1\sim V_2$, if $V_1\prec V_2$ and $V_2\prec
V_1$.
\end{definition}

\begin{remark}
(1) Recall that if $V_1\sim V_2$ then $V_1\prec_\Phi V_2$ and
$V_2\prec_\Psi V_1$ for some diffeomorphisms $\Phi, \Psi$, but
perhaps $\Psi\neq \Phi^{-1}$.

(2) As in the case of conformal relations, one can consider, for
practical purposes, a single differentiable manifold $M$ in which
two time-oriented Lorentzian metrics $\g_1$, $\g_2$ are defined, and
study when the timecones of $\g_2 (\equiv \Phi^*\g_2)$ are wider
than the cones of $\g_1$ (and with agreeing time-orientations), i.e.
if the identity in $M$ is a causal mapping. Nevertheless, the
notation $\g_1 \prec \g_2$ means also the possibility that the
timecones of $\Phi^*\g_2$ are  (non-necessarily strictly) wider than
the cones of $\g_1$ for some $\Phi$.

(3) Even though the time-orientations can be usually handled in a
simple way, their role cannot be overlooked. In fact, it is not
difficult to find a Lorentzian manifold such that the two
spacetimes obtained by choosing different time-orientations are
not isocausal (see Fig. \ref{notimesim}).

(4) One can check that, {\em locally all the spacetimes are
isocausal} \cite[Theorem 4.4]{garciaparrado05} (but, obviously,
not necessarily conformal). This supports that the notion of
``Causality'' (which is appealing as a global concept) deals with
properties invariant by isocausality, not only by conformal
diffeomorphisms.

\begin{figure}[ht]
\begin{center} \psfrag{X}{$x$} \psfrag{T}{$t$} \psfrag{G}{$\gamma$}
\psfrag{A}{$\,\,x=-1$}\psfrag{B}{$\,x=1$}
 \includegraphics[width=7cm]{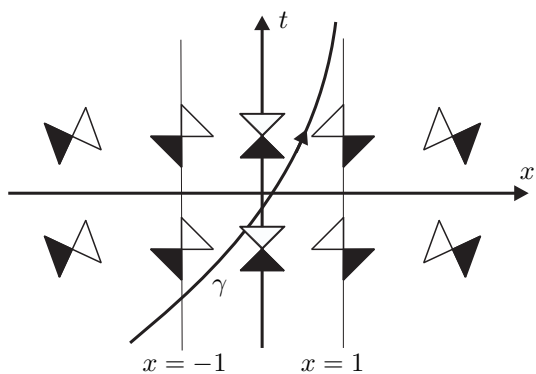}
\end{center}
\caption{This spacetime, which admits a ``black hole region'' $-1
\leq x \leq 1$, is not isocausal (nor, thus, conformal) with the
one obtained by reversing its time-orientation (which does not
admit such a region).} \label{notimesim}
\end{figure}

\end{remark}
Now, it is easy to check the following result:  \begin{theorem} If
$V_1\prec V_2$ and $V_2$ is globally hyperbolic, causally stable,
strongly causal, distinguishing, causal, chronological, or not
totally vicious, then so is $V_1$. \label{preservation}
\end{theorem}
\begin{proof}
This is an exercise recalling that: (i)  if $\bm{v}$ is causal (or
$\bm{\gamma}$ is a closed timelike or causal curve) then
$d\Phi(\bm{v})$ is causal (or $\Phi \circ \bm{\gamma}$ is a closed
timelike or causal curve), and (ii) if $d\Phi^{-1}(v')$ is
non-causal (or $t'\circ \phi$ is a time function; $\phi^{-1}(S')$
is a Cauchy hypersurface, etc.) then $v'$ is non-causal (or $t'$
is a time function; $S'$ is Cauchy, etc.), see
\cite{garciaparrado03} for details.
\end{proof}
The conditions appearing in this  result comprise all the levels
in  the  standard hierarchy of causality, except causally
continuous and causally simple. Nevertheless,  these levels are
not necessarily preserved. In fact, there is an explicit
counterexample \cite[Section 3.2]{garciaparrado05b} 
which shows that  $V_2$ may be causally simple and $V_1$
non-causally continuous, with $V_1\prec V_2$.

Now, fix a manifold $M$, and define the  {\em isocausal structure}
\index{isocausal structure} of the spacetime $(M,g)$ as its
equivalence class
 coset$(\g)$ in the quotient set Con$(M)/\sim$
($\equiv$ Lor$(M)/\sim$).  A partial order $\preceq$ in
Con$(M)/\sim$ can be defined by
$$
\mbox{coset}(\g_1)\preceq\mbox{coset}(\g_2) \Leftrightarrow
(M,\g_1)\prec(M,\g_2).
$$
 Isocausal structures can be naturally grouped in sets
totally ordered by ``$\preceq$'',  in the form
$$
\underbrace{\dots\preceq\mbox{coset}(\g_1)\dots\preceq
\mbox{coset}(\tilde\g_1)\dots}_{\mbox{glob. hyp.}}\preceq
\underbrace{\dots\preceq\mbox{coset}(\g_2)\preceq\dots}_{\mbox{causally
stable}}\preceq
\underbrace{\dots\mbox{coset}(\g_m)\preceq\dots}_{\dots\,\,
\dots}.
$$
Of course, some of the groups in a totally ordered chain may be
empty; for example, if $M$ were compact no chain would contain
chronological spacetimes. Furthermore the relation ``$\preceq$''
is not a total order and so a globally hyperbolic spacetime need
not be related to, say, a causally stable spacetime (see
\cite{garciaparrado05b} for exhaustive examples). Nevertheless,
except for the two excluded levels (causal continuity and
simplicity) relation $\preceq$ yields a refinement of the standard
causal hierarchy, introducing further relations between elements
of each level.

\subsection{ Some  examples}

In order to study the possible isocausality of two spacetimes,
there are two basic naive ideas (see \cite{garciaparrado05b}):

\ben \item In order to prove $V_1 \prec V_2$. 
Try to find an explicit causal mapping. For example, consider two
Generalized Robertson-Walker (GRW) spacetimes on the same manifold
$M$, that is $M$ is a warped product $I\times_{f_i} S$, where
$I\subset \R$ is an interval, $S$ is a  manifold endowed with a
positive definite Riemannian metric $g_S$ and, with natural
identifications:
$$
g_i= -dt^2 + f^2_i(t) g_S .
$$
Now, assume that $S$ is compact (i.e., the GRW spacetime is
closed) and $I$ is unbounded. Then, it is not difficult  to check
that they are isocausal if both warping functions $f_i$ are
bounded away from 0 and $\infty$, that is $ 0< \mbox{ Inf}(f_i)
\leq \mbox{ Sup}(f_i) <\infty$ for $ i=1,2.$ In fact, a causal
mapping type $(t,x)\rightarrow (\varphi(t),x)$ can be found
easily.

\item In order to prove $V_1 \not\prec V_2$. 
Try to find a causal invariant which would be transferred by the
causal mapping (or its inverse), but  not shared by both spacetimes.
In fact, this is the reason why $V_1 \not\prec V_2$ if $V_2$
lies higher than $V_1$ 
in the standard ladder of causality (with causal continuity and
simplicity removed). In this sense, criteria  as the following are
useful:

{\bf Criterion}. {\em Assume that $V_1\prec V_2$ and that $V_1$
admits $j$ inextendible future-directed causal curves (or, in
general, $j$ submanifolds at no point spacelike and closed as
subsets of $V_1$) $\gamma_i, i=1, \dots , j$ satisfying either of
the following conditions ($\downarrow$ will denote the common
chronological past of all the points of the corresponding subset):

\begin{enumerate}
\item $V_1= I^+({\bm \gamma}_i) \cup {\bm \gamma}_i \cup I^-({\bm
\gamma}_i)$. \item ${\bm \gamma}_i\subset\downarrow {\bm
\gamma}_{i+1}$, $\forall i=1, \dots , j-1$, $j>1$.
\end{enumerate}
 Then so does $V_2$.}

In fact, if $\Phi: V_1 \rightarrow V_2$ is the causal mapping, the
sets $\Phi({\bm \gamma}_i)$, $i=1,\dots,j$ satisfy condition 1 in
$V_2$ whenever ${\bm \gamma}_i$, $i=1,\dots,j$ do in $V_1$. To
prove the second point use the straightforward property:
$$
\Phi(\downarrow A)\subset\downarrow\Phi(A),\ A\subset V_1,
$$
as required.

 As a simple application, it is easy to show
that there are infinitely many  rectangles of $\L^2$, in standard
Cartesian coordinates $(t,x)$, which are not isocausal %
(see \cite[Figure 5]{garciaparrado05b})\footnote{ Notice also
that, as these rectangles are neither conformal, this also
re-proves the existence of infinitely many different
simply-connected conformal Lorentz surfaces \index{Lorentz
surfaces} (in contrast with the Riemannian case), stressed by
Weinstein \cite{weinstein96}.}. \een

By using these type of arguments one can study the isocausal
structure of GRW spacetimes, obtaining as a typical result:

\begin{theorem} \label{tclasifGRW} Consider any GRW spacetime $V= I\times_f
S, I \subset \R$ with $S$ diffeomorphic to a $(n-1)-$sphere. Then
$V$ is isocausal to one and only one of the following types of
product spacetimes:
\begin{enumerate}
\item $\R \times  \S^{n-1}$, i.e., Einstein static universe, with
metric
$$
g=-dt^2+ g_0,
$$
where $g_0$ represents the metric of the unit ($n-1$)-dimensional
sphere. \item $]0,\infty[ \times \S^{n-1}$ with metric as in the
case 1.
\item $]-\infty,0[ \times \S^{n-1}$. The metric is as in the case 1.

\item $]0,L[ \times \S^{n-1}$, for some $L>0$.
\end{enumerate}
 Moreover, causal
structures belonging to the above cases can be sorted as follows
$$
\mbox{\em coset}(\g_{4}(L))\preceq\left\{\begin{array}{c}
              \mbox{\em coset}(\g_{2})\\
               \mbox{\em coset}(\g_{3})\end{array}\right\}
\preceq\mbox{\em coset}(\g_{1}),
$$
where the roman subscripts mean that the representing metric
belongs to the corresponding point of the above description.
\end{theorem}

Finally, it is worth pointing out the following question regarding
stability \index{stable isocausality} 
(recall Section
\ref{subs-stably}), also stressed in \cite{garciaparrado05b}. In
the three first cases of Theorem \ref{tclasifGRW}, all the
spacetimes are isocausal to a fixed one and, thus, the isocausal
structure is  $C^0$-stable
 in the set of all the metrics on
$I\times S$. Nevertheless, in the last case there are different
isocausal structures. And, in fact, classical de Sitter spacetime
$\S^n_1$ lies in this case and has   a {\em $C^r$-unstable}
isocausal structure  for any $r\geq 0$ (very roughly, a criterion
as the one explained above is applicable to $\S^n_1$, but the
number of curves $\gamma_i$ in this criterion varies under
appropriate arbitrarily small $C^r$ perturbations). In contrast
with this case, the isocausal structure of Lorentz-Minkowski
$\L^n$ is again stable. In fact, a simple computation shows that
any $\g$ on $\R^n$ becomes isocausal to $\L^n$ if it satisfies:
(i) $\partial_t$ is a $\g$-timelike vector field, and (ii) there
exists $0 < \theta_- \leq \theta_+ <\pi/2$ such that the Euclidean
angle $\theta$ (for the usual Euclidean metric in $\R^n$) of any
$\g$-lightlike tangent vector and $\partial_t$ satisfies:
$\theta_- \leq \theta \leq \theta_+$. Summing up:

\begin{theorem} $\empty$
\begin{itemize}
\item[(1)] The isocausal structure of Lorentz Minkowski $\L^n$ is
stable in the $C^0$ (and, thus, in any $C^r$) topology.
\item[(2)] The isocausal structure of $\S_1^n$ is unstable in  any
$C^r$ topology.
\end{itemize}
\end{theorem}
The first result goes in the same direction that Christodoulou and
Klainerman's landmark result \cite{christodoulou93}, who proved
the nonlinear stability of four-dimensional Minkowski spacetime (a
small amount of gravitational radiation added in the initial data
of $\L^n$  will disperse to  infinity without any singularities or
black holes being formed). The second one suggests that the
isocausal structure of de Sitter spacetime cannot be regarded as a
physically reasonable one.

\section*{Acknowledgments}

Careful reading by Prof. J.M.M. Senovilla and A.N. Bernal is
acknowledged. One of the authors (EM) partially supported by GNFM of
INDAM and by MIUR under project PRIN 2005 from Universit\`a di
Camerino; the other author (MS) partially supported by MEC-FEDER
Grant MTM2007-60731 and  J. Andal. Grant P06-FQM-01951.

\section*{References}

\renewcommand{\refname}{}    

\frenchspacing


%
{\footnotesize
\begin{theindex}

  \item 2-dimensional spacetime, 16, 38

  \indexspace

  \item  curvature tensor, 15

  \indexspace

  \item acausal subset, 43
  \item achronal subset, 43
  \item admissible measures, 30
  \item Alexandrov's topology, 27
  \item arbitrarily small, 8
  \item Avez-Seifert result, 50, 51

  \indexspace

  \item Carter's ``virtuosity'', 39
  \item Cauchy function, 44
  \item Cauchy horizon, 43
  \item Cauchy hypersurface, 43
  \item Cauchy temporal function, 43
  \item Cauchy time function, 44
  \item causal boundary, 12
  \item causal cones, 4
  \item causal diamonds, 43
  \item causal future, 7
  \item causal ladder, 18, 19
  \item causal mapping, 53
  \item causal relations, 7
  \item causal spacetimes, 21
  \item causal structure, 6
  \item causal vector or curve, 3
  \item Causality Theory, 18
  \item causally continuous spacetimes, 39
  \item causally convex neighborhood, 8
  \item causally related events, 7
  \item causally simple spacetimes, 40
  \item chronological future, 7
  \item chronological spacetimes, 21
  \item chronologically related events, 7
  \item closed curve, 18
  \item closed timelike geodesic, 52
  \item compact spacetime, 19, 21, 52
  \item conformal class of metrics, 7
  \item conformal Killing vector field, 19
  \item conformal spacetimes, 6
  \item conjugated events, 14
  \item connecting curve, 7
  \item conventions, 7
  \item convex neighborhood, 8

  \indexspace

  \item distinguishes $p$ in $U$ (future, past), 22, 23
  \item distinguishing spacetime, 22, 23
  \item distinguishing subsequence (limit of curves), 29
  \item domain of dependence, 43

  \indexspace

  \item endpoint, 7
  \item Euler characteristic, 4, 38
  \item event, 5

  \indexspace

  \item fine topology on the metric space, 36
  \item future reflecting spacetimes, 34
  \item future set, 10
  \item future-directed vector, 4

  \indexspace

  \item generalized time function, 35
  \item geodesic ray, 30
  \item Geroch's Cauchy time function, 47
  \item globally hyperbolic neighborhood, 9
  \item globally hyperbolic neighborhoods, 9
  \item globally hyperbolic spacetimes, 42
  \item grandfather's paradox, 18, 43

  \indexspace

  \item Hausdorff (topology), 28
  \item hierarchy of spacetimes, 18
  \item horismos, 7
  \item horismotically related events, 7, 15

  \indexspace

  \item imprisoned curve, 29
  \item inextendible curve, 7
  \item inner continuity, 32
  \item interval topology on the metric space, 36
  \item invariant under conformal transformations, 17
  \item isocausal ladder, 53
  \item isocausal spacetimes, 53
  \item isocausal structure, 54

  \indexspace

  \item Jacobi class, 16
  \item Jacobi equation, 15
  \item Jacobi field, 15

  \indexspace

  \item K-relation, 11
  \item Killing vector field, 19

  \indexspace

  \item lightlike geodesic, 14
  \item lightlike vector or curve, 3
  \item Lipschitzianity of causal curves, 25
  \item loop, 18
  \item Lorentz surfaces, 56
  \item Lorentz-Minkowski spacetime, 9
  \item Lorentzian distance, 12
  \item lorentzian length, 12
  \item Lorentzian manifold, 3
  \item lower semi-continuous (Lor. distance), 13

  \indexspace

  \item maximizing lightlike curve, 14
  \item multiplicity of Jacobi fields, 17

  \indexspace

  \item naked singularities, 43
  \item non-imprisoning spacetime, 28
  \item non-totally vicious spacetimes, 18
  \item nonspacelike vector, 3
  \item normal neighborhood, 8
  \item null vector, 3

  \indexspace

  \item orientable manifold, 5
  \item orthonormal basis, 4
  \item outer continuity, 32

  \indexspace

  \item partially imprisoned curve, 29
  \item past reflecting spacetimes, 34
  \item past set, 10
  \item past-directed vector, 4
  \item periodic curve, 18
  \item periodic timelike geodesic, 52
  \item pointwise conformal metrics, 6
  \item pointwise conformal spacetimes, 6
  \item pregeodesic, 14

  \indexspace

  \item quasi-limit, 29
  \item quotient Jacobi equation, 16
  \item quotient space of vector fields $Q$, 16

  \indexspace

  \item reflecting spacetimes, 34
  \item regular measures, 31
  \item reversed triangle inequality, 13

  \indexspace

  \item scalar product, 4
  \item set of parts ${\cal P}(M)$, 22
  \item simple neighborhood, 8
  \item smoothability problems, 45
  \item Sobolev space of causal curves, 25
  \item space of causal curves, 49
  \item spacelike  vector or curve, 3
  \item spacetime, 5
  \item stable isocausality, 56
  \item stable spacetime property, 37
  \item stably causal spacetimes, 36
  \item stably chronological spacetimes, 37
  \item static spacetime, 51
  \item stationary spacetime, 19
  \item strictly causally related events, 7
  \item Strong Cosmic Censorship Hypothesis, 18
  \item strongly causal spacetimes, 26

  \indexspace

  \item temporal function, 35
  \item time function, 35
  \item time-orientable double covering, 4
  \item time-orientable manifold, 4, 5
  \item time-orientation, 4
  \item time-separation, 12
  \item timelike vector or curve, 3
  \item totally vicious, 19

  \indexspace

  \item volume functions, 30

\end{theindex}


\end{document}